\documentclass{article}
\usepackage{amssymb}

\input{tcilatex}
\begin{document}

\begin{center}
\bigskip {\Large GRASSMANN\ PHASE\ SPACE\ THEORY}\bigskip

for the\bigskip

{\Large BEC/BCS CROSSOVER }\bigskip

in\bigskip

{\Large COLD\ FERMIONIC\ ATOMIC\ GASES}

\bigskip

B. J. Dalton and N. M. Kidwani\medskip

\bigskip Centre for Quantum Technology Theory, Swinburne University of
Technology, Melbourne, Victoria 3122, Australia\bigskip
\end{center}

\section{Abstract}

Grassmann Phase Space Theory (GPST) is applied to the BEC/BCS crossover in
cold fermionic atomic gases and used to determine the evolution (over either
time or temperature) of the Quantum Correlation Functions (QCF) that
specify: (a) the positions of the spin up and spin down fermionic atoms in a
single Cooper pair and (b) the positions of the two spin up and two spin
down fermionic atoms in two Cooper pairs The first of these QCF is relevant
to describing the change in size of a Cooper pair, as the fermion-fermion
coupling constant is changed through the crossover from a small Cooper pair
on the BEC side to a large Cooper pair on the BCS side. The second of these
QCF is important for describing the correlations between the positions of
the fermionic atoms in two Cooper pairs, which is expected to be small at
the BEC or BCS sides of the crossover, but is expected to be significant in
the strong interaction unitary regime, where the size of a Cooper pair is
comparable to the separation between Cooper pairs. In GPST the QCF\ are
ultimately given via the stochastic average of products of Grassmann
stochastic momentum fields, and GPST shows that the stochastic average of
the products of Grassmann stochastic momentum fields at a later time (or
lower temperature) is related linearly to the stochastic average of the
products of Grassmann stochastic momentum fields at an earlier time (or
higher temperature), and that the matrix elements involved in the linear
relations are all c-numbers. Expressions for these matrix elements
corresponding to a small time increment (or a small temperature change) have
been obtained analytically, providing the formulae needed for numerical
studies of the evolution that are planned for a future publication. Various
initial conditions are considered, including those for a non-interacting
fermionic gas at zero temperature and a high temperature gas (where the
effect of the interactions can be ignored in the initial state). These would
be relevant for studying the time evolution of the creation of a Cooper pair
when the interaction is switched on via Feshbach resonance methods, or the
evolution as the temperature is lowered for particular choices of the
fermion-fermion coupling constant, such as where the coupling constant is
very large - corresponding to the unitary regime. Full derivations of the
expressions have been presented in the Appendices.\pagebreak

\section{Introduction}

\label{Section - Introduction}

Research into ultracold atomic gases \cite{Chin16a} has been a major
activity since the 1990's when Bose-Einstein condensation was achieved for
bosonic atoms. Non-interacting untrapped bosonic atoms at zero temperature
form a BEC, with a macroscopic occupancy of the lowest single particle
energy state, since there is no Pauli exclusion principle that forbids
multiple occupancy. Since the 2000's ultracold gases of fermionic atoms have
been prepared, and here quite different effects occur. Non-interacting
untrapped fermionic atoms at zero temperature form a Fermi gas, with each
energy state only being occupied by atoms with different spins due to the
Pauli exclusion principle. Consequently, energy states are filled up to the
Fermi energy $E_{F}$, whose associated wave number $k_{F}$ is proportional
to the inverse of the average separation between the atoms. In both cases
the single particle states are plane waves with momentum $\hslash k$.

When interactions are taken into account there are features in common though
-- for both types of atom the interaction potential only has a range b of $%
O(10^{-9}m)$, whilst the typical separation between atoms in these dilute
gases is of $O(5\times 10^{-7}m)$ making it unlikely that more than two
atoms interact at once. The interaction between a pair of atoms determines
the scattering length $a$ -- which is related to the collision cross
section. However, recent advances involving Feshbach resonances have made it
possible to modify the scattering length \cite{Chin10a} (for example by
varying a static magnetic field that is coupled to the atoms so that an
unbound state in an open scattering channel coincides with a bound state in
a closed channel) so that its value ranges from negative to positive across
the resonance -- where (if decay processes are ignored) it would become
infinite. At this resonance (or unitary) situation the strong interaction
regime applies, with different effects to the weak interaction regimes well
away from resonance on either side.

One of the interesting phenomena involving interacting ultracold fermionic
atomic gases is the BEC/BCS\ crossover \cite{Ketterle08a}, \cite{Zwerger12a}%
, \cite{Randeria14a}, \cite{Strinati18a}. If the atoms are essentially
unconfined (as in a weak trapping potential) and at zero temperature the
weak interaction regime differs for bosons and fermions. For the simplest
bosonic case of spin $0$ atoms the picture is that each atom behaves
essentially independently, with the effect of the interatomic potentials
being reflected in a mean field based on the average density of all the
other atoms. The behaviour is treated using the Gross-Pitaevskii equation 
\cite{Pitaevski03a}. However, for the simplest fermionic case of spin $%
{\frac12}%
$ atoms there is the possibility of two atoms with opposite spins pairing up
and behaving as a single entity. The behaviour can be described using the
BCS-Leggett theory \cite{Leggett80a}, \cite{Ketterle08a}, (also a mean field
theory). Here though there are two extreme possibilities depending on
whether the two-body scattering length $a$ is positive or negative. For $%
1/(k_{F}a)<<0$ superfluid BCS behaviour occurs based on loosely bound Cooper
pairs of atoms with opposite spins and momenta (analogous to the Cooper
pairs of electrons in metallic superconductors). For $1/(k_{F}a)>>0$ a BEC
forms based on tightly bound pairs of atoms with opposite spins and momenta
that constitute a spin $0$ bosonic molecule. These regimes involve weakly
interacting fermions. By using Fesbach resonance methods, the BEC/BCS
crossover from the BEC side to the BCS side can be achieved experimentally,
including the intermediate strong interaction (or unitary) regime, where the
scattering length becomes very large. The behaviour of the quantum
correlation functions (QCF) describing the positions of the fermionic atoms
in the Cooper pairs can be measured using methods such as Bragg spectroscopy 
\cite{Torma15a}.

However, in terms of presenting a challenge for developing theories of
interacting ultracold fermionic atomic gases it is the strong interaction
regime (where the scattering length $|a|>>1/k_{F}$), that is currently of
great interest. Here the pair size is on the order of the interparticle
separation $1/k_{F}$ \cite{Ketterle08a}. There are a number of methods that
can be used to treat interacting fermi gases. These include: (a)
Diagrammatic methods \cite{Strinati12a}, \cite{Nozieres85a}, \cite{Tajima17a}
(b) Field theory techniques \cite{Nishida07a}, \cite{Mulkerin22a} (c)
Quantum Monte Carlo methods \cite{Bulgac12a}, \cite{Pang16a}, \cite%
{Richie20a}, \cite{Jensen20a} (d) Density functional methods \cite{Kohn99a}
as well as (e) BEC-Leggett theory \cite{Leggett80a}, \cite{Ketterle08a} \
and (f) Variational approaches \cite{Balian81a}. There are many variants
within each category. Many experimental studies have also been carried out,
for example \cite{Hoinka17a}, \cite{Horikoshi17a} and references therein.
Detailed descriptions of these theoretical methods and how well they agree
with experiment are presented \ in these references and in review articles
such as Refs. \cite{Ketterle08a}, \cite{Zwerger12a}, \cite{Randeria14a}, 
\cite{Strinati18a}, \cite{Leggett80a}, \cite{Torma15a}. Generally speaking
the current theories do not always agree quantitatively with experiment,
though qualitative agreement occurs. All of these theoretical methods have
advantages and disadvantages. BEC-Leggett\ theory provides a good simple
over-view of the crossover, but it is based on a variational approach in
which the quantum state is assumed to be pairs of opposite spins each
described by the same single pair wavefunction. This may not be a good
approximation in the unitary regime. For example fig 1 in Ref. \cite%
{Mulkerin22a}, shows that BEC-Leggett theory disagrees quantitatively with
experiment for thermodynamic quantities such as the chemical potential.
Quantum Monte-Carlo methods are also based around assumed choices of
variational functions, and in general the fermion sign problem leads to
difficulties. To quote Ref. \cite{Pang16a} "A typical wavefunction or
density matrix of a many-fermion system has a complex nodal structure that
prevents an easy stochastic process from driving the system to its ground
state or desired excited state. This problem, originating from the Pauli
exclusion principle of fermions, is known as the fermion-sign problem,
because crossing a node of the wavefunction or density matrix changes its
sign, and thus disallows a straightforward weight interpretation of the
wavefunction or density matrix." Field theory methods involve considering
re-normalisation issues, which makes the approach rather complicated.
Diagrammatic methods involve trying to decide which Feynman diagrams are
important and which can be neglected. Density functional methods have been
mainly used in atomic physics. Even more sophistocated theories such as the
beyond Gaussian pair fluctuation theory in Ref \cite{Mulkerin22a} does not
agree with experiment in the intermediate temperature range $%
0.1T_{F}<T<0.5T_{F}$ \ for the unitary fermi gas, though quantitative
agreement occurs in other temperature ranges. Hence there is scope for
developing other methods for treating the many-body fermionic systems that
are involved in ultracold fermionic atomic gases.

Phase space methods (which were developed to treat topics in quantum optics)
can also be applied to treat problems in ultracold atomic gases. These
methods are described in a number of textbooks \cite{GardinerZoller}, \cite%
{WallsMilburn}, \cite{BarnettRadmore}, \cite{Dalton15a}. Essentially the
quantum density operator is represented by a distribution function
(functional) which depends on phase space variables (fields) representing
mode (field) annihilation, creation operators, and the average value of
physical quantities are then given by phase space integrals (functional
integrals) involving the distribution function and with physical quantities
replaced by functions of the phase space variables. Several distinct
distribution functions have been introduced, based on whether the physical
quantities involve normally ordered, symmetrically ordered or anti-normally
ordered expressions involving the mode (field) annihilation, creation
operaors. These are the Glauber-Sudarshan $P$, Wigner $W$ and Husimi $Q$
functions respectively. Often the phase space variables (field) for
annihilation, creation operators are related by complex conjugation, but
treatments involving double phase space distributions where this does not
apply also exist \cite{DrummondGardiner}. The evolution of the density
operator is mapped onto a Fokker-Planck equation (functional FPE) for the
distribution function (functional). It is then common to introduce
stochastic phase space variables (fields) which satisfy Ito (or
Stratonovich) stochastic differential (field) equations that are equivalent
to the FPE (FFPE) and which involve stochastic Wiener increments, such that
the phase space integrals that give the average value for physical
quantities can then be replaced stochastic averages of functions of the
now-stochastic phase space variables (fields). One of the general advantages
of phase space theory is that it does not require any presumptions regarding
what the actual behaviour of the system will turn out to be. By contrast,
BEC-Leggett theory \cite{Leggett80a}, \cite{Ketterle08a} is based on an
assumed state vector in which all the fermions are in Cooper pairs.

Since quantum optics involves the physics of photons - which are bosons - it
is not surprising that the application of phase space methods in ultracold
atomic gases has mainly been for bosonic atoms such as in the treatment of
Bose-Einsten Condensates. However, although less common, phase space methods
can also be applied to treat fermionic atoms - though the treatment must
take into account the fact that the mode (or field) annihilation, creation
operators obey anti-commutation rules rather than the commutation rules that
apply in the bosonic case. This means that the square of a fermionic mode
annihilation or creation operator is zero, so the question then arises as to
what type of phase space variable is best suited for the fermionic
situation? For bosons the natural choice is to use c-numbers for the phase
space variables or fields - since they commute on multiplication and their
square is non-zero. For fermions it may seem more natural to use numbers
that have properties such as the square being zero and having
anti-commutation properties in multiplication - just as for the mode or
field operators they represent. Such quantities are known in mathematics as
Grassmann numbers, and the first phase space theory for fermions was
formulated by Cahill and Glauber in terms of Grassmann phase space variables 
\cite{CahillGlauber} - Grassmann Phase Space Theory (GPST). One of the first
applications followed shortly later \cite{Plimak01a} and was to momentum
correlations between fermions of opposite spins in a $1D$ trap.

There are however phase space theories for fermions which are based on
c-number quantities. One approach involves introducing Gaussian projectors
specified by c-number phase space variables \cite{Corney06a}, where for $n$
modes these are $1+n(2n-1)$ variables that determine the $2n\times 2n$
covariance matrix involved in defining the Gaussian projector. Numerical
applications \cite{Corney04a} of the Gaussian operator approach have been
made to the Hubbard model for the case of $16$ sites, and to the
dissociation of a bosonic system of two fermionic atoms. These cases are all
microscopic and involve small numbers of fermions and modes. In the
treatment by Rosales et al \cite{Rosales15a} the density operator is
represented by a $Q$ distribution function of the c-number variables, where $%
Q$ is a \ positive quantity that can be interpreted as a probability. The $Q$
function is defined on a phase space equivalent to the space of real,
anti-symmetric matrices. The approach has recently been applied to Majorana
fermions \cite{Joseph18a}, and the Fermi-Hubbard Hamiltonian \cite{Joseph21a}%
. Here the mode annihilation $\widehat{c}_{i}$ and creation $\widehat{c}%
_{i}^{\dag }$ operators being replaced by Hermitian Marjorana operators $(%
\widehat{c}_{i}+\widehat{c}_{i}^{\dag })$ and $-i(\widehat{c}_{i}-\widehat{c}%
_{i}^{\dag })$. In the Gaussian phase space theory approach, Fokker-Planck
equations for the distribution function are also replaced by stochastic
equations for phase space variables, which in this case parameterise a $%
2n\times 2n$ covariance matrix.

Further developments of Grassmann phase space theory (GPST)\ for fermions
are set out in following textbook \cite{Dalton15a} and articles \cite%
{Dalton16a}, \cite{Dalton17a}, \cite{Dalton20a}. \ The same general phase
space theory features apply as described previously. It is convenient to
represent the quantum density operator for the system by a Grassmann
(un-normalised B) distribution function \cite{Plimak01a} in a phase space,
but again the phase space variables replacing the fermion annihilation and
creation operators are Grassmann variables. For $n$ modes there are $2n$
phase space variables if a double space is used. Here the projector used in
the expression for the density operator is that involving fermion (Bargmann)
coherent states, which are given by similar expressions to the boson
(Bargmann) coherent states that are used in phase space theories for bosonic
systems. Experimentally measureable quantities for topics such as the
BEC/BCS crossover are given by quantum correlation functions (QCF), and
these can be related theoretically to Grassmann phase space integrals
involving the distribution function and with the fermion operators being
replaced by Grassmann phase space variables. As previously, evolution
equations (over time or temperature) for the density operator lead to
Fokker-Planck equations for the distribution function. These in turn are
replaced by Ito stochastic equations for stochastic Grassmann phase space
variables, and QCF now given by stochastic averages of products of these
stochastic variables. A key point is that the stochastic averages for QCF at
a slightly later time (or slightly lower temperature) can be shown to be
related linearly to such stochastic averages at an earlier time via matrices
that only involve c-numbers. Even though these matrices involve stochastic
quantities such as Wiener increments, they are not dependent on Grassmann
variables, which enables computations to be carried out without having to
represent Grassmann variables on the computer, and this makes Grassmannn
phase space theory numerically computable. The initial stochastic averages
of products are obtained from the initial density operator. It should also
be noted that although the distribution function (functionals) are only
quasi-probabilities (since they are Grassmann functions and not c-number
functions), the expressions for QCF and the mean values of physical
quantities has the same formal structure as in standard probability theory -
with the physical quantities given by Grassmann variables being weighted by
the (quasi) probability and then integrated over - albeit involving
Grasssmann integrals rather than c-number integrals. However, in a further
development \cite{Polyakov16a} an underlying c-number interpretation of
Grassmann phase space representations has been constructed. The related
equations for stochastic Grassmann variables have the same form as in Ref. 
\cite{Dalton16a}. In the present paper Grassmann phase space theory is used
to develop coupled c-number equations for the QCF \ themselves after the
stochastic Grassmann variables have been averaged out.

The utility of Grassmann phase space theory has been successfully tested 
\cite{Kidwani20a} on a simple fermion system that can be treated exactly by
standard methods, namely determining the coherence between two Cooper pair
states functions for interacting spin 
$\frac12$
fermions. Here the numerical application of the theory agrees in detail with
the analytic expression for the evolution of the coherence. However, the
GPST\ work of Ref.\cite{Plimak01a} involved a system consisting of many
fermions in a 1D harmonic trap, and GPST was used numerically to calculate
momentum correlation functions. Hence it is reasonable to apply the
Grassmann phase space theory approach on bigger fermionic systems, of which
the BEC/BCS crossover topic is an example.

In the present paper, we now apply the Grassmann phase space theory to
treating the BEC/BCS crossover - a much more complex system involving many
more fermions and modes. The QCF that are studied are those specifying the
position of the spin up and spin down fermions in a single Cooper pair, and
those specifying the positions of the two spin up and two spin down fermions
in two Cooper pairs. We focus particularly on treating how the size of a
Cooper pair changes through the crossover (BEC\ theory \cite{Ketterle08a}
indicates it changes smoothly), and on how the correlations between two
Cooper pairs change through the crossover (BEC-Leggett\ theory \cite%
{Ketterle08a} indicates there is little correlation at the BEC or BCS ends,
but strong correlation should occur in the unitary regime). Both time
evolution and temperature evolution will be studied, the former from an
initial state at near zero temperature for non-interacting fermions with the
fermions confined within the Fermi sphere, the latter from an initial \
state at high temperature where the fermion interactions can be ignored.
GPST may be useful in describing the pseudo-gap regime (see Ref \cite%
{Randeria14a}) between the transition temperature $T_{c}$ for superfluid
behaviour and the temperature $T^{\ast }$ below which Cooper pairs are
thought to form, as it directly focuses on the QCF describing the relative
positions of spin up and spin down fermions. This QCF should show evidence
of "preformed" Cooper pairs in the temperature regime $T_{c}<T<T^{\ast }$.

The present paper sets out the theory for the equations that determine the
above QCF, starting with the Ito stochastic equations for stochastic
Grassmann phase space variables. The stochastic averages that determine the
matrices that relate the stochastic averages for QCF at a later time (or
temperature) to the stochastic averages at an earlier time have been carried
out analytically, leading to standard non-stochastic matrices. The latter
matrices can then be used to study numerically the evolution of the QCF over
a finite time or temperature interval. Such calculations will be presented
in a subsequent paper.

The plan of the paper is as follows. In Section \ref{Section - Phase Space
Theory} we present the basic equations for GPST in the context of the
BEC/BCS crossover both for time an temperature evolution, stating from the
Hamiltonian, proceding via the FFPE for the Grassmann distribution
functional and ending with the Ito SFE both for stochastic position fields
and stochastic momentum fields. In Section \ref{Section - Quantum
Correlation Functions} we present the general expressions for the QCF for a
single Cooper pair and for two Cooper pairs. In Section \ref{Section - One
Cooper Pair First Order Changes to QCF} we present the results for the
temperature and time evolution for the QCF for a single Cooper pair, correct
to first order and with the stochastic averages carried out analytically. In
Section \ref{Section - Two Cooper Pairs First Order Changes to QCF}, we
present the analogous results for \ the QCF describing \ two Cooper pairs.
Equations for carrying out the time or temperature evolution over a finite
interval based on the matrix $M$ obtained in the previous two Sections are
set out in Section \ \ref{Section - Solution of Evolution Equations for QCF}%
. The introduction of dimensionless variables and numerical issues are
discussed. Finally, a Summary and Conclusions from the work is presented in
Section \ref{Section - Summary and Conclusions}. Detailed derivations are
set out in Appendices \ref{Appendix A - Details GFT} and \ref{Appendix B -
Srochastic Averaging and QCF Eqns}.

\pagebreak

\section{Grassman Phase Space Theory for Fields}

\label{Section - Phase Space Theory}

In this Section we set out the basic equations for applying Grassmann phase
space theory to the topic of the BEC/BCS crossover in ultracold atomic fermi
gases. The general methodology and a full description of Grassmann algebra
and calculus is set out in the following textbook \cite{Dalton15a} and
articles \cite{Dalton16a}, \cite{Dalton17a}, \cite{Dalton20a}\ . We begin by
writing the Hamiltonian involved in terms of fermion \ field operators, then
introduce the quantum density operator and its $B$ representation
distribution functional involving fermion Grassmann field functions,
followed by the two basic equations the density operator satisfies for
describing time and temperature evolution. The quantum correlation functions
(QCF) relevant to \ fermion position measurements for the size of and
correlation between Cooper pairs are then defined. We set out the functional
Fokker-Planck equations (FFPE) for the distribution functional in the two
types of evolution, and then introduce the corresponding Ito stochastic
field equations that are equivalent to the FFPE. We finally obtain the Ito
stochastic equations for momentum fields in \ the case of untrapped
fermions. \medskip

\subsection{Hamiltonian and Field Operators}

We consider a trapped Fermi gas of spin $1/2$ fermionic atoms with spin
conserving collisions of zero range between pairs of atoms .Here there are
two distinct internal states corresponding to spin up and spin down atoms,
designated by $\alpha =u(\uparrow ),d(\downarrow )$, with $-\alpha $
referring to the opposite spin state. The Fermi gas is isolated from the
environment so no relaxation effects are involved. This model was also
considered by Plimak et al \cite{Plimak01a}.

The Hamiltonian is written in terms of the field operators $\hat{\Psi}%
_{\alpha }(r)$, $\hat{\Psi}_{\alpha }(r)^{\dag }$ as

\begin{eqnarray}
\hat{H}_{f}\mbox{\rule{-0.5mm}{0mm}} &=&\mbox{\rule{-1mm}{0mm}}\int %
\mbox{\rule{-1mm}{0mm}}dr\left( \sum_{\alpha }\frac{\hbar ^{2}}{2m}\nabla 
\hat{\Psi}_{\alpha }(r)^{\dag }\mbox{\rule{-0.5mm}{0mm}}\cdot %
\mbox{\rule{-0.5mm}{0mm}}\nabla \hat{\Psi}_{\alpha }(\ r)+\sum_{\alpha }\hat{%
\Psi}_{\alpha }(r)^{\dag }V_{\alpha }\hat{\Psi}_{\alpha }(r)\right. 
\nonumber \\
&&\hspace{2cm}\left. +\frac{g}{2}\sum_{\alpha }\hat{\Psi}_{\alpha }(r)^{\dag
}\hat{\Psi}_{-\alpha }(r)^{\dag }\hat{\Psi}_{-\alpha }(r)\hat{\Psi}_{\alpha
}(r)\right)  \label{Eq.HamiltonianFieldModel} \\
&=&\widehat{{\small K}}+\widehat{{\small V}}+\widehat{{\small U}}
\label{Eq.HamiltTerms}
\end{eqnarray}%
and is the sum of kinetic energy, trap potential energy and collision
interaction energy terms. Here $m$ is the mass of the fermionic atom, $%
V_{\alpha }$ is the trap potential for spin $\alpha $ and $g$ is the
fermion-fermion interaction constant - which can be parameterised in terms
of an effective $s$-wave scattering amplitude $a_{s}$ via $g=4\pi a_{s}\hbar
^{2}/m$. The exact scattering amplitude is given in Ref. \cite{Ketterle08a}
(see Eq. (94)). For simplicity $r$ refers to position in 3D. The Hamiltonian
commutes with the number operator $\widehat{N}$, where 
\begin{equation}
\widehat{N}=\sum_{\alpha }\int \mbox{\rule{-1mm}{0mm}}dr\,\left( \hat{\Psi}%
_{\alpha }(r)^{\dag }\,\hat{\Psi}_{\alpha }(r)\right)  \label{Eq.NumberOpr}
\end{equation}%
\smallskip\ 

\subsection{B Distribution Functional}

For fermionic systems with spin components $\alpha $ the field operators $%
\hat{\Psi}_{\alpha }(r)$, $\hat{\Psi}_{\alpha }(r)^{\dag }$ may be expanded
in a complete orthonormal set of single fermion mode functions $\phi
_{\alpha i}(r)$ for each component $\alpha $ as 
\begin{equation}
\widehat{\psi }_{\alpha }(r)=\dsum\limits_{i}\widehat{c}_{\alpha i}\;\phi
_{\alpha i}(r)\qquad \hat{\Psi}_{\alpha }(r)^{\dag }=\dsum\limits_{i}%
\widehat{c}_{\alpha i}^{\dag }\;\phi _{\alpha i}^{\ast }(r)
\label{Eq.FermionFieldOprExpn}
\end{equation}%
where $\widehat{c}_{\alpha i}$, $\widehat{c}_{\alpha i}^{\dag }$ are
fermionic annihilation, creation operators for the mode $\alpha i$. The
orthogonality and completeness properties are%
\begin{equation}
\int dr\,\phi _{\alpha i}^{\ast }(r)\phi _{\alpha j}(r)=\delta _{ij}\qquad
\sum_{i}\phi _{\alpha i}(r)\phi _{\alpha i}^{\ast }(s)=\delta (r-s)
\label{Eq.OrthogComplete}
\end{equation}

In Grassmann phase space field theory a quantum density operator $\widehat{%
\rho }$ can be represented by its $B$ distribution functional $B[\ \psi (r)]$
in which the field operators $\hat{\Psi}_{\alpha }(r)$, $\hat{\Psi}_{\alpha
}(r)^{\dag }$\ are represented via Grassmann fields $\psi _{\alpha }(r)$\
and $\psi _{\alpha }^{+}(r)$\ respectively, where $\psi (r)\equiv \{\psi
_{\alpha }(r),\psi _{\alpha }^{+}(r)\}$. Here the Grassmann fields are
expanded as 
\begin{equation}
\psi _{\alpha }(r)=\dsum\limits_{i}g_{\alpha i}\;\phi _{\alpha i}(r)\qquad
\psi _{\alpha }^{+}(r)=\dsum\limits_{i}g_{\alpha i}^{+}\;\phi _{\alpha
i}^{\ast }(r)  \label{Eq.GrassmannFields}
\end{equation}%
where $g_{\alpha i}$, $g_{\alpha i}^{+}$ are Grassmann phase space variables
for the mode $\alpha i$.

The canonical form of a density operator for the $B$ distribution is written
in terms of Bargmann coherent state projectors $\widehat{\Lambda }[\psi (r)]$
in terms of a Grassmann functional integral as 
\begin{equation}
\widehat{\rho }=\dint \dprod\limits_{\alpha }d\psi _{\alpha }^{+}\,d\psi
_{\alpha }\,B[\ \psi (r)]\,\widehat{\Lambda }[\psi (r)]
\label{Eq.CanonBDensityOpr}
\end{equation}%
where 
\begin{equation}
\widehat{\Lambda }[\psi (r)]=\left\vert \psi (r)\right\rangle
_{B}\;\left\langle \psi ^{+}(r)^{\ast }\right\vert _{B}
\label{Eq.BargmannCohStateProj}
\end{equation}%
and the Bargmann states are given by 
\begin{eqnarray}
\left\vert \psi (r)\right\rangle _{B} &=&\exp (\dprod\limits_{\alpha
}\dsum\limits_{i}\widehat{c}_{\alpha i}^{\dag }\;g_{\alpha i})\;\left\vert
0\right\rangle =\dprod\limits_{\alpha i}(\left\vert 0_{\alpha
i}\right\rangle -g_{\alpha i}\left\vert 1_{\alpha i}\right\rangle ) 
\nonumber \\
\left\vert \psi ^{+}(r)^{\ast }\right\rangle _{B} &=&\exp
(\dprod\limits_{\alpha }\dsum\limits_{i}\widehat{c}_{\alpha i}^{\dag
}\;(g_{\alpha i}^{+})^{\ast })\;\left\vert 0\right\rangle
=\dprod\limits_{\alpha i}(\left\vert 0_{\alpha i}\right\rangle -(g_{\alpha
i}^{+})^{\ast }\left\vert 1_{\alpha i}\right\rangle )
\label{Eq.BargmannCohStates}
\end{eqnarray}%
with $\left\vert 0\right\rangle $ being the vacuum state with no fermions
and $\left\vert 0_{\alpha i}\right\rangle $, $\left\vert 1_{\alpha
i}\right\rangle $ being one fermion states with $0$, $1$ fermion in the $%
\alpha i$ mode.

The Bargmann states are eigenstates for the field operators 
\begin{equation}
\widehat{\psi }_{\alpha }(r)\,\left\vert \psi (r)\right\rangle _{B}=\psi
_{\alpha }(r)\,\left\vert \psi (r)\right\rangle _{B}\qquad _{B}\left\langle
\psi ^{+}(r)^{\ast }\right\vert \,\hat{\Psi}_{\alpha }(r)^{\dag
}=_{B}\left\langle \psi ^{+}(r)^{\ast }\right\vert \,\psi _{\alpha }^{+}(r)
\label{Eq.EigenstatesFieldOprs}
\end{equation}%
and are not normalised to unity. In fact 
\begin{equation}
_{B}\left\langle \chi ^{+}(r)^{\ast }|\,\psi (r)\right\rangle _{B}=\exp
\dsum\limits_{\alpha i}h_{\alpha i}^{+}\,g_{\alpha i}
\label{Eq.ScalarProdBargStates}
\end{equation}%
where the Grassman field $\chi ^{+}(r)$ involves phase space amplitudes $%
h_{\alpha i}^{+}$. The Grassmann functional integral $\dint
\dprod\limits_{\alpha }d\psi _{\alpha }^{+}\,d\psi _{\alpha }\equiv \dint
d\psi ^{+}\,d\psi $ involving the functional $B[\ \psi (r)]$ is equivalent
to the multiple Grassmann phase space integral over the phase space
variables $\dprod\limits_{\alpha }dg_{\alpha 1}^{+}dg_{\alpha
2}^{+}....\,dg_{\alpha n}^{+}dg_{\alpha n}....dg_{\alpha 2}dg_{\alpha 1}$ of
the distribution function $B(g)$ (where $g\equiv \{g_{\alpha i},g_{\alpha
i}^{+}\}$) which is equivalent to the functional $B[\ \psi (r)]$.\medskip

\subsection{Evolution of Density Operator}

There are two cases of interest. The first is where the density operator
describes time evolution in a non-equilibrium system, in which case $%
\widehat{\rho }$ satisfies the Liouville-von Neumann equation 
\begin{equation}
\frac{\partial }{\partial t}\widehat{\rho }=\frac{-i}{\hbar }[\widehat{H}%
_{f}\,,\widehat{\rho }]  \label{Eq.LVN}
\end{equation}%
involving the Hamiltonian $\widehat{H}_{f,}$ and the second where the
density operator describes temperature evolution in an equilibrium system,
in which case the density operator is given by the expression for the
canonical ensemble 
\begin{eqnarray}
\widehat{\rho } &=&\exp (-\beta \widehat{H}_{f,})/Z
\label{Eq.DensOprCanonEnsemble} \\
Z &=&Tr\exp (-\beta \widehat{H}_{f})  \label{Eq.PartitionFn}
\end{eqnarray}%
Here $Z$ is the partition function and $\beta =1/(k_{B}T)$, where $T$ is the
temperature. The numerator $\widehat{\sigma }=\exp (-\beta $ $\widehat{H}%
_{f,})$ satisfies a Matsubara equation. 
\begin{equation}
\frac{\partial }{\partial \beta }\widehat{\sigma }=\frac{-1}{2}[\widehat{H}%
_{f}\,,\widehat{\sigma }]_{+}  \label{Eq.Matsubara}
\end{equation}%
In both cases $\widehat{\rho }$ and $\widehat{\sigma }$ can be represented
via a $B$ distribution, as in Eq.(\ref{Eq.CanonBDensityOpr}). \medskip

\subsection{Quantum Correlation Functions and B Distribution Functional}

Physical quantities of interest can be expressed as quantum correlation
functions (referred to as pair-correlation functions in the BEC/BCS
crossover community, though QCF in the present paper may inovlve more than
one pair of fermions) involving the field operators $\hat{\Psi}_{\alpha }(r)$%
, $\hat{\Psi}_{\alpha }(r)^{\dag }$ and the density operator $\widehat{\rho }
$ and $\widehat{\sigma }$. In the case of the BEC/.BCS crossover we are
interested in the probability of finding a fermion of type $\alpha _{1}$ at
position $r_{1}$, a fermion of type $\alpha _{2}$ at position $r_{2}$, ...,
and a fermion of type $\alpha _{p}$ at position $r_{p}$. This probability is
given by 
\begin{equation}
P(\alpha _{1}r_{1},\alpha _{2}r_{2}\cdots \alpha _{p}r_{p})=Tr(\widehat{\rho 
}\,\Lambda (\alpha _{1}r_{1},\alpha _{2}r_{2}\cdots \alpha _{p}r_{p}))
\label{Eq.JointPositionProb}
\end{equation}%
where $\Lambda (\alpha _{1}r_{1},\alpha _{2}r_{2}\cdots \alpha _{p}r_{p})$
is the projector 
\begin{eqnarray}
&&\Lambda (\alpha _{1}r_{1},\alpha _{2}r_{2}\cdots \alpha _{p}r_{p}) 
\nonumber \\
&=&\hat{\Psi}_{\alpha _{1}}(r_{1})^{\dag }\cdots \hat{\Psi}_{\alpha
_{p}}(r_{p})^{\dag }\,\left\lfloor 0\right\rangle \left\langle 0\right\vert
\,\hat{\Psi}_{\alpha _{p}}(r_{p})\cdots \hat{\Psi}_{\alpha _{1}}(r_{1})
\label{Eq.PositionProj}
\end{eqnarray}

It can be shown (see Ref. \cite{Dalton15a}, see Sects. 7.8.2, 12.4.2) that
the required probabillity is given by a phase space functional integral
involving the $B$ distribution functional. This is an \ example of a QCF. We
have 
\begin{eqnarray}
&&P(\alpha _{1}r_{1},\alpha _{2}r_{2}\cdots \alpha _{p}r_{p})  \nonumber \\
&=&\int \int d\psi ^{+}d\psi \,\psi _{\alpha _{p}}(r_{p})\cdots \psi
_{\alpha _{1}}(\ r_{1})B[\psi (r),\psi ^{+}(r)]\psi _{\alpha
_{1}}^{+}(r_{1})\cdots \psi _{\alpha _{p}}^{+}(r_{p}))  \nonumber \\
&&  \label{Eq.QCFandGrassmannFnalInteg}
\end{eqnarray}%
for both the $\widehat{\rho }$ and $\widehat{\sigma }$ cases. \medskip

\subsection{Functional Fokker-Planck Equations}

The functional Fokker-Planck equation for the $B$ distribution functional $%
B[\ \psi (r)]$ is derived starting from Eq. (\ref{Eq.CanonBDensityOpr}) via
using the correspondence rules in conjunction with functional
differentiation rules. We have for both the $\widehat{\rho }$ and $\widehat{%
\sigma }$ cases%
\begin{eqnarray}
\hat{\rho} &\Rightarrow &\widehat{{\small \Psi }}_{\alpha }(r)\,\hat{\rho}%
\quad B[\psi ]\Rightarrow \psi _{\alpha }(r)\,B[\psi ]\qquad \qquad 
\nonumber \\
\hat{\rho} &\Rightarrow &\hat{\rho}\,\widehat{{\small \Psi }}_{\alpha
}(r)\qquad B[\psi ]\Rightarrow B[\psi ]\,(+\frac{\overleftarrow{\delta }}{%
\delta \psi _{\alpha }^{+}(r)})  \nonumber \\
\hat{\rho} &\Rightarrow &\widehat{{\small \Psi }}_{\alpha }^{\dag }(r)\,\hat{%
\rho}\quad B[\psi ]\Rightarrow (+\frac{\overrightarrow{\delta }}{\delta \psi
_{\alpha }(r)})\,B[\psi ]  \nonumber \\
\qquad \hat{\rho} &\Rightarrow &\hat{\rho}\,\widehat{{\small \Psi }}_{\alpha
}^{\dag }(r)\quad B[\psi ]\Rightarrow \,B[\psi ]\,\psi _{\alpha }^{+}(r)
\label{Eq.FermionFieldOprsCorresRules}
\end{eqnarray}%
The correspondence rules are derived in Ref. \cite{Dalton15a} (see Sects.
13.1.2, 13.1.3). The first and fourth are based on the eigenvalue equations
in Eq. (\ref{Eq.EigenstatesFieldOprs}). The functional Fokker-Planck
equation for the time evolution of the $B$ distribution functional $B[\ \psi
(r)]$ for the density operator $\hat{\rho}$ due to the Hamiltonian in Eq. (%
\ref{Eq.HamiltonianFieldModel}) is derived in Ref. \cite{Dalton15a} (see
Sects. I.2). An analogous treatment applies for the temperature evolution of
the $B$ distribution functional $B[\ \psi (r)]$ for the density operator
numerator $\ \widehat{\sigma }$ - the evolution being governed by Eq. (\ref%
{Eq.Matsubara}). \smallskip

\subsubsection{Time Evolution Case}

In the case of the density operator $\widehat{\rho }$ satisfying the
Liouville-von Neumann equation the time evolution of the $B$ distribution
functional is as follows. There are terms corresponding to the kinetic $(K)$%
, trap potential $(V)$ and fermion-fermion interaction $(U)$ terms in the
Hamiltonian. The latter term has been written as a double space integral via
the introduction of a Dirac delta function to comply with the requirements
for deriving the Ito stochastic equations \cite{Dalton20a} from the general
theory \cite{Dalton15a}, \cite{Dalton16a}, \cite{Dalton17a} .

The kinetic energy term is 
\begin{eqnarray}
&&\left( \frac{\partial }{\partial t}B[\psi (r)]\right) _{K}  \nonumber \\
&=&-\,\frac{i}{\hbar }\int ds\,\left[ \left\{ \left( \frac{-\hbar ^{2}}{2m}%
\nabla ^{2}{\small \psi }_{u}{\small (s)\,}B[\ \psi (r)]\right) \frac{%
\overleftarrow{\delta }}{\delta \psi _{u}(s)}\right\} \right.  \nonumber \\
&&\left. +\left\{ \left( \frac{-\hbar ^{2}}{2m}\nabla ^{2}{\small \psi }_{d}%
{\small (s)\,}B[\ \psi (r)]\right) \frac{\overleftarrow{\delta }}{\delta
\psi _{d}(s)}\right\} \right.  \nonumber \\
&&\left. -\left\{ \left( \frac{-\hbar ^{2}}{2m}\nabla ^{2}{\small \psi }%
_{u}^{+}{\small (s)\,}B[\ \psi (r)]\right) \frac{\overleftarrow{\delta }}{%
\delta \psi _{u}^{+}(s)}\right\} \right.  \nonumber \\
&&\left. -\left\{ \left( \frac{-\hbar ^{2}}{2m}\nabla ^{2}{\small \psi }%
_{d}^{+}{\small (s)\,}B[\ \psi (r)]\right) \frac{\overleftarrow{\delta }}{%
\delta \psi _{d}^{+}(s)}\right\} \right]  \nonumber \\
&&  \label{Eq.FFPEKineticFermiFldModel}
\end{eqnarray}%
and only contributes to the drift term.

The trap potential energy term is 
\begin{eqnarray}
&&\left( \frac{\partial }{\partial t}B[\psi (r)]\right) _{V}  \nonumber \\
&=&-\,\frac{i}{\hbar }\int {\small ds}\,\left[ \left\{ \left( V_{u}\psi
_{u}(s)\,B[\ \psi (r)]\right) \frac{\overleftarrow{\delta }}{\delta \psi
_{u}(s)}\right\} +\left\{ \left( V_{d}\psi _{d}(s)\,B[\ \psi (r)]\right) 
\frac{\overleftarrow{\delta }}{\delta \psi _{d}(s)}\right\} \right. 
\nonumber \\
&&\left. -\left\{ \left( V_{u}\psi _{u}^{+}(s)\,B[\ \psi (r)]\right) \frac{%
\overleftarrow{\delta }}{\delta \psi _{u}^{+}(s)}\right\} -\left\{ \left(
V_{d}\psi _{d}^{+}(s)\,B[\ \psi (r)]\right) \frac{\overleftarrow{\delta }}{%
\delta \psi _{d}^{+}(s)}\right\} \right]  \nonumber \\
&&  \label{Eq.FFPEPotentialFermiFldModel}
\end{eqnarray}%
and also only contributes to the drift term.

The fermion-fermion interaction term is 
\begin{eqnarray}
&&\left( \frac{\partial }{\partial t}B[\psi (r)]\right) _{U}  \nonumber \\
&{\small =}&{\small -\,}\frac{i}{\hbar }\frac{g}{2}\int \int {\small ds}%
\,dr\,\left[ \left\{ \psi _{u}(s)\,\delta (r-s)\,\psi _{d}(r)\,B[\psi (r)]%
\frac{\overleftarrow{\delta }}{\delta \psi _{d}(r)}\frac{\overleftarrow{%
\delta }}{\delta \psi _{u}(s)}\right\} \right.  \nonumber \\
&&\left. +\left\{ \psi _{d}(s)\,\delta (r-s)\,\psi _{u}(r)\,B[\psi (r)]\frac{%
\overleftarrow{\delta }}{\delta \psi _{u}(r)}\frac{\overleftarrow{\delta }}{%
\delta \psi _{d}(s)}\right\} \right.  \nonumber \\
&&\left. -\left\{ \psi _{u}^{+}(s)\,\delta (r-s)\,\psi _{d}^{+}(r)\,B[\psi
(r)]\frac{\overleftarrow{\delta }}{\delta \psi _{d}^{+}(r)}\frac{%
\overleftarrow{\delta }}{\delta \psi _{u}^{+}(s)}\right\} \right.  \nonumber
\\
&&\left. -\left\{ \psi _{d}^{+}(s)\,\delta (r-s)\,\psi _{u}^{+}(r)\,B[\psi
(r)]\frac{\overleftarrow{\delta }}{\delta \psi _{u}^{+}(r)}\frac{%
\overleftarrow{\delta }}{\delta \psi _{d}^{+}(s)}\right\} \right]
\label{Eq.FFPEInteractFermiFldModel}
\end{eqnarray}%
and is the only contribution to the diffusion term.\smallskip

\subsubsection{Temperature Evolution Case}

In the case of the density operator $\widehat{\sigma }$ satisfying the
Matsubara equation the temperature evolution of the $B$ distribution
functional is as follows.

The kinetic energy term is 
\begin{eqnarray}
&&\left( \frac{\partial }{\partial \beta }B[\psi (r)]\right) _{K}  \nonumber
\\
&=&-\frac{1}{2}\int ds\,\left[ \left\{ \left( \frac{-\hbar ^{2}}{2m}\nabla
^{2}{\small \psi }_{u}{\small (s)\,}B[\ \psi (r)]\right) \frac{%
\overleftarrow{\delta }}{\delta \psi _{u}(s)}\right\} \right.  \nonumber \\
&&\left. +\left\{ \left( \frac{-\hbar ^{2}}{2m}\nabla ^{2}{\small \psi }_{d}%
{\small (s)\,}B[\ \psi (r)]\right) \frac{\overleftarrow{\delta }}{\delta
\psi _{d}(s)}\right\} \right.  \nonumber \\
&&\left. +\left\{ \left( \frac{-\hbar ^{2}}{2m}\nabla ^{2}{\small \psi }%
_{u}^{+}{\small (s)\,}B[\ \psi (r)]\right) \frac{\overleftarrow{\delta }}{%
\delta \psi _{u}^{+}(s)}\right\} \right.  \nonumber \\
&&\left. +\left\{ \left( \frac{-\hbar ^{2}}{2m}\nabla ^{2}{\small \psi }%
_{d}^{+}{\small (s)\,}B[\ \psi (r)]\right) \frac{\overleftarrow{\delta }}{%
\delta \psi _{d}^{+}(s)}\right\} \right]  \nonumber \\
&&  \label{Eq.KEMatsub}
\end{eqnarray}%
and only contributes to the drift term.

The trap potential energy term is 
\begin{eqnarray}
&&\left( \frac{\partial }{\partial \beta }B[\psi (r)]\right) _{V}  \nonumber
\\
&=&-\frac{1}{2}\int {\small ds}\,\left[ +\left\{ \left( V_{u}\psi
_{u}(s)\,B[\ \psi (r)]\right) \frac{\overleftarrow{\delta }}{\delta \psi
_{u}(s)}\right\} +\left\{ \left( V_{d}\psi _{d}(s)\,B[\ \psi (r)]\right) 
\frac{\overleftarrow{\delta }}{\delta \psi _{d}(s)}\right\} \right. 
\nonumber \\
&&\left. +\left\{ \left( V_{u}\psi _{u}^{+}(s)\,B[\ \psi (r)]\right) \frac{%
\overleftarrow{\delta }}{\delta \psi _{u}^{+}(s)}\right\} +\left\{ \left(
V_{d}\psi _{d}^{+}(s)\,B[\ \psi (r)]\right) \frac{\overleftarrow{\delta }}{%
\delta \psi _{d}^{+}(s)}\right\} \right]  \nonumber \\
&&  \label{Eq.PEMatsub}
\end{eqnarray}%
and also only contributes to the drift term.

The fermion-fermion interaction term is 
\begin{eqnarray}
&&\left( \frac{\partial }{\partial \beta }B[\psi (r)]\right) _{U}  \nonumber
\\
&{\small =}&{\small \,-\;}\frac{g}{4}\int \int {\small ds}\,dr\,\left[
\left\{ \psi _{u}(s)\,\delta (r-s)\,\psi _{d}(r)\,B[\psi (r)]\frac{%
\overleftarrow{\delta }}{\delta \psi _{d}(r)}\frac{\overleftarrow{\delta }}{%
\delta \psi _{u}(s)}\right\} \right.  \nonumber \\
&&\left. +\left\{ \psi _{d}(s)\,\delta (r-s)\,\psi _{u}(r)\,B[\psi (r)]\frac{%
\overleftarrow{\delta }}{\delta \psi _{u}(r)}\frac{\overleftarrow{\delta }}{%
\delta \psi _{d}(s)}\right\} \right.  \nonumber \\
&&\left. +\left\{ \psi _{u}^{+}(s)\,\delta (r-s)\,\psi _{d}^{+}(r)\,B[\psi
(r)]\frac{\overleftarrow{\delta }}{\delta \psi _{d}^{+}(r)}\frac{%
\overleftarrow{\delta }}{\delta \psi _{u}^{+}(s)}\right\} \right.  \nonumber
\\
&&\left. +\left\{ \psi _{d}^{+}(s)\,\delta (r-s)\,\psi _{u}^{+}(r)\,B[\psi
(r)]\frac{\overleftarrow{\delta }}{\delta \psi _{u}^{+}(r)}\frac{%
\overleftarrow{\delta }}{\delta \psi _{d}^{+}(s)}\right\} \right]
\label{Eq.IntnMatsub}
\end{eqnarray}%
and is the only contribution to the diffusion term. Note the differences
between the Liouville-von Neumann and Matusubara cases. Apart from the
overall multiplier being related to $-i/\hbar $ rather than $-1/2$, the
pairs of terms involving $\psi _{\alpha }(r)$ and the pairs of terms
involving $\psi _{\alpha }^{+}(r)$ have the same sign in the Matsubara case
and opposite signs in the other case.$\,$\medskip

\subsection{Ito Stochastic Position Field Equations}

\label{SubSect - Ito Stochastic Position Field Eqns}

The Ito stochastic field equations are derived from the FFPE via using the
first order functional derivative (or drift) terms to give the classical (or
non-noise) terms in the Ito SFE and via using the second order (or
diffusion) terms to determine the noise (or fluctuation) terms in the Ito
SFE. The latter terms involve Gaussian-Markoff stochastic noise quantities
and require the Takagi factorisation \cite{Takagi25} of the diffusion matrix.

The basic principle is that the time (or temperature) evolution of the $B$
distribution functional is equivalent to the time (or temperature) evolution
of stochastic Grassmann fields $\widetilde{\psi }_{\alpha }(r,t),\widetilde{%
\psi }_{\alpha }^{+}(r,t)$ (or $\widetilde{\psi }_{\alpha }(r,\beta ),%
\widetilde{\psi }_{\alpha }^{+}(r,\beta )$) which now replace the Grassman
phase space fields $\psi _{\alpha }(r),\psi _{\alpha }^{+}(r)$, and where
the evolution is now described by Ito stochastic field equations. These are
chosen so that the stochastic averages of the $\widetilde{\psi }_{\alpha
}(r,t),\widetilde{\psi }_{\alpha }^{+}(r,t)$ (or $\widetilde{\psi }_{\alpha
}(r,\beta ),\widetilde{\psi }_{\alpha }^{+}(r,\beta )$) gives the same
result as the Grassmann functional integrals involving the $\psi _{\alpha
}(r),\psi _{\alpha }^{+}(r)$. In particular, for the time evolution case the
joint position probabilities $P(\alpha _{1}r_{1},\alpha _{2}r_{2}\cdots
\alpha _{p}r_{p})$ are now given by 
\begin{eqnarray}
&&P(\alpha _{1}r_{1},\alpha _{2}r_{2}\cdots \alpha _{p}r_{p})  \nonumber \\
&=&\mbox{Tr}(\hat{\rho}\,\hat{\Psi}_{\alpha _{1}}(r_{1})^{\dag }\cdots \hat{%
\Psi}_{\alpha _{p}}(r_{p})^{\dag }\,\left\lfloor 0\right\rangle \left\langle
0\right\vert \,\hat{\Psi}_{\alpha _{p}}(r_{p})\cdots \hat{\Psi}_{\alpha
_{1}}(r_{1})).\quad  \nonumber \\
&=&\int \int d\psi ^{+}d\psi \,\psi _{\alpha _{p}}(r_{p})\cdots \psi
_{\alpha _{1}}(\ r_{1})B[\psi (r),\psi ^{+}(r)]\psi _{\alpha
_{1}}^{+}(r_{1})\cdots \psi _{\alpha _{p}}^{+}(r_{p}))  \nonumber \\
&=&\overline{\psi _{\alpha _{p}}(r_{p})\cdots \psi _{\alpha _{1}}(\
r_{1})\;\psi _{\alpha _{1}}^{+}(r_{1})\cdots \psi _{\alpha _{p}}^{+}(r_{p}})
\label{Eq.QCFStochastAver}
\end{eqnarray}%
where the bar indicates a stochastic average. The same expressions apply for
the temperature evolution case ,but with $t$ being replaced by $\beta $.

To set out the FFPE and the related Ito SFE it is useful in this Section to
modify the notation for the field functions. The indices $A,B$ are inroduced
where each takes on the values $1,2$ and where (which $\alpha ,\beta $
specifying $u$ or $d$ as before) $\psi _{\alpha 1}(s)\equiv \psi _{\alpha
}(s)$ and $\psi _{\alpha 2}(s)\equiv \psi _{\alpha }^{+}(s)$. Similarly, $%
\widetilde{\psi }_{\alpha 1}(s,t)\equiv \widetilde{\psi }_{\alpha }(s,t)$
and $\widetilde{\psi }_{\alpha 2}(s,t)\equiv \widetilde{\psi }_{\alpha
}^{+}(s,t)$ for the stochastic fields.

For the case of time evolution we may write the functional Fokker-Planck
equation in the general form 
\begin{eqnarray}
\frac{\partial }{\partial t}B[\psi ] &=&-\tsum\limits_{\alpha A}\dint
dr\,(A_{\alpha A}[\psi (r),r]\,B[\psi ])\frac{\overleftarrow{\delta }}{%
\delta \psi _{\alpha A}(r)}  \nonumber \\
&&{\small +}\frac{1}{2}\tsum\limits_{\alpha A,\beta B}\diint
ds\,dr\,(D_{\alpha A\,\beta B}[\psi (s),s;\psi (r),r]\,B[\psi ])\frac{%
\overleftarrow{\delta }}{\delta \psi _{\beta B}(r)}\frac{\overleftarrow{%
\delta }}{\delta \psi _{\alpha A}(s)}  \nonumber \\
&&  \label{Eq.FermionFFPE}
\end{eqnarray}%
and then the Ito stochastic field equations can be written as the sum of a
classical and a noise contribution for the change in $\widetilde{\psi }%
_{\alpha A}$ as 
\begin{eqnarray}
\delta \widetilde{\psi }_{\alpha A}(r,t) &\equiv &\widetilde{\psi }_{\alpha
A}(r,t+\delta t)-\widetilde{\psi }_{\alpha A}(r,t)  \nonumber \\
&=&A_{\alpha A}[\widetilde{\psi }(r,t),r]\,\delta
t+\tsum\limits_{a}B_{a}^{\alpha A}[\widetilde{\psi }(r,t),r]\;\delta \omega
_{a}(t_{+})  \label{Eq.ItoStochFieldEqn} \\
&=&\left( \delta \widetilde{\psi }_{\alpha A}({\small r})\right)
_{class}+\left( \delta \widetilde{\psi }_{\alpha A}({\small r})\right)
_{noise}  \label{Eq.ClassicaNoiseTerms}
\end{eqnarray}%
where the matrix $B$ is linked to the diffusion matrix $D$ via Takagi
factorisation \cite{Takagi25} as (see Ref \cite{Dalton15a}, p312) 
\begin{eqnarray}
(BB^{T})_{\alpha A\,\beta B} &=&\dsum\limits_{a}B_{a}^{\alpha A}[\widetilde{%
\psi }(s,t),s]\,B_{a}^{\beta B}[\widetilde{\psi }(r,t),r]  \nonumber \\
&=&D_{\alpha A\,\beta B}[\widetilde{\psi }(s,t),s;\widetilde{\psi }(r,t),r]
\label{Eq.NoiseFieldResult}
\end{eqnarray}%
The drift and diffusion matrices can be read off from Eqs. (\ref%
{Eq.FFPEKineticFermiFldModel}), (\ref{Eq.FFPEPotentialFermiFldModel}) and (%
\ref{Eq.FFPEInteractFermiFldModel}). In determining the matrix $B$ the delta
function $\delta (r-s)=\frac{1}{V}\sum\limits_{q}\exp iq\cdot (r-s)$
expressed in terms of box normalised plane waves, with $V=L^{3}$ and (for
3D) $q\equiv \{q_{x}(\frac{2\pi }{L}),q_{y}(\frac{2\pi }{L}),q_{z}(\frac{%
2\pi }{L})\}$ - where $q_{x},q_{y},q_{z}$ are all integers.is used in the
diffusion matrix. The details for determining the matrix $B$ are set out in
Appendix \ref{Appendix A - Details GFT}.

The Wiener increments $\delta \omega _{a}(t_{+})$ are defined in terms of
Gaussian-Markoff random noise terms $\Gamma _{a}(t)$ as 
\begin{equation}
\delta \omega _{a}(t_{+})=\int_{t}^{t+\delta t}dt_{1}\,\Gamma _{a}(t_{1})
\label{Eq.WienerIncrem}
\end{equation}%
Properties of the Gaussian-Markoff random noise terms are set out in
Appendix \ref{Appendix A - Details GFT}. From these results we can then show
that the Wiener increments have the following properties 
\begin{eqnarray}
\overline{\delta \omega _{{\small a}}(t_{+})}\, &=&0  \nonumber \\
\overline{\delta \omega _{{\small a}}(t_{+})\delta \omega _{b}(t_{+})}\,
&=&\delta _{{\small ab}}\;\delta t  \nonumber \\
\overline{\delta \omega _{a}(t_{+})\delta \omega _{b}(t_{+})\delta \omega
_{c}(t_{+})} &=&0  \nonumber \\
\overline{\delta \omega _{a}(t_{+})\delta \omega _{b}(t_{+})\delta \omega
_{c}(t_{+})\delta \omega _{d}(t_{+})} &=&\overline{\delta \omega _{{\small a}%
}(t_{+})\delta \omega _{b}(t_{+})}\;\overline{\delta \omega _{{\small c}%
}(t_{+})\delta \omega _{d}(t_{+})}  \nonumber \\
&&+\overline{\delta \omega _{{\small a}}(t_{+})\delta \omega _{c}(t_{+})}\;\,%
\overline{\delta \omega _{{\small b}}(t_{+})\delta \omega _{d}(t_{+})} 
\nonumber \\
&&+\overline{\delta \omega _{{\small a}}(t_{+})\delta \omega _{d}(t_{+})}\;\,%
\overline{\delta \omega _{{\small b}}(t_{+})\delta \omega _{c}(t_{+})} 
\nonumber \\
&&  \label{Eq.PropsWienerIncrem}
\end{eqnarray}%
showing that the stochastic averages of a single $\delta \omega _{a}$ is
zero and the stochastic average of the product of two $\delta \omega _{a}$'s
is zero if they are different and equal to $\delta t$ if they are the same.
In addition, the stochastic averages of products of odd numbers of $\delta
\omega _{a}$\ are zero and stochastic averages of products of even numbers
of $\delta \omega _{a}$\ are the sums of products of stochastic averages of
pairs of $\delta \omega _{a}$.

For the case of temperature evolution, the same forms apply but with $t$
being replaced by $\beta $.\smallskip

\subsubsection{Time Evolution Case}

For the time evolution case the stochastic fields at time $t+\delta t$ can
be related to the stochastic fields at time $t$ as follows. We have now
reverted to the $u,d,u^{+},d^{+}$ notation.

We have (see Sect. \ref{SubSection - B Matrix Time Evn} for details)%
\begin{eqnarray}
&&\left[ 
\begin{tabular}{l}
\hline
$\widetilde{\psi }_{u}(s,t+\delta t)$ \\ \hline
$\widetilde{\psi }_{d}(s,t+\delta t)$ \\ \hline
$\widetilde{\psi }_{u}^{+}(s,t+\delta t)$ \\ \hline
$\widetilde{\psi }_{d}^{+}(s,t+\delta t)$ \\ \hline
\end{tabular}%
\right]  \nonumber \\
&=&\left[ 
\begin{tabular}{|l|l|l|l}
\hline
$\Theta _{u,u}(\delta t)$ & $\Theta _{u,d}(\delta t)$ & $0$ & $0$ \\ \hline
$\Theta _{d,u}(\delta t)$ & $\Theta _{d,d}(\delta t)$ & $0$ & $0$ \\ \hline
$0$ & $0$ & $\Theta _{u,u}^{+}(\delta t)$ & $\Theta _{u,d}^{+}(\delta t)$ \\ 
\hline
$0$ & $0$ & $\Theta _{d,u}^{+}(\delta t)$ & $\Theta _{d,d}^{+}(\delta t)$ \\ 
\hline
\end{tabular}%
\right] \;\left[ 
\begin{tabular}{l}
\hline
$\widetilde{\psi }_{u}(s,t)$ \\ \hline
$\widetilde{\psi }_{d}(s,t)$ \\ \hline
$\widetilde{\psi }_{u}^{+}(s,t)$ \\ \hline
$\widetilde{\psi }_{d}^{+}(s,t)$ \\ \hline
\end{tabular}%
\right]  \nonumber \\
&&  \label{Eq.ItoSFETimeEvoln}
\end{eqnarray}%
where the $\Theta $ quantities are%
\begin{eqnarray}
&&\left[ \Theta (\delta t)\right]  \nonumber \\
&=&\left[ 
\begin{tabular}{|l|l|}
\hline
$1+\frac{i}{\hbar }\left\{ -\frac{\hbar ^{2}}{2m}\nabla ^{2}+V_{u}\right\}
\delta t$ & $+\lambda \sum\limits_{q}\exp (+iqs)\left\{ \delta \widetilde{%
\omega }_{u,d}^{q}+i\delta \widetilde{\omega }_{d,u}^{q}\right\} $ \\ \hline
$+\lambda \sum\limits_{q}\exp (-iqs)\left\{ \delta \widetilde{\omega }%
_{u,d}^{q}-i\delta \widetilde{\omega }_{d,u}^{q}\right\} $ & $1+\frac{i}{%
\hbar }\left\{ -\frac{\hbar ^{2}}{2m}\nabla ^{2}+V_{d}\right\} \delta t$ \\ 
\hline
\end{tabular}%
\right]  \nonumber \\
&&\left[ \Theta ^{+}(\delta t)\right]  \nonumber \\
&=&\left[ 
\begin{tabular}{|l|l|}
\hline
$1-\frac{i}{\hbar }\left\{ -\frac{\hbar ^{2}}{2m}\nabla ^{2}+V_{u}\right\}
\delta t$ & $+\lambda \sum\limits_{q}\exp (-iqs)\left\{ \delta \widetilde{%
\omega }_{u+,d+}^{q}+i\delta \widetilde{\omega }_{d+,u+}^{q}\right\} $ \\ 
\hline
$+\lambda \sum\limits_{q}\exp (+iqs)\left\{ -\delta \widetilde{\omega }%
_{u+,d+}^{q}+i\delta \widetilde{\omega }_{d+,u+}^{q}\right\} $ & $1-\frac{i}{%
\hbar }\left\{ -\frac{\hbar ^{2}}{2m}\nabla ^{2}+V_{d}\right\} \delta t$ \\ 
\hline
\end{tabular}%
\right]  \nonumber \\
&&  \label{Eq.ThetaMatrixTimeEvn}
\end{eqnarray}%
where $\lambda =\sqrt{\frac{-\,ig}{2\hslash V}}$. These quantities involve
the Laplacian, the trap potentials, the time interval $\delta t$ as well as
four stochastic Wiener increments (\ref{Eq.WienerIncrem}) $\delta \widetilde{%
\omega }_{u,d}^{q}$, $\delta \widetilde{\omega }_{d,u}^{q}$, $\delta 
\widetilde{\omega }_{u+,d+}^{q}$ and $\widetilde{\omega }_{d+,u+}^{q}$ for
each $q$. The latter satisfy conditions analogous to (\ref%
{Eq.PropsWienerIncrem}). Note the sum over $q$ and that there is no change
in the position coordinate $s$. This is a consequence of the zero range
interaction. \bigskip

\subsubsection{Temperature Evolution Case}

For the temperature evolution case the stochastic fields at temperature $%
\beta +\delta \beta $ can be related to the stochastic fields at temperature 
$\beta $ as follows. We have also now reverted to the $u,d,u^{+},d^{+}$%
notation.

We have \ (see Sect. \ref{SubSection - B Matrix Temp Evn} for details)%
\begin{eqnarray}
&&\left[ 
\begin{tabular}{l}
\hline
$\widetilde{\psi }_{u}(s,\beta +\delta \beta )$ \\ \hline
$\widetilde{\psi }_{d}(s,\beta +\delta \beta )$ \\ \hline
$\widetilde{\psi }_{u}^{+}(s,\beta +\delta \beta )$ \\ \hline
$\widetilde{\psi }_{d}^{+}(s,\beta +\delta \beta )$ \\ \hline
\end{tabular}%
\right]  \nonumber \\
&=&\left[ 
\begin{tabular}{|l|l|l|l}
\hline
$\Theta _{u,u}(\delta \beta )$ & $\Theta _{u,d}(\delta \beta )$ & $0$ & $0$
\\ \hline
$\Theta _{d,u}(\delta \beta )$ & $\Theta _{d,d}(\delta \beta )$ & $0$ & $0$
\\ \hline
$0$ & $0$ & $\Theta _{u,u}^{+}(\delta \beta )$ & $\Theta _{u,d}^{+}(\delta
\beta )$ \\ \hline
$0$ & $0$ & $\Theta _{d,u}^{+}(\delta \beta )$ & $\Theta _{d,d}^{+}(\delta
\beta )$ \\ \hline
\end{tabular}%
\right] \;\left[ 
\begin{tabular}{l}
\hline
$\widetilde{\psi }_{u}(s,\beta )$ \\ \hline
$\widetilde{\psi }_{d}(s,\beta )$ \\ \hline
$\widetilde{\psi }_{u}^{+}(s,\beta )$ \\ \hline
$\widetilde{\psi }_{d}^{+}(s,\beta )$ \\ \hline
\end{tabular}%
\right]  \nonumber \\
&&  \label{Eq.ItoSFETempEvn}
\end{eqnarray}%
where the $\Theta $ quantities are%
\begin{eqnarray}
&&\left[ \Theta (\delta \beta )\right]  \nonumber \\
&=&\left[ 
\begin{tabular}{|l|l|}
\hline
$1+\frac{1}{2}\left\{ -\frac{\hbar ^{2}}{2m}\nabla ^{2}+V_{u}\right\} \delta
\beta $ & $+\eta \sum\limits_{q}\exp (+iqs)\left\{ \delta \widetilde{\omega }%
_{u,d}^{q}+i\delta \widetilde{\omega }_{d,u}^{q}\right\} $ \\ \hline
$+\eta \sum\limits_{q}\exp (-iqs)\left\{ \delta \widetilde{\omega }%
_{u,d}^{q}-i\delta \widetilde{\omega }_{d,u}^{q}\right\} $ & $1+\frac{1}{2}%
\left\{ -\frac{\hbar ^{2}}{2m}\nabla ^{2}+V_{d}\right\} \delta \beta $ \\ 
\hline
\end{tabular}%
\right]  \nonumber \\
&&\left[ \Theta ^{+}(\delta \beta )\right]  \nonumber \\
&=&\left[ 
\begin{tabular}{|l|l|}
\hline
$1+\frac{1}{2}\left\{ -\frac{\hbar ^{2}}{2m}\nabla ^{2}+V_{u}\right\} \delta
\beta $ & $+\eta \sum\limits_{q}\exp (-iqs)\left\{ \delta \widetilde{\omega }%
_{u+,d+}^{q}+i\delta \widetilde{\omega }_{d+,u+}^{q}\right\} $ \\ \hline
$+\eta \sum\limits_{q}\exp (+iqs)\left\{ \delta \widetilde{\omega }%
_{u+,d+}^{q}-i\delta \widetilde{\omega }_{d+,u+}^{q}\right\} $ & $1+\frac{1}{%
2}\left\{ -\frac{\hbar ^{2}}{2m}\nabla ^{2}+V_{d}\right\} \delta \beta $ \\ 
\hline
\end{tabular}%
\right]  \nonumber \\
&&  \label{Eq.ThetaMatrixTempEvn}
\end{eqnarray}%
where $\eta =\sqrt{\frac{-\,g}{4V}}$. These quantities involve the
Laplacian, the trap potentials, the temperature interval $\delta \beta $ as
well as four stochastic Wiener increments (\ref{Eq.WienerIncrem}) $\delta 
\widetilde{\omega }_{u,d}^{q}$, $\delta \widetilde{\omega }_{d,u}^{q}$, $%
\delta \widetilde{\omega }_{u+,d+}^{q}$ and $\widetilde{\omega }_{d+,u+}^{q}$
for each $q$. The latter satisfy conditions analogous to (\ref%
{Eq.PropsWienerIncrem}). Note the sum over $q$ and that there is no change
in the position coordinate $s$. This is a consequence of the zero range
interaction. Note the sign differences for both the classical and noise
terms between \ the time and temperature evolution cases. There is also a
change in the constants for these terms. \medskip

\subsection{Ito Stochastic Momentum Field Equations}

For the free field situation where $V_{u}=V_{d}=0$, it is convenient to
introduce stochastic momentum fields $\tilde{\phi}_{\alpha }(k),\,\tilde{\phi%
}_{\alpha }^{+}(k)$, which are related to the stochastic position fields $%
\tilde{\psi}_{\alpha }(s),\tilde{\psi}_{\alpha }^{+}(s)$ via spatial Fourier
transforms involving box normalised plane waves $\exp (ik\cdot s)/\sqrt{V}$.
The wave numbers $k$ are discrete and given by $k\equiv \{k_{x}(\frac{2\pi }{%
L}),k_{y}(\frac{2\pi }{L}),k_{z}(\frac{2\pi }{L})\}$ - where $%
k_{x},k_{y},k_{z}$ are all integers and $V=L^{3}$. For ease of notation, we
leave the $t$ or $\beta $ dependence implicit. 
\begin{eqnarray}
\tilde{\psi}_{\alpha }(s) &=&\frac{1}{\sqrt{V}}\dsum\limits_{k}\,\exp
(ik\cdot s)\,\tilde{\phi}_{\alpha }(k)  \nonumber \\
\tilde{\psi}_{\alpha }^{+}(s) &=&\frac{1}{\sqrt{V}}\dsum\limits_{k}\,\exp
(-ik\cdot s)\,\tilde{\phi}_{\alpha }^{+}(k)  \label{Eq.FourierStochFields}
\end{eqnarray}%
The inverse transformation is given as Eq. (\ref{Eq.StochMtmFldsInverse}) in
Appendix \ref{Appendix A - Details GFT}. The use of stochastic momentum
fields enables the $\nabla ^{2}$ terms to be replaced by $k^{2}$ when the
stochastic position fields are replaced by stochastic momentum fields in
Eqs. (\ref{Eq.ItoSFETimeEvoln}) and (\ref{Eq.ItoSFETempEvn}).\smallskip

\subsubsection{Time Evolution Case - Momentum Fields}

For the time evolution case the stochastic momentum fields at time $t+\delta
t$ can be related to the stochastic momentum fields at time $t$ as follows.
To obtain this result we substitute the expressions (\ref%
{Eq.FourierStochFields}) for the stochastic position fields into Eq. (\ref%
{Eq.ItoSFETimeEvoln}).

We have%
\begin{eqnarray}
&&\left[ 
\begin{tabular}{l}
\hline
$\widetilde{\phi }_{u}(k,t+\delta t)$ \\ \hline
$\widetilde{\phi }_{d}(k,t+\delta t)$ \\ \hline
$\widetilde{\phi }_{u}^{+}(k,t+\delta t)$ \\ \hline
$\widetilde{\phi }_{d}^{+}(k,t+\delta t)$ \\ \hline
\end{tabular}%
\right]  \nonumber \\
&=&\left[ 
\begin{tabular}{|l|l|l|l}
\hline
$F_{u,u}(k,l;q,\delta t)$ & $F_{u,d}(k,l;q,\delta t)$ & $0$ & $0$ \\ \hline
$F_{d,u}(k,l;q,\delta t)$ & $F_{d,d}(k,l;q,\delta t)$ & $0$ & $0$ \\ \hline
$0$ & $0$ & $F_{u,u}^{+}(k,l;q,\delta t)$ & $F_{u,d}^{+}(k,l;q,\delta t)$ \\ 
\hline
$0$ & $0$ & $F_{d,u}^{+}(k,l;q,\delta t)$ & $F_{d,d}^{+}(k,l;q,\delta t)$ \\ 
\hline
\end{tabular}%
\right] \;  \nonumber \\
&&\times \left[ 
\begin{tabular}{l}
\hline
$\widetilde{\phi }_{u}(l,t)$ \\ \hline
$\widetilde{\phi }_{d}(l,t)$ \\ \hline
$\widetilde{\phi }_{u}^{+}(l,t)$ \\ \hline
$\widetilde{\phi }_{d}^{+}(l,t)$ \\ \hline
\end{tabular}%
\right]  \nonumber \\
&&  \label{Eq.ItoSMtmTimeEvln}
\end{eqnarray}%
where the $F$ quantities are%
\begin{eqnarray}
&&\left[ F(k,l;q,\delta t)\right]  \nonumber \\
&=&\left[ 
\begin{tabular}{|l|l|}
\hline
$\delta _{q,0}\;\delta _{k,l}\left( 1+\frac{i}{\hbar }\left\{ \frac{\hbar
^{2}k^{2}}{2m}\right\} \delta t\right) $ & $+\lambda \;\delta
_{(k-q),l}\;\left\{ \delta \widetilde{\omega }_{u,d}^{q}+i\delta \widetilde{%
\omega }_{d,u}^{q}\right\} $ \\ \hline
$+\lambda \;\delta _{(k+q),l}\;\left\{ \delta \widetilde{\omega }%
_{u,d}^{q}-i\delta \widetilde{\omega }_{d,u}^{q}\right\} $ & $\delta
_{q,0}\;\delta _{k,l}\left( 1+\frac{i}{\hbar }\left\{ \frac{\hbar ^{2}k^{2}}{%
2m}\right\} \delta t\right) $ \\ \hline
\end{tabular}%
\right]  \nonumber \\
&&\left[ F^{+}(k,l;q,\delta t)\right]  \nonumber \\
&=&\left[ 
\begin{tabular}{|l|l|}
\hline
$\delta _{q,0}\;\delta _{k,l}\left( 1-\frac{i}{\hbar }\left\{ \frac{\hbar
^{2}k^{2}}{2m}\right\} \delta t\right) $ & $+\lambda \;\delta
_{(k+q),l}\;\left\{ \delta \widetilde{\omega }_{u+,d+}^{q}+i\delta 
\widetilde{\omega }_{d+,u+}^{q}\right\} $ \\ \hline
$+\lambda \;\delta _{(k-q),l}\;\left\{ -\delta \widetilde{\omega }%
_{u+,d+}^{q}+i\delta \widetilde{\omega }_{d+,u+}^{q}\right\} $ & $\delta
_{q,0}\;\delta _{k,l}\left( 1-\frac{i}{\hbar }\left\{ \frac{\hbar ^{2}k^{2}}{%
2m}\right\} \delta t\right) $ \\ \hline
\end{tabular}%
\right]  \nonumber \\
&&  \label{Eq.FTimeEvoln}
\end{eqnarray}%
where $\lambda =\sqrt{\frac{-\,ig}{2\hbar V}}$. In each row of the right
side of Eq.(\ref{Eq.ItoSFETimeEvoln}) the sum over repeated indices $q,l$ is
assumed \smallskip

\subsubsection{Temperature Evolution Case - Momentum Fields}

For the temperature evolution case the stochastic momentum fields at
temperature $\beta +\delta \beta $ can be related to the stochastic momentum
fields at temperature $\beta $ as follows. To obtain this result we
substitute the expressions (\ref{Eq.FourierStochFields}) for the stochastic
position fields into Eq. (\ref{Eq.ItoSFETempEvn}).

We have%
\begin{eqnarray}
&&\left[ 
\begin{tabular}{l}
\hline
$\widetilde{\phi }_{u}(k,\beta +\delta \beta )$ \\ \hline
$\widetilde{\phi }_{d}(k,\beta +\delta \beta )$ \\ \hline
$\widetilde{\phi }_{u}^{+}(k,\beta +\delta \beta )$ \\ \hline
$\widetilde{\phi }_{d}^{+}(k,\beta +\delta \beta )$ \\ \hline
\end{tabular}%
\right]  \nonumber \\
&=&\left[ 
\begin{tabular}{|l|l|l|l}
\hline
$F_{u,u}(k,l;q,\delta \beta )$ & $F_{u,d}(k,l;q,\delta \beta )$ & $0$ & $0$
\\ \hline
$F_{d,u}(k,l;q,\delta \beta )$ & $F_{d,d}(k,l;q,\delta \beta )$ & $0$ & $0$
\\ \hline
$0$ & $0$ & $F_{u,u}^{+}(k,l;q,\delta \beta )$ & $F_{u,d}^{+}(k,l;q,\delta
\beta )$ \\ \hline
$0$ & $0$ & $F_{d,u}^{+}(k,l;q,\delta \beta )$ & $F_{d,d}^{+}(k,l;q,\delta
\beta )$ \\ \hline
\end{tabular}%
\right]  \nonumber \\
&&\times \;\left[ 
\begin{tabular}{l}
\hline
$\widetilde{\phi }_{u}(l,\beta )$ \\ \hline
$\widetilde{\phi }_{d}(l,\beta )$ \\ \hline
$\widetilde{\phi }_{u}^{+}(l,\beta )$ \\ \hline
$\widetilde{\phi }_{d}^{+}(l,\beta )$ \\ \hline
\end{tabular}%
\right]  \nonumber \\
&&  \label{Eq.ItoSMtmFTemper}
\end{eqnarray}%
where the $F$ quantities are%
\begin{eqnarray}
&&\left[ F(k,l;q,\delta \beta )\right]  \nonumber \\
&=&\left[ 
\begin{tabular}{|l|l|}
\hline
$\delta _{q,0}\;\delta _{k,l}\left( 1+\frac{1}{2}\left\{ \frac{\hbar
^{2}k^{2}}{2m}\right\} \delta \beta \right) $ & $+\eta \;\delta
_{(k-q),l}\;\left\{ \delta \widetilde{\omega }_{u,d}^{q}+i\delta \widetilde{%
\omega }_{d,u}^{q}\right\} $ \\ \hline
$+\eta \;\delta _{(k+q),l}\;\left\{ \delta \widetilde{\omega }%
_{u,d}^{q}-i\delta \widetilde{\omega }_{d,u}^{q}\right\} $ & $\delta
_{q,0}\;\delta _{k,l}\left( 1+\frac{1}{2}\left\{ \frac{\hbar ^{2}k^{2}}{2m}%
\right\} \delta \beta \right) $ \\ \hline
\end{tabular}%
\right]  \nonumber \\
&&\left[ F^{+}(k,l;q,\delta \beta )\right]  \nonumber \\
&=&\left[ 
\begin{tabular}{|l|l|}
\hline
$\delta _{q,0}\;\delta _{k,l}\left( 1+\frac{1}{2}\left\{ \frac{\hbar
^{2}k^{2}}{2m}\right\} \delta \beta \right) $ & $+\eta \;\delta
_{(k+q),l}\;\left\{ \delta \widetilde{\omega }_{u+,d+}^{q}+i\delta 
\widetilde{\omega }_{d+,u+}^{q}\right\} $ \\ \hline
$+\eta \;\delta _{(k-q),l}\;\left\{ \delta \widetilde{\omega }%
_{u+,d+}^{q}-i\delta \widetilde{\omega }_{d+,u+}^{q}\right\} $ & $\delta
_{q,0}\;\delta _{k,l}\left( 1+\frac{1}{2}\left\{ \frac{\hbar ^{2}k^{2}}{2m}%
\right\} \delta \beta \right) $ \\ \hline
\end{tabular}%
\right]  \nonumber \\
&&  \label{Eq.FTempEvol}
\end{eqnarray}%
where $\eta =\sqrt{\frac{-\,g}{4V}}$. In each row of the right side of Eq.(%
\ref{Eq.ItoSMtmFTemper}) the sum over repeated indices $q,l$ is
assumed.\medskip \pagebreak

\section{Quantum Correlation Functions for Cooper Pairs}

\label{Section - Quantum Correlation Functions}

In this Section we derive the expressions for the QCF's involving in
determining the size of a single Cooper pair and determining the correlation
between the positions of the fermions in two Cooper pairs.

For the case of a single Cooper pair where the spin down fermion is at
position $r_{1}$ and the spin up fermion is at position $r_{2}$ the QCF\
that can be used to describe the size of the Cooper pair is 
\begin{equation}
X(d\,r_{1},u\,r_{2})=\mbox{Tr}(\,\hat{\rho}\,\hat{\Psi}_{u}(r_{2})^{\dag }\,%
\hat{\Psi}_{d}(r_{1})^{\dag }\,\left\lfloor 0\right\rangle \left\langle
0\right\vert \,\hat{\Psi}_{d}(r_{1})\,\hat{\Psi}_{u}(r_{2})\,).
\label{Eq.QCFOneCooperPair}
\end{equation}

For the case of two Cooper pairs where the spin down fermions are at
positions $r_{1}$, $r_{2}$ and the spin up fermions are at positions $r_{3}$%
, $r_{4}$ the QCF\ that can be used to describe the correlation of the
positions of the fermions in the two Cooper pairs is 
\begin{eqnarray}
&&X(d\,r_{1},d\,r_{2,}\,u\,r_{3},u\,r_{4})=  \nonumber \\
&&\mbox{Tr}(\hat{\rho}\,\,\hat{\Psi}_{u}(r_{4})^{\dag }\,\hat{\Psi}%
_{u}(r_{3})^{\dag }\,\hat{\Psi}_{d}(r_{2})^{\dag }\,\hat{\Psi}%
_{d}(r_{1})^{\dag }\left\lfloor 0\right\rangle \left\langle 0\right\vert \,%
\hat{\Psi}_{d}(r_{1})\,\hat{\Psi}_{d}(r_{2})\,\hat{\Psi}_{u}(r_{3})\,\hat{%
\Psi}_{u}(r_{4})\,).  \nonumber \\
&&  \label{Eq.QCFTwoCooperPair}
\end{eqnarray}

\medskip

\subsection{One Cooper Pair: QCF}

For the case of time evolution the stochastic expressions for $%
X(d\,r_{1},u\,r_{2})$ can be written in terms of stochastic position or
momentum fields as%
\begin{eqnarray}
X(d\,r_{1},u\,r_{2};t) &=&\overline{\widetilde{\psi }_{d}(r_{1},t)\,%
\widetilde{\psi }_{u}(r_{2},t)\;\widetilde{\psi }_{u}^{+}(r_{2},t)\,%
\widetilde{\psi }_{d}^{+}(r_{1},t)}  \label{Eq.QCFOneCooperTimeEvn} \\
&=&\frac{1}{(\sqrt{V})^{4}}\dsum\limits_{k_{1},k_{2},k_{3},k_{4}}\exp
i\{(k_{1}-k_{4})\cdot r_{1}\}\;\exp i\{(k_{2}-k_{3})\cdot r_{2}\}  \nonumber
\\
&&\times \overline{\widetilde{\phi }_{d}(k_{1},t)\,\widetilde{\phi }%
_{u}(k_{2},t)\;\widetilde{\phi }_{u}^{+}(k_{3},t)\,\widetilde{\phi }%
_{d}^{+}(k_{4},t)}  \label{Eq.QCFOneCooperMtmTimeEvn}
\end{eqnarray}%
The case of temperature evolution leads to the same forms with $t$ replaced
by $\beta $ in Eq.(\ref{Eq.QCFOneCooperTimeEvn}) and in Eq. (\ref%
{Eq.QCFOneCooperMtmTimeEvn}), along with the additional factor $1/Z$ on the
right side of Eq. (\ref{Eq.QCFOneCooperMtmTimeEvn}) to allow for $\widehat{%
\rho }$ being replaced by $\widehat{\sigma }$ when temperature evolution is
involved.

The momentum stochastic correlation function of interest is 
\begin{equation}
X(d\,k_{1},u\,k_{2},u^{+}\,k_{3},d^{+}\,k_{4})=\overline{\widetilde{\phi }%
_{d}(k_{1})\,\widetilde{\phi }_{u}(k_{2})\;\widetilde{\phi }_{u}^{+}(k_{3})\,%
\widetilde{\phi }_{d}^{+}(k_{4})}  \label{Eq.MtmStochCFOneCooper}
\end{equation}%
where we have left the $t$ or $\beta $ dependences implicit to shorten the
notation.\bigskip

\subsection{Two Cooper Pairs: QCF}

For the case of time evolution the stochastic expressions for $%
X(d\,r_{1},d\,r_{2,}\,u\,r_{3},u\,r_{4})$ can be written in terms of
stochastic position or momentum fields as%
\begin{eqnarray}
&&X(d\,r_{1},d\,r_{2,}\,u\,r_{3},u\,r_{4};t)  \nonumber \\
&=&\overline{\widetilde{\psi }_{d}(r_{1},t)\,\widetilde{\psi }_{d}(r_{2},t)\,%
\widetilde{\psi }_{u}(r_{3},t)\,\widetilde{\psi }_{u}(r_{4},t)\;\widetilde{%
\psi }_{u}^{+}(r_{4},t)\,\widetilde{\psi }_{u}^{+}(r_{3},t)\,\widetilde{\psi 
}_{d}^{+}(r_{2},t)\,\widetilde{\psi }_{d}^{+}(r_{1},t)}  \nonumber \\
&&  \label{Eq.QCFTwoCooperTimeEvn} \\
&=&\frac{1}{(\sqrt{V})^{8}}\dsum\limits_{k_{1},k_{2},k_{3},k_{4}}\dsum%
\limits_{k_{5},k_{6},k_{7},k_{8}}\exp i\{(k_{1}-k_{8})\cdot r_{1}\}\;\exp
i\{(k_{2}-k_{7})\cdot r_{2}\}  \nonumber \\
&&\times \exp i\{(k_{3}-k_{6})\cdot r_{3}\}\;\exp i\{(k_{4}-k_{5})\cdot
r_{4}\}  \nonumber \\
&&\times \overline{\widetilde{\phi }_{d}(k_{1},t)\,\widetilde{\phi }%
_{d}(k_{2},t)\,\widetilde{\phi }_{u}(k_{3},t)\,\widetilde{\phi }%
_{u}(k_{4},t)\;\widetilde{\phi }_{u}^{+}(k_{5},t)\,\widetilde{\phi }%
_{u}^{+}(k_{6},t)\,\widetilde{\phi }_{d}^{+}(k_{7},t)\,\widetilde{\phi }%
_{d}^{+}(k_{8},t)}  \nonumber \\
&&  \label{Eq.QCFTwoCooperMtmTimeEvn}
\end{eqnarray}%
The case of temperature evolution leads to the same forms with $t$ replaced
by $\beta $ in Eq.(\ref{Eq.QCFTwoCooperTimeEvn}) and in Eq. (\ref%
{Eq.QCFTwoCooperMtmTimeEvn}), along with the additional factor $1/Z$ on the
right side of Eq. (\ref{Eq.QCFTwoCooperMtmTimeEvn}) to allow for $\widehat{%
\rho }$ being replaced by $\widehat{\sigma }$ when temperature evolution is
involved.

The momentum stochastic correlation function of interest is 
\begin{eqnarray}
&&X(d\,k_{1},d\,k_{2},u\,k_{3},u%
\,k_{4},u^{+}k_{5},u^{+}k_{6},d^{+}k_{7},d^{+}k_{8})  \nonumber \\
&=&\overline{\widetilde{\phi }_{d}(k_{1})\,\widetilde{\phi }_{d}(k_{2})\,%
\widetilde{\phi }_{u}(k_{3})\,\widetilde{\phi }_{u}(k_{4})\;\widetilde{\phi }%
_{u}^{+}(k_{5})\,\widetilde{\phi }_{u}^{+}(k_{6})\,\widetilde{\phi }%
_{d}^{+}(k_{7})\,\widetilde{\phi }_{d}^{+}(k_{8})}
\label{Eq.MtmStochCFTwoCooper}
\end{eqnarray}%
where we have left the $t$ or $\beta $ dependences implicit to shorten the
notation.

\subsection{Number of Fermions - Partition Function Factor}

The mean number of fermions can be obtained by taking the mean of the number
operator (\ref{Eq.NumberOpr}). The mean equals $N$, the fixed number of
fermions. By expanding the position field operators in terms of plane waves
and carrying out the space integrals we find that for temperature evolution 
\begin{eqnarray}
\left\langle \widehat{N}\right\rangle &=&N  \nonumber \\
&=&\frac{1}{Z}\sum_{\alpha }\int dr\mathbf{\,}\mbox{Tr}\hat{\Psi}_{\alpha
}(r)\,\widehat{\sigma }\,\hat{\Psi}_{\alpha }(r)^{\dag }  \nonumber \\
&=&\frac{1}{Z}\sum_{\alpha }\sum_{k}\overline{\widetilde{\phi }_{\alpha
}(k,\beta )\,\widetilde{\phi }_{\alpha }^{+}(k,\beta )}
\label{Eq.FermionNumber}
\end{eqnarray}%
showing there is a relationship between the partition function $Z$, the
number of fermions $N$ and QCF of the form $X(\alpha k,\alpha k)=\overline{%
\widetilde{\phi }_{\alpha }(k,\beta )\,\widetilde{\phi }_{\alpha }^{+}(k%
\mathbf{,}\beta )}$. In applying the theory to temperature evolution we have 
\begin{equation}
\frac{1}{Z}=\frac{N}{\left( \sum_{\alpha }\sum_{k}X(\alpha k,\alpha
k)\right) }  \label{Eq.ResultPartitionFn}
\end{equation}%
so the $1/Z$ factors in the results for $X(d\,r_{1},u\,r_{2})$ and $%
X(d\,r_{1},d\,r_{2,}\,u\,r_{3},u\,r_{4})$ may be replaced by the last
expression. Hence for example Eq.(\ref{Eq.QCFOneCooperMtmTimeEvn}) gives 
\begin{eqnarray}
&&\frac{X(d\,r_{1},u\,r_{2};\beta )}{N}  \nonumber \\
&=&\left\{ 
\begin{array}{c}
\frac{1}{(\sqrt{V})^{4}}\dsum\limits_{k_{1},k_{2},k_{3},k_{4}}\exp
i\{(k_{1}-k_{4})\cdot r_{1}\}\;\exp i\{(k_{2}-k_{3})\cdot r_{2}\} \\ 
\times \overline{\widetilde{\phi }_{d}(k_{1},\beta )\,\widetilde{\phi }%
_{u}(k_{2},\beta )\;\widetilde{\phi }_{u}^{+}(k_{3},\beta )\,\widetilde{\phi 
}_{d}^{+}(k_{4},\beta )}%
\end{array}%
\right\}  \nonumber \\
&&{\Large \div }\left\{ \sum_{\alpha }\sum_{k}\overline{\widetilde{\phi }%
_{\alpha }(k,\beta )\,\widetilde{\phi }_{\alpha }^{+}(k,\beta )}\right\}
\label{Eq.QCFOneCooperTempEvnNoPartitionFn}
\end{eqnarray}%
so only the stochastic momentum QCF are involved. \pagebreak

\section{Single Cooper Pair - First Order Changes to QCF}

\label{Section - One Cooper Pair First Order Changes to QCF}

In this Section we will derive expressions for the first order changes in
the QCF for a single Cooper pair based on stochastic momentum fields by
carrying out the stochastic averages analytically \ The results will be
correct to order $\delta t$ or $\delta \beta $ for time or temperature
evolution respectively.

\subsection{One Cooper Pair - Time Evolution}

The first order change in the QCF $\overline{\widetilde{\phi }_{d}(k_{1})\,%
\widetilde{\phi }_{u}(k_{2})\;\widetilde{\phi }_{u}^{+}(k_{3})\,\widetilde{%
\phi }_{d}^{+}(k_{4})}$ due to time evolution between $t$ and $t+\delta t$
is derived in Appendix \ref{Appendix B - Srochastic Averaging and QCF Eqns}
and the result is 
\begin{eqnarray}
&&\overline{\widetilde{\phi }_{d}(k_{1},t+\delta t)\,\widetilde{\phi }%
_{u}(k_{2},t+\delta t)\;\widetilde{\phi }_{u}^{+}(k_{3},t+\delta t)\,%
\widetilde{\phi }_{d}^{+}(k_{4},t+\delta t)}  \nonumber \\
&=&\dsum\limits_{l_{1}l_{2}l_{3}l_{4}}\delta _{k_{1},l_{1}}\,\delta
_{k_{2},l_{2}}\,\delta _{k_{3},l_{3}}\,\delta _{k_{4},l_{4}}\left( 1+\frac{i%
}{\hbar }\left\{ \frac{\hbar ^{2}k_{1}^{2}}{2m}+\frac{\hbar ^{2}k_{2}^{2}}{2m%
}-\frac{\hbar ^{2}k_{3}^{2}}{2m}-\frac{\hbar ^{2}k_{4}^{2}}{2m}\right\}
\delta t\right)  \nonumber \\
&&\times \overline{\widetilde{\phi }_{d}(l_{1},t)\,\widetilde{\phi }%
_{u}(l_{2},t)\;\widetilde{\phi }_{u}^{+}(l_{3},t)\,\widetilde{\phi }%
_{d}^{+}(l_{4},t)}  \nonumber \\
&&+2\lambda ^{2}\,\dsum\limits_{l_{1}l_{2}l_{3}l_{4}}\{\delta
_{k_{1},l_{1}}\,\delta _{k_{2},l_{2}}\,\delta
_{(k_{3}+k_{4}),(l_{3}+l_{4})}\}\,\delta t\times \overline{\widetilde{\phi }%
_{d}(l_{1},t)\,\widetilde{\phi }_{u}(l_{2},t)\;\widetilde{\phi }%
_{u}^{+}(l_{3},t)\,\widetilde{\phi }_{d}^{+}(l_{4},t)}  \nonumber \\
&&\mathbf{-}2\lambda ^{2}\,\dsum\limits_{l_{1}l_{2}l_{3}l_{4}}\{\delta
_{(k_{1}+k_{2}),(l_{1}+l_{2})}\,\delta _{k_{3},l_{3}}\,\delta
_{k_{4},l_{4}}\}\,\delta t\times \overline{\widetilde{\phi }_{d}(l_{1},t)\,%
\widetilde{\phi }_{u}(l_{2},t)\;\widetilde{\phi }_{u}^{+}(l_{3},t)\,%
\widetilde{\phi }_{d}^{+}(l_{4},t)}  \nonumber \\
&&  \label{Eq.QCFOnePairTimeEvnFirstOrderChange}
\end{eqnarray}%
where $\lambda =\sqrt{\frac{-\,ig}{2\hbar V}}$. The key feature of this last
result is that the

$\overline{\widetilde{\phi }_{d}(k_{1})\,\widetilde{\phi }_{u}(k_{2})\;%
\widetilde{\phi }_{u}^{+}(k_{3})\,\widetilde{\phi }_{d}^{+}(k_{4})}$ at time 
$t+\delta t$ depend linearly on the $\overline{\widetilde{\phi }_{d}(l_{1})\,%
\widetilde{\phi }_{u}(l_{2})\;\widetilde{\phi }_{u}^{+}(l_{3})\,\widetilde{%
\phi }_{d}^{+}(l_{4})}$ at time $t$, and the change in these QCF is
proportional to $\delta t$. Furthermore, the QCF are c-numbers so their time
dependence can be calculated on a computer. The Grassmann variables are no
longer present. The behaviour depends of course on the initial conditions
for $\overline{\widetilde{\phi }_{d}(k_{1})\,\widetilde{\phi }_{u}(k_{2})\;%
\widetilde{\phi }_{u}^{+}(k_{3})\,\widetilde{\phi }_{d}^{+}(k_{4})}$ at $t=0$%
, and this will be discussed below.\smallskip

\subsection{Initial Condition - Time Evolution from t=0}

A natural initial condition at $t=0$ would be to assume the fermionic atoms
were non-interacting, and to turn on the interaction via Feshbach resonance
methods. The choice of the interaction constant could span the range for the
BEC/BCS crossover. If the $N$ fermionic atoms are non-interacting the
initial state could be that for zero temperature, in which case all plane
wave modes up to the Fermi surface would be occupied by one spin down atom
and one spin up atom. The density operator is based on a pure state $%
\left\vert \Psi \right\rangle $ 
\begin{eqnarray}
\widehat{\rho } &=&\left\vert \Psi \right\rangle \left\langle \Psi
\right\vert  \nonumber \\
\left\vert \Psi \right\rangle &=&\dprod\limits_{k,(|k|\leq k_{F})}\left\vert
1_{ku}\right\rangle \otimes \dprod\limits_{k,(|k|\leq k_{F})}\left\vert
1_{kd}\right\rangle  \label{Eq.InitialStateTimeEvoln}
\end{eqnarray}%
where $k_{F}=(3\pi ^{2}N/V)^{1/3\text{ \thinspace }}$is the radius of the
Fermi sphere.

Hence%
\begin{eqnarray}
&&\widehat{\Phi }_{d}(k_{1})\,\widehat{\Phi }_{u}(k_{2})\,\left\vert \Psi
\right\rangle  \nonumber \\
&=&\dprod\limits_{k\neq k_{1},(|k|\leq k_{F})}\left\vert 1_{ku}\right\rangle
\,\left\vert 0_{k_{1}u}\right\rangle \otimes \dprod\limits_{k\neq
k_{2},(|k|\leq k_{F})}\left\vert 1_{kd}\right\rangle \,\left\vert
0_{k_{2}u}\right\rangle \qquad k_{1,}k_{2}\quad inside\;FS  \nonumber \\
&=&0\qquad k_{1,}k_{2}\quad outside\;FS  \label{Eq.InitialStateMtmOprs}
\end{eqnarray}%
so that for $t=0$%
\begin{eqnarray}
&&\overline{\widetilde{\phi }_{d}(k_{1})\,\widetilde{\phi }_{u}(k_{2})\;%
\widetilde{\phi }_{u}^{+}(k_{3})\,\widetilde{\phi }_{d}^{+}(k_{4})} 
\nonumber \\
&=&\mbox{Tr}(\widehat{\Phi }_{d}(k_{1})\,\widehat{\Phi }_{u}(k_{2})\,\hat{%
\rho}\,\widehat{\Phi }_{u}(k_{3})^{\dag }\,\widehat{\Phi }_{d}(k_{4})^{\dag
}).  \nonumber \\
&=&\delta _{k_{1},k_{4}}\;\delta _{k_{2},k_{3}}\qquad
k_{1,}k_{2},k_{3},k_{4}\quad inside\;FS  \nonumber \\
&=&0\qquad otherwise  \label{Eq.InitialNonZeroQCFTimeEvn}
\end{eqnarray}

Thus from Eq.(\ref{Eq.QCFOnePairTimeEvnFirstOrderChange}) the non-zero $%
\overline{\widetilde{\phi }_{d}(k_{1})\,\widetilde{\phi }_{u}(k_{2})\;%
\widetilde{\phi }_{u}^{+}(k_{3})\,\widetilde{\phi }_{d}^{+}(k_{4})}$ at $%
t=\delta t$ that are present in first order must originate from $\overline{%
\widetilde{\phi }_{d}(l_{1},0)\,\widetilde{\phi }_{u}(l_{2},0)\;\widetilde{%
\phi }_{u}^{+}(l_{2},0)\,\widetilde{\phi }_{d}^{+}(l_{1},0)}=1$ with $%
l_{1},l_{2}$ not outside the Fermi sphere. These are $\overline{\widetilde{%
\phi }_{d}(l_{1},\delta t)\,\widetilde{\phi }_{u}(l_{2},\delta t)\;%
\widetilde{\phi }_{u}^{+}(k_{3},\delta t)\,\widetilde{\phi }%
_{d}^{+}(k_{4},\delta t)}$ where $k_{3}+k_{4}=l_{2}+l_{1}$ from the first $%
2\lambda ^{2}$ term and $\overline{\widetilde{\phi }_{d}(k_{1},\delta t)\,%
\widetilde{\phi }_{u}(k_{2},\delta t)\;\widetilde{\phi }_{u}^{+}(l_{2},%
\delta t)\,\widetilde{\phi }_{d}^{+}(l_{1},\delta t)}$ where $%
k_{1}+k_{2}=l_{1}+l_{2}$ from the second $2\lambda ^{2}$ term \ These
restrictions represent momentum conservation.

A specific case of interest is where $l_{1}=l_{4}=+k$ and $l_{2}=l_{3}=-k$,
where $k$ is on \ the Fermi sphere. The term $\overline{\widetilde{\phi }%
_{d}(+k,\delta t)\,\widetilde{\phi }_{u}(-k,\delta t)\;\widetilde{\phi }%
_{u}^{+}(+k,\delta t)\,\widetilde{\phi }_{d}^{+}(-k,\delta t)}$ could
originate from the first $2\lambda ^{2}\times $ $\overline{\widetilde{\phi }%
_{d}(+k,0)\,\widetilde{\phi }_{u}(-k,0)\;\widetilde{\phi }_{u}^{+}(-k,0)\,%
\widetilde{\phi }_{d}^{+}(+k,0)}$ term but would not originate from the
second $2\lambda ^{2}$ $\times $ $\overline{\widetilde{\phi }_{d}(+k,0)\,%
\widetilde{\phi }_{u}(-k,0)\;\widetilde{\phi }_{u}^{+}(-k,0)\,\widetilde{%
\phi }_{d}^{+}(+k,0)}$ term \ Hence there would be no cancelation of the
contributions. On the other hand, for the case where $l_{1}=+k$, $l_{2}=-k$, 
$l_{3}=-k$ and $l_{4}=+k$ where $k$ is on the Fermi sphere, the term $%
\overline{\widetilde{\phi }_{d}(+k,\delta t)\,\widetilde{\phi }%
_{u}(-k,\delta t)\;\widetilde{\phi }_{u}^{+}(-k,\delta t)\,\widetilde{\phi }%
_{d}^{+}(+k,\delta t)}$ could originate from the first $2\lambda ^{2}\times $
$\overline{\widetilde{\phi }_{d}(+k,0)\,\widetilde{\phi }_{u}(-k,0)\;%
\widetilde{\phi }_{u}^{+}(-k,0)\,\widetilde{\phi }_{d}^{+}(+k,0)}$ term and
could also originate from the second $2\lambda ^{2}$ $\times $ $\overline{%
\widetilde{\phi }_{d}(+k,0)\,\widetilde{\phi }_{u}(-k,0)\;\widetilde{\phi }%
_{u}^{+}(-k,0)\,\widetilde{\phi }_{d}^{+}(+k,0)}$ term. However in this case
the two contributions cancel out. $\medskip $

\subsection{One Cooper Pair - Temperature Evolution}

The first order change in the QCF $\overline{\widetilde{\phi }_{d}(k_{1})\,%
\widetilde{\phi }_{u}(k_{2})\;\widetilde{\phi }_{u}^{+}(k_{3})\,\widetilde{%
\phi }_{d}^{+}(k_{4})}$ due to temperature evolution between $\beta $ and $%
\beta +\delta \beta $ is derived in Appendix \ref{Appendix B - Srochastic
Averaging and QCF Eqns} and the result is 
\begin{eqnarray}
&&\overline{\widetilde{\phi }_{d}(k_{1},\beta +\delta \beta )\,\widetilde{%
\phi }_{u}(k_{2},\beta +\delta \beta )\;\widetilde{\phi }_{u}^{+}(k_{3},%
\beta +\delta \beta )\,\widetilde{\phi }_{d}^{+}(k_{4},\beta +\delta \beta )}
\nonumber \\
&=&\dsum\limits_{l_{1}l_{2}l_{3}l_{4}}\delta _{k_{1},l_{1}}\,\delta
_{k_{2},l_{2}}\,\delta _{k_{3},l_{3}}\,\delta _{k_{4},l_{4}}\left( 1+\frac{1%
}{2}\left\{ \frac{\hbar ^{2}k_{1}^{2}}{2m}+\frac{\hbar ^{2}k_{2}^{2}}{2m}+%
\frac{\hbar ^{2}k_{3}^{2}}{2m}+\frac{\hbar ^{2}k_{4}^{2}}{2m}\right\} \delta
\beta \right)  \nonumber \\
&&\times \overline{\widetilde{\phi }_{d}(l_{1},\beta )\,\widetilde{\phi }%
_{u}(l_{2},\beta )\;\widetilde{\phi }_{u}^{+}(l_{3},\beta )\,\widetilde{\phi 
}_{d}^{+}(l_{4},\beta )}  \nonumber \\
&&-2\eta ^{2}\,  \nonumber \\
&&\times \dsum\limits_{l_{1}l_{2}l_{3}l_{4}}\{\delta _{k_{1},l_{1}}\,\delta
_{k_{2},l_{2}}\,\delta _{(k_{3}+k_{4}),(l_{3}+l_{4})}\}\,\delta \beta \times 
\overline{\widetilde{\phi }_{d}(l_{1},\beta )\,\widetilde{\phi }%
_{u}(l_{2},\beta )\;\widetilde{\phi }_{u}^{+}(l_{3},\beta )\,\widetilde{\phi 
}_{d}^{+}(l_{4},\beta )}  \nonumber \\
&&-2\eta ^{2}\,  \nonumber \\
&&\times \dsum\limits_{l_{1}l_{2}l_{3}l_{4}}\{\delta
_{(k_{1}+k_{2}),(l_{1}+l_{2})}\,\delta _{k_{3},l_{3}}\,\delta
_{k_{4},l_{4}}\}\,\delta \beta \times \overline{\widetilde{\phi }%
_{d}(l_{1},\beta )\,\widetilde{\phi }_{u}(l_{2},\beta )\;\widetilde{\phi }%
_{u}^{+}(l_{3},\beta )\,\widetilde{\phi }_{d}^{+}(l_{4},\beta )}  \nonumber
\\
&&  \label{Eq.QCFOnePairTempEvnFirstOrderChangeA}
\end{eqnarray}%
where $\eta =\sqrt{\frac{-g}{4V}}$. Similar comments to the time evolution
case apply here as well. The initial conditions are discussed below.\medskip

\subsection{Initial Condition - Temperature Evolution from $\protect\beta =0$%
}

A natural initial condition at $\beta =0$ would be to assume the fermionic
atoms were non-interacting, and to turn on the interaction via Feshbach
resonance methods. The choice of the interaction constant could span the
range for the BEC/BCS crossover. If the $N$ fermionic atoms are
non-interacting the initial state could be that for a very high temperature,
in which case a large number of plane wave modes could be occupied, both
inside and outside the Fermi surface by at most one spin down atom and one
spin up atom. The density operator is based on a mixed state. Suppose we
consider the case where there are $n$ different modes $k$, where $|k|$ would
range from being $\ll k_{F}$ to being $\gg k_{F}$. The one-fermion states
are $\left\vert 1_{ku}\right\rangle $ or $\left\vert 1_{kd}\right\rangle $
if occupied, and $\left\vert 0_{ku}\right\rangle $ or $\left\vert
0_{kd}\right\rangle $ if unoccupied. So a typical $N$ fermion state will be $%
\left\vert \lambda \right\rangle =\dprod\limits_{k}\;\left\vert \mu
_{ku}\right\rangle \left\vert \nu _{kd}\right\rangle $ where $\mu ,\nu =0,1$
only and with $\lambda \equiv \{\mu _{ku},\nu _{kd}\}$. The number of
distinct states $\left\vert \lambda \right\rangle $ is given by the number
of ways of choosing $N$ occupied one fermion states out of the set of $2n$
possible one fermion states, and this is given by $\mathcal{N=}%
^{2n}C_{N}=2n!/[(N)!(2n-N)!]\approx (2n)^{N}/N!.$since we have $n\gg N$. At
high temperatures each of these states $\left\vert \lambda \right\rangle $
is equally probable, so the density operator is given by 
\begin{eqnarray}
\widehat{\rho } &=&\frac{1}{\mathcal{N}}\dsum\limits_{\lambda }\left\vert
\lambda \right\rangle \left\langle \lambda \right\vert  \nonumber \\
&\approx &\frac{N!}{(2n)^{N}}\dsum\limits_{\lambda }\left\vert \lambda
\right\rangle \left\langle \lambda \right\vert
\label{Eq.InitialDensOprTempEvoln}
\end{eqnarray}

We then want to consider the quantity 
\begin{eqnarray}
&&\mbox{Tr}(\widehat{\Phi }_{d}(k_{1})\,\widehat{\Phi }_{u}(k_{2})\,\hat{\rho%
}\,\widehat{\Phi }_{u}(k_{3})^{\dag }\,\widehat{\Phi }_{d}(k_{4})^{\dag }) 
\nonumber \\
&=&\frac{1}{\mathcal{N}}\dsum\limits_{\lambda }\mbox{Tr}(\widehat{\Phi }%
_{d}(k_{1})\,\widehat{\Phi }_{u}(k_{2})\,(\left\vert \lambda \right\rangle
\left\langle \lambda \right\vert )\,\widehat{\Phi }_{u}(k_{3})^{\dag }\,%
\widehat{\Phi }_{d}(k_{4})^{\dag })
\end{eqnarray}%
so to get a non-zero contribution for a given $\lambda $ we need $\widehat{%
\Phi }_{d}(k_{1})\,\widehat{\Phi }_{u}(k_{2})\,\left\vert \lambda
\right\rangle \neq 0$ and $\widehat{\Phi }_{d}(k_{4})\,\widehat{\Phi }%
_{u}(k_{3})\,\left\vert \lambda \right\rangle \neq 0$. Noting that $\widehat{%
\Phi }_{\alpha }(k)\left\vert 1_{k\alpha }\right\rangle $ $=\left\vert
0_{k\alpha }\right\rangle \,$and $\widehat{\Phi }_{\alpha }(k)\left\vert
0_{k\alpha }\right\rangle $ $=0$ we see that to get a no-zero contribution
for a given $\lambda $ we require $k_{1}=k_{4}$ and $k_{2}=k_{3}$ which
restricts $\lambda $ to be of the form $\left\vert \lambda
^{\#}\right\rangle =\left\vert 1_{k_{1}d}\right\rangle \left\vert
1_{k_{2}u}\right\rangle \dprod\limits_{k\neq k_{1,}k_{2}}\;\left\vert \mu
_{ku}\right\rangle \left\vert \nu _{kd}\right\rangle $. Similarly to the
previous situation, there are $\mathcal{N}^{\#}=^{2n-2}C_{N-2}\approx
(2n)^{(N-2)}/(N-2)!$ different states $\left\vert \lambda ^{\#}\right\rangle 
$ (since this equals the number of ways of choosing $N-2$ occupied one
fermion states from $2n-2$ possible remaining one fermion states). In this
case we have 
\begin{eqnarray}
&&\mbox{Tr}(\widehat{\Phi }_{d}(k_{1})\,\widehat{\Phi }_{u}(k_{2})\,\hat{\rho%
}\,\widehat{\Phi }_{u}(k_{3})^{\dag }\,\widehat{\Phi }_{d}(k_{4})^{\dag }) 
\nonumber \\
&=&\frac{1}{\mathcal{N}}\delta _{k_{1,}k_{4}}\,\delta
_{k_{2,}k_{3}}\,\dsum\limits_{\lambda ^{\#}}\mbox{Tr}(\,\left\vert
0_{k_{1}d}\right\rangle \left\vert 0_{k_{2}u}\right\rangle \left(
\dprod\limits_{k\neq k_{1,}k_{2}}\;\left\vert \mu _{ku}\right\rangle
\left\vert \nu _{kd}\right\rangle \otimes \left\langle \mu _{ku}\right\vert
\left\langle \nu _{kd}\right\vert \right) \,\left\langle
0_{k_{1}d}\right\vert \left\langle 0_{k_{2}u}\right\vert \,)  \nonumber \\
&=&\frac{1}{\mathcal{N}}\delta _{k_{1,}k_{4}}\,\delta
_{k_{2,}k_{3}}\,\dsum\limits_{\lambda ^{\#}}\,1=\frac{\mathcal{N}^{\#}}{%
\mathcal{N}}\delta _{k_{1,}k_{4}}\,\delta _{k_{2,}k_{3}}\,  \nonumber \\
&\approx &\frac{1}{4}\left( \frac{N}{n}\right) ^{2}\delta
_{k_{1,}k_{4}}\,\delta _{k_{2,}k_{3}}
\end{eqnarray}

Writing $\widehat{\sigma }=Z\,\widehat{\rho }$ we then see that at $\beta =0$
\begin{eqnarray}
&&\overline{\widetilde{\phi }_{d}(k_{1})\,\widetilde{\phi }_{u}(k_{2})\;%
\widetilde{\phi }_{u}^{+}(k_{3})\,\widetilde{\phi }_{d}^{+}(k_{4})} 
\nonumber \\
&=&\mbox{Tr}(\widehat{\Phi }_{d}(k_{1})\,\widehat{\Phi }_{u}(k_{2})\,%
\widehat{\sigma }\,\widehat{\Phi }_{u}(k_{3})^{\dag }\,\widehat{\Phi }%
_{d}(k_{4})^{\dag }).  \nonumber \\
&\approx &\frac{Z}{4}\left( \frac{N}{n}\right) ^{2}\delta
_{k_{1,}k_{4}}\,\delta _{k_{2,}k_{3}}  \label{Eq.InitialNonZeroQCFTempEvn}
\end{eqnarray}%
Thus from Eq.(\ref{Eq.QCFOnePairTempEvnFirstOrderChangeA}) the non-zero $%
\overline{\widetilde{\phi }_{d}(k_{1})\,\widetilde{\phi }_{u}(k_{2})\;%
\widetilde{\phi }_{u}^{+}(k_{3})\,\widetilde{\phi }_{d}^{+}(k_{4})}$ at $%
\beta =\delta \beta $ that are present in first order must originate from $%
\overline{\widetilde{\phi }_{d}(l_{1},0)\,\widetilde{\phi }_{u}(l_{2},0)\;%
\widetilde{\phi }_{u}^{+}(l_{2},0)\,\widetilde{\phi }_{d}^{+}(l_{1},0)}%
\approx \frac{Z}{4}\left( \frac{N}{n}\right) ^{2}$ with $l_{1},l_{2}$ not
restricted to the Fermi sphere. These are $\overline{\widetilde{\phi }%
_{d}(l_{1},\delta \beta )\,\widetilde{\phi }_{u}(l_{2},\delta \beta )\;%
\widetilde{\phi }_{u}^{+}(k_{3},\delta \beta )\,\widetilde{\phi }%
_{d}^{+}(k_{4},\delta \beta )}$ where $k_{3}+k_{4}=l_{2}+l_{1}$ from the
first $2\eta ^{2}$ term and $\overline{\widetilde{\phi }_{d}(k_{1},\delta
\beta )\,\widetilde{\phi }_{u}(k_{2},\delta \beta )\;\widetilde{\phi }%
_{u}^{+}(l_{2},\delta \beta )\,\widetilde{\phi }_{d}^{+}(l_{1},\delta \beta )%
}$ where $k_{1}+k_{2}=l_{1}+l_{2}$ from the second $2\eta ^{2}$ term \ These
restrictions represent momentum conservation.

A specific case of interest is where $l_{1}=l_{4}=+k$ and $l_{2}=l_{3}=-k$,
where $k$ need not be on \ the Fermi sphere. The term $\overline{\widetilde{%
\phi }_{d}(+k,\delta \beta )\,\widetilde{\phi }_{u}(-k,\delta \beta )\;%
\widetilde{\phi }_{u}^{+}(+k,\delta \beta )\,\widetilde{\phi }%
_{d}^{+}(-k,\delta \beta )}$ could originate from the first $2\eta
^{2}\times $ $\overline{\widetilde{\phi }_{d}(+k,0)\,\widetilde{\phi }%
_{u}(-k,0)\;\widetilde{\phi }_{u}^{+}(-k,0)\,\widetilde{\phi }_{d}^{+}(+k,0)}
$ term but would not originate from the second $2\eta ^{2}$ $\times $ $%
\overline{\widetilde{\phi }_{d}(+k,0)\,\widetilde{\phi }_{u}(-k,0)\;%
\widetilde{\phi }_{u}^{+}(-k,0)\,\widetilde{\phi }_{d}^{+}(+k,0)}$ term \
Hence there would be no cancelation of the contributions. On the other hand,
for the case where $l_{1}=+k$, $l_{2}=-k$, $l_{3}=-k$ and $l_{4}=+k$ where $%
k $ need not be on the Fermi sphere, the term $\overline{\widetilde{\phi }%
_{d}(+k,\delta \beta )\,\widetilde{\phi }_{u}(-k,\delta \beta )\;\widetilde{%
\phi }_{u}^{+}(-k,\delta \beta )\,\widetilde{\phi }_{d}^{+}(+k,\delta \beta )%
}$ could originate from the first $2\eta ^{2}\times $ $\overline{\widetilde{%
\phi }_{d}(+k,0)\,\widetilde{\phi }_{u}(-k,0)\;\widetilde{\phi }%
_{u}^{+}(-k,0)\,\widetilde{\phi }_{d}^{+}(+k,0)}$ term and could also
originate from the second $2\eta ^{2}$ $\times $ $\overline{\widetilde{\phi }%
_{d}(+k,0)\,\widetilde{\phi }_{u}(-k,0)\;\widetilde{\phi }_{u}^{+}(-k,0)\,%
\widetilde{\phi }_{d}^{+}(+k,0)}$ term. However in this temperature
evolution case the two contributions do not cancel out - in contrast to the
time evolution case. \pagebreak

\section{Two Cooper Pairs - First Order Changes to QCF}

\label{Section - Two Cooper Pairs First Order Changes to QCF}

In this Section we will derive expressions for the first order changes in
the QCF for two Cooper pairs based on stochastic momentum fields by carrying
out the stochastic averages analytically \ The results will be correct to
order $\delta t$ or $\delta \beta $ for time or temperature evolution
respectively.

\subsection{Two Cooper Pairs - Time Evolution}

For the case of time evolution the QCF for two Cooper pairs at time $%
t+\delta t$ is related to that at time $t$ via 
\begin{eqnarray}
&&\left[ 
\begin{array}{c}
\widetilde{\phi }_{d}(k_{1},t+\delta t)\,\widetilde{\phi }%
_{d}(k_{2},t+\delta t)\;\widetilde{\phi }_{u}(k_{3},t+\delta t)\,\widetilde{%
\phi }_{u}(k_{4},t+\delta t) \\ 
\times \;\widetilde{\phi }_{u}^{+}(k_{5},t+\delta t)\,\widetilde{\phi }%
_{u}^{+}(k_{6},t+\delta t)\widetilde{\phi }_{d}^{+}(k_{7},t+\delta t)\,%
\widetilde{\phi }_{d}^{+}(k_{8},t+\delta t)%
\end{array}%
\right] _{StochAver}  \nonumber \\
&&  \nonumber \\
&=&\dsum\limits_{l_{1}}\dsum\limits_{l_{2}}\dsum\limits_{l_{3}}\dsum%
\limits_{l_{4}}\dsum\limits_{l_{5}}\dsum\limits_{l_{6}}\dsum\limits_{l_{7}}%
\dsum\limits_{l_{8}}  \nonumber \\
&&\left[ 
\begin{array}{c}
(\delta _{k_{1},l_{1}}\,\delta _{k_{2},l_{2}}\,\delta _{k_{3},l_{3}}\,\delta
_{k_{4},l_{4}}\,\delta _{k_{5},l_{5}}\,\delta _{k_{6},l_{6}}\,\delta
_{k_{7},l_{7}}\,\delta _{k_{8},l_{8}}) \\ 
\times \left( 1+\frac{i}{\hbar }\left\{ \frac{\hbar ^{2}k_{1}^{2}}{2m}+\frac{%
\hbar ^{2}k_{2}^{2}}{2m}+\frac{\hbar ^{2}k_{3}^{2}}{2m}+\frac{\hbar
^{2}k_{4}^{2}}{2m}-\frac{\hbar ^{2}k_{5}^{2}}{2m}-\frac{\hbar ^{2}k_{6}^{2}}{%
2m}-\frac{\hbar ^{2}k_{7}^{2}}{2m}-\frac{\hbar ^{2}k_{8}^{2}}{2m}\right\}
\delta t\right) \\ 
\\ 
+(-2\lambda ^{2})\,\delta _{k_{1},l_{1}}\delta _{k_{2},l_{2}}\delta
_{k_{3},l_{3}}\delta _{k_{4},l_{4}}\,\delta t \\ 
\times \left( 
\begin{array}{c}
\delta _{k_{6},l_{5}}\delta _{k_{8},l_{8}}\;\delta
_{(k_{5}+k_{7}),(l_{6}+l_{7})}+\delta _{k_{6},l_{5}}\delta
_{k_{7},l_{7}}\;\delta _{(k_{5}+k_{8}),(l_{6}+l_{8})} \\ 
+\delta _{k_{5},l_{6}}\delta _{k_{8},l_{8}}\;\delta
_{(k_{6}+k_{7}),(l_{5}+l_{7})}+\delta _{k_{5},l_{6}}\delta
_{k_{7},l_{7}}\;\delta _{(k_{6}+k_{8}),(l_{5}+l_{8})}%
\end{array}%
\right) \\ 
\\ 
+(2\lambda ^{2})\,\delta _{k_{5},l_{5}}\,\delta _{k_{6},l_{6}}\,\delta
_{k_{7},l_{7}}\,\delta _{k_{8},l_{8}}\,\delta t \\ 
\times \left( 
\begin{array}{c}
\delta _{k_{1},l_{1}}\delta _{k_{3},l_{4}}\;\delta
_{(k_{2}+k_{4}),(l_{2}+l_{3})}+\delta _{k_{1},l_{1}}\delta
_{k_{4},l_{3}}\;\delta _{(k_{2}+k_{3}),(l_{2}+l_{4})} \\ 
+\delta _{k_{2},l_{2}}\delta _{k_{3},l_{4}}\;\delta
_{(k_{1}+k_{4}),(l_{1}+l_{3})}+\delta _{k_{2},l_{2}}\delta
_{k_{4},l_{3}}\;\delta _{(k_{1}+k_{3}),(l_{1}+l_{4})}%
\end{array}%
\right)%
\end{array}%
\right]  \nonumber \\
&&  \nonumber \\
&&\times \left[ 
\begin{array}{c}
\widetilde{\phi }_{d}(l_{1},t)\,\widetilde{\phi }_{d}(l_{2},t)\;\widetilde{%
\phi }_{u}(l_{3},t)\,\widetilde{\phi }_{u}(l_{4},t) \\ 
\times \;\widetilde{\phi }_{u}^{+}(l_{5},t)\,\widetilde{\phi }%
_{u}^{+}(l_{6},t)\widetilde{\phi }_{d}^{+}(l_{7},t)\,\widetilde{\phi }%
_{d}^{+}(l_{8},t)%
\end{array}%
\right] _{StochAver}  \nonumber \\
&&  \label{Eq.QCFOTwoPairsTempEvnFirstOrderChangeA}
\end{eqnarray}%
where $\lambda =\sqrt{\frac{-\,ig}{2\hbar V}}$. Thus the first order change
in the QCF for two Cooper pairs depends linearly on $\delta t$. The
derivation is given in Sect.\ref{SubSection - Two Cooper Pairs Time Evoln} \
Note the signs for the $2\lambda ^{2}$ terms are opposite and that half of
the kinetic energy terms $\frac{\hbar ^{2}k^{2}}{2m}$ have opposite signs.
Similar intial conditions to the single Cooper pair time evolution apply.
\smallskip

\subsection{Two Cooper Pairs - Temperature Evolution}

For the case of temperature evolution the QCF for two Cooper pairs at
temperature $\beta +\delta \beta $ is related to that at temperature $\beta $
via 
\begin{eqnarray}
&&\left[ 
\begin{array}{c}
\widetilde{\phi }_{d}(k_{1},\beta +\delta \beta )\,\widetilde{\phi }%
_{d}(k_{2},\beta +\delta \beta )\;\widetilde{\phi }_{u}(k_{3},\beta +\delta
\beta )\,\widetilde{\phi }_{u}(k_{4},\beta +\delta \beta ) \\ 
\times \;\widetilde{\phi }_{u}^{+}(k_{5},\beta +\delta \beta )\,\widetilde{%
\phi }_{u}^{+}(k_{6},\beta +\delta \beta )\widetilde{\phi }%
_{d}^{+}(k_{7},\beta +\delta \beta )\,\widetilde{\phi }_{d}^{+}(k_{8},\beta
+\delta \beta )%
\end{array}%
\right] _{StochAver}  \nonumber \\
&&  \nonumber \\
&=&\dsum\limits_{l_{1}}\dsum\limits_{l_{2}}\dsum\limits_{l_{3}}\dsum%
\limits_{l_{4}}\dsum\limits_{l_{5}}\dsum\limits_{l_{6}}\dsum\limits_{l_{7}}%
\dsum\limits_{l_{8}}  \nonumber \\
&&\left[ 
\begin{array}{c}
(\delta _{k_{1},l_{1}}\,\delta _{k_{2},l_{2}}\,\delta _{k_{3},l_{3}}\,\delta
_{k_{4},l_{4}}\,\delta _{k_{5},l_{5}}\,\delta _{k_{6},l_{6}}\,\delta
_{k_{7},l_{7}}\,\delta _{k_{8},l_{8}}) \\ 
\times \left( 1+\frac{1}{2}\left\{ \frac{\hbar ^{2}k_{1}^{2}}{2m}+\frac{%
\hbar ^{2}k_{2}^{2}}{2m}+\frac{\hbar ^{2}k_{3}^{2}}{2m}+\frac{\hbar
^{2}k_{4}^{2}}{2m}+\frac{\hbar ^{2}k_{5}^{2}}{2m}+\frac{\hbar ^{2}k_{6}^{2}}{%
2m}+\frac{\hbar ^{2}k_{7}^{2}}{2m}+\frac{\hbar ^{2}k_{8}^{2}}{2m}\right\}
\delta \beta \right) \\ 
\\ 
+(+2\eta ^{2})\,\delta _{k_{1},l_{1}}\delta _{k_{2},l_{2}}\delta
_{k_{3},l_{3}}\delta _{k_{4},l_{4}}\,\delta \beta \\ 
\times \left( 
\begin{array}{c}
\delta _{k_{6},l_{5}}\delta _{k_{8},l_{8}}\;\delta
_{(k_{5}+k_{7}),(l_{6}+l_{7})}+\delta _{k_{6},l_{5}}\delta
_{k_{7},l_{7}}\;\delta _{(k_{5}+k_{8}),(l_{6}+l_{8})} \\ 
+\delta _{k_{5},l_{6}}\delta _{k_{8},l_{8}}\;\delta
_{(k_{6}+k_{7}),(l_{5}+l_{7})}+\delta _{k_{5},l_{6}}\delta
_{k_{7},l_{7}}\;\delta _{(k_{6}+k_{8}),(l_{5}+l_{8})}%
\end{array}%
\right) \\ 
\\ 
+(2\eta ^{2})\,\delta _{k_{5},l_{5}}\,\delta _{k_{6},l_{6}}\,\delta
_{k_{7},l_{7}}\,\delta _{k_{8},l_{8}}\,\delta \beta \\ 
\times \left( 
\begin{array}{c}
\delta _{k_{1},l_{1}}\delta _{k_{3},l_{4}}\;\delta
_{(k_{2}+k_{4}),(l_{2}+l_{3})}+\delta _{k_{1},l_{1}}\delta
_{k_{4},l_{3}}\;\delta _{(k_{2}+k_{3}),(l_{2}+l_{4})} \\ 
+\delta _{k_{2},l_{2}}\delta _{k_{3},l_{4}}\;\delta
_{(k_{1}+k_{4}),(l_{1}+l_{3})}+\delta _{k_{2},l_{2}}\delta
_{k_{4},l_{3}}\;\delta _{(k_{1}+k_{3}),(l_{1}+l_{4})}%
\end{array}%
\right)%
\end{array}%
\right]  \nonumber \\
&&  \nonumber \\
&&\times \left[ 
\begin{array}{c}
\widetilde{\phi }_{d}(l_{1},\beta )\,\widetilde{\phi }_{d}(l_{2},\beta )\;%
\widetilde{\phi }_{u}(l_{3},\beta )\,\widetilde{\phi }_{u}(l_{4},\beta ) \\ 
\times \;\widetilde{\phi }_{u}^{+}(l_{5},\beta )\,\widetilde{\phi }%
_{u}^{+}(l_{6},\beta )\widetilde{\phi }_{d}^{+}(l_{7},\beta )\,\widetilde{%
\phi }_{d}^{+}(l_{8},\beta )%
\end{array}%
\right] _{StochAver}  \nonumber \\
&&  \label{Eq.QCFOTwoPairsTempEvnFirstOrderChangeB}
\end{eqnarray}%
where $\eta =\sqrt{\frac{-\,g}{4V}}$. Thus the first order change in the QCF
for two Cooper pairs depends linearly on $\delta \beta $. The derivation is
similar to the case of time evolution. Note the signs for the $2\eta ^{2}$
terms are now the same and that all the kinetetic energy terms $\frac{\hbar
^{2}k^{2}}{2m}$ also have the same sign. Similar intial conditions to the
single Cooper pair temperature evolution also apply.\pagebreak

\section{Solution of Evolution Equations for QCF}

\label{Section - Solution of Evolution Equations for QCF}

The evolution equations for the QCF can be written in terms of the
definitions in Eqs.(\ref{Eq.MtmStochCFOneCooper}) and (\ref%
{Eq.MtmStochCFTwoCooper}) for the one Cooper pair QCF $X(d\,k_{1},u%
\,k_{2},u^{+}\,k_{3},d^{+}\,k_{4})$ and the two Cooper pair QCF $%
X(d\,k_{1},d\,k_{2},u\,k_{3},u%
\,k_{4},u^{+}k_{5},u^{+}k_{6},d^{+}k_{7},d^{+}k_{8})$ in the forms 
\begin{eqnarray}
&&\frac{\partial }{\partial \xi }X(d\,k_{1},u\,k_{2},u^{+}\,k_{3},d^{+}%
\,k_{4})  \nonumber \\
&=&\dsum%
\limits_{l_{1}l_{2}l_{3}l_{4}}M(k_{1},k_{2},k_{3},k_{4};l_{1},l_{2},l_{3},l_{4})\,X(d\,l_{1},u\,l_{2},u^{+}\,l_{3},d^{+}\,l_{4})
\label{Eq.EvnEqnOneCooper} \\
&&\frac{\partial }{\partial \xi }X(d\,k_{1},d\,k_{2},u\,k_{3},u%
\,k_{4},u^{+}k_{5},u^{+}k_{6},d^{+}k_{7},d^{+}k_{8})  \nonumber \\
&=&\dsum%
\limits_{l_{1}l_{2}l_{3}l_{4}}M(k_{1},k_{2},k_{3},k_{4},k_{5},k_{6},k_{7},k_{8};l_{1},l_{2},l_{3},l_{4},l_{5},l_{6},l_{7,}l_{8})\,
\nonumber \\
&&\times
X(d\,l_{1},d\,l_{2},u\,l_{3},u%
\,l_{4},u^{+}l_{5},u^{+}l_{6},d^{+}l_{7},d^{+}l_{8})
\label{Eq.EvnEqnTwoCooper}
\end{eqnarray}%
where $\xi $ is either $t$ or $\beta $ for time or temperature evolution,
and the matrices $M$ can be read off from Eq. (\ref%
{Eq.QCFOnePairTimeEvnFirstOrderChange}), (\ref%
{Eq.QCFOnePairTempEvnFirstOrderChangeA}) or (\ref%
{Eq.QCFOTwoPairsTempEvnFirstOrderChangeA}). The elements of $M$ are
designated $M_{\{k\};\{l\}}$, where for one Cooper pair $\{k\}\equiv
\{k_{1},k_{2},k_{3},k_{4}\}$, $\{l\}\equiv \{l_{1},l_{2},l_{3},l_{4}\}$,
whilst for two Cooper pairs $\{k]\equiv
\{k_{1},k_{2},k_{3},k_{4},k_{5},k_{6},k_{7},k_{8}\}$, $\{l\}\equiv
\{l_{1},l_{2},l_{3},l_{4},l_{5},l_{6},l_{7,}l_{8}\}$. Clearly the sets $%
\{k\} $, $\{l\}$ respectively define the rows, columns of $M$ .

For the cases of time evolution it can be seen by inspection that $M$ $=iH$,
where $H$ is a real, symmetrric matrix - $H_{\{k\};\{l\}\text{ }%
}=H_{\{l\};\{k\}\text{ }}$. Note that $\lambda ^{2}=\frac{-ig}{2\hslash V}$'
The time evolution equations are of the form 
\begin{equation}
\frac{\partial }{\partial t}X(t)=iH\;X(t)  \label{Eq.GenTimeEvolnEqns}
\end{equation}%
and can be solved using the normalised, orthogonal column eigenvectors $%
x_{i} $ of $H$. Thus with real eigenvalues $\lambda _{i}$ 
\begin{equation}
H\;x_{i}=\lambda _{i}\,x_{i}\qquad x_{i}^{T}\;x_{j}=\delta _{ij}
\label{Eq.EigenvectorsH}
\end{equation}%
the solution is of the form 
\begin{eqnarray}
X(t) &=&\dsum\limits_{i}c_{i}(0)\;\exp (i\,\lambda _{i}\,t)\;x_{i}
\label{Eq.SolnQCFTime} \\
c_{i}(0) &=&x_{i}^{T}\;X(0)
\end{eqnarray}%
where the initial condition is specified by the coefficients $c_{i}(0)$
which are determined from the initial column vector $X(0)$ .

For the cases of temperature evolution it can be seen by inspection that $M$ 
$=J$, where $J$ is a real, symmetrric matrix - $J_{\{k\};\{l\}\text{ }%
}=J_{\{l\};\{k\}\text{ }}$. Note that $\eta ^{2}=\frac{-g}{4V}$. ' The
temperature evolution equations are of the form 
\begin{equation}
\frac{\partial }{\partial \beta }X(\beta )=J\;X(\beta )
\label{Eq.GenTempEvnEqns}
\end{equation}%
and can be solved using the normalised, orthogonal column eigenvectors $%
y_{i} $ of $J$. Thus with real eigenvalues $\theta _{i}$ 
\begin{equation}
J\;y_{i}=\theta _{i}\,y_{i}\qquad y_{i}^{T}\;y_{j}=\delta _{ij}
\label{Eq.EigenVectorsJ}
\end{equation}%
the solution is of the form 
\begin{eqnarray}
X(\beta ) &=&\dsum\limits_{i}c_{i}(0)\;\exp (\theta _{i}\,\beta )\;y_{i}
\label{Eq.SolnQCFTemp} \\
c_{i}(0) &=&y_{i}^{T}\;X(0)
\end{eqnarray}%
where the initial condition is specified by the coefficients $c_{i}(0)$
which are determined from the initial column vector $X(0)$ .

This method of solution requires first calculating the matrix $M$ and the
initial column vector $X(0)$, then determining the eigenvectors and
eigenvalues of $H$ or $J$. The alternative is a step-by-step development of
the QCF using Eq.(\ref{Eq.QCFOnePairTimeEvnFirstOrderChange}), (\ref%
{Eq.QCFOnePairTempEvnFirstOrderChangeA}) or (\ref%
{Eq.QCFOTwoPairsTempEvnFirstOrderChangeA}) \smallskip

\subsection{Dimensionless Variables}

It is first convenient to introduce dimensionless variables. The wave vector 
$k_{F}$, energy $E_{F}$ and temperature $T_{F}$ associated with the Fermi
surface for non-interacting fermions are all related to the number density $%
N/V$ as follows:%
\begin{eqnarray}
k_{F} &=&(3\pi ^{2}\frac{N}{V})^{1/3} \\
E_{F} &=&\frac{\hslash ^{2}k_{F}^{2}}{2m}=k_{B}T_{F}
\end{eqnarray}%
This leads to scales $t_{F}$ for time, $\beta _{F}$ for the inverse
temperature $1/(k_{B}T)$ and $k_{F}$ for the wave vector given by 
\begin{eqnarray}
t_{F} &=&\frac{2\pi \hslash }{E_{F}}\qquad t=T\,t_{F} \\
\beta _{F} &=&\frac{1}{k_{B}T_{F}}=\frac{1}{E_{F}}\qquad \beta =B\,\beta _{F}
\\
k &=&K\,k_{F}
\end{eqnarray}%
so $T$ is the dimensionless time, $B$ is dimensionless inverse temperature
and $K$ is the dimensionless wave vector.

As the coupling constant $g$ is given by 
\begin{equation}
g=\frac{4\pi a_{s}\,\hslash ^{2}}{m}
\end{equation}%
we can write the quantities $2\lambda ^{2}$ and $2\eta ^{2}$ that occur in
the time evolution and temperature evolution equations as 
\begin{eqnarray}
2\lambda ^{2} &=&\frac{(-ig)}{\hslash V}=-i\frac{16}{3}\frac{1}{N}\frac{1}{%
t_{F}}(a_{s}k_{F}) \\
2\eta ^{2} &=&\frac{(-g)}{2V}=-\frac{4}{3\pi }\frac{1}{N}\frac{1}{\beta _{F}}%
(a_{s}k_{F})
\end{eqnarray}

Hence we see that using $\frac{\hbar ^{2}k^{2}}{2m}=K^{2}\frac{2\pi \hslash 
}{t_{F}}=K^{2}\frac{1}{\beta _{F}}$ the time and temperature evolution
equations Eqs. (\ref{Eq.QCFOnePairTimeEvnFirstOrderChange}), (\ref%
{Eq.QCFOnePairTempEvnFirstOrderChangeA}), (\ref%
{Eq.QCFOTwoPairsTempEvnFirstOrderChangeA}) and (\ref%
{Eq.QCFOTwoPairsTempEvnFirstOrderChangeB}) can be written in terms of
dimensionless time $T,$ dimensionless inverse temperature $B$ and
dimensionless wave vectors $K,L$. The dimensionless quantities $N$ and $%
(a_{s}k_{F})$ play the role of parameters. The unitary case is when $%
1/(a_{s}k_{F})=0$, corresponding to $a_{s}\rightarrow \infty $.\smallskip

\subsection{Numerical Issues}

In the simplest situation where the BEC/BCS crossover is for a $1D$ system,
we can estimate the size of the matrices $J$ or $H$ in Section \ref{Section
- Solution of Evolution Equations for QCF} for both the one and two Cooper
pair QCF calculations in terms of the number of modes $n$ that need to be
taken into account. As mentioned previously, not all occupied modes are
expected to be important - as it is likely that only $k$ values near the
Fermi surface will be affected. Thus $n\ll N$, where $N$ is the total number
of fermions. For the one Cooper pair QCF the quantities $%
k_{1},k_{2},k_{3},k_{4},l_{1},l_{2},l_{3},l_{4}$ each take on $n$ values, so
the matrices $J$ or $H$ will involve $n^{8}$ elements. For the two Cooper
pair QCF the quantities $%
k_{1},k_{2},k_{3},k_{4},k_{5},k_{6},k_{7},k_{8},l_{1},l_{2},l_{3},l_{4},l_{5},l_{6},l_{7,}l_{8} 
$ each take on $n$ values, so the matrices $J$ or $H$ will involve $n^{16}$
elements. Fortunately, due to the presence od many Kronecka deltas in the
formulae for the matrix elements, most elements are zero so both $J$ or $H$
are very sparse matrices - which means that after the non-zero elements have
been identified once, the calculation can start by putting all elements as
zero and only evaluate those that are non-zero. Actually, because of
selection rules such as $k_{1}=l_{1}$, $k_{2}=l_{2}$ , $%
k_{3}+k_{4}=l_{3}+l_{4}$ that apply in the one Cooper pair case, the number
of non-zero elements are of order $n^{5}$ rather than $n^{8}.$Similarly, in
the two Cooper pair case the number of non-zero elements is of order $n^{9}$
rather than $n^{16}$. Nevertheless, these can still be large numbers even
though each sparse matrix can have its rows and columns reordered via a
unitary transformation to place all the non-zero elements in the top-left
corner before determining the eigenvalues and orthogonal eigenvectors (most
of the eigenvalues will then be zero corresponding to the reordered matrix
having most rows and columns zero). These eigenvectors could then be
transformed back via the inverse transformation to give the eigenvectors of
the original sparse matrix. However, for $n=10$ the one Cooper pair case
would involve calculating of order $10^{5}$ non-zero elements, though for
the two Cooper pair case of order $10^{9}$ elements woukl be required. The
evolution over time or temperature is then determined from the eigenvalues
and eigenvalues at just the final time or temperature. Such calculations
should be feasible on supercomputers.

By comparison, if the Gaussian operator fermion phase space theory were
applied, the one Cooper pair case would be treatable via stochastic
equations involving the $2n\times 2n$ covariance matrix - involving $4n^{2}$
elements. As these elements each involves stochastic Wiener increments, the
covariance matrix would be recalculated at each time or temperature step, so
if there are $\nu $ steps, the total number of covariance matrix elements
calculated would be $4\nu n^{2}$. However, as the evolution has to be
carried out for a number $\xi $ stochastic trajectories, the total number of
covariance matrix elements calculated would be $4\xi \nu n^{2}$. For $n=10$, 
$\nu =100$ and $\xi =100$ this is of order $10^{6}$, which is much the same
order as for GPST. So there is no obvious reason why the Gaussian operator
fermion phase space theory is faster than GPST.

\pagebreak

\section{Summary}

\label{Section - Summary and Conclusions}

In the present paper it has been shown how Grassmann Phase Space Theory
(GSPT) can be applied to the BEC/BCS crossover in cold fermionic atomic
gases and used to determine the evolution (over either time or temperature)
of the Quantum Correlation Functions (QCF) that specify: (a) the positions
of the spin up and spin down fermionic atoms in a single Cooper pair and (b)
the positions of the two spin up and two spin down fermionic atoms in two
Cooper pairs. The first of these QCF is relevent to describing the change in
size of a Cooper pair, as the fermion-fermion coupling constant is changed
through the crossover from a small Cooper pair on the BEC side to a large
Cooper pair on the BCS side. The second of these QCF is important for
describing the correlations berween the positions of the fermionic atoms in
two Cooper pairs, which is expected to be small at the BEC or BCS sides of
the crossover, but is expected to be significant in the strong interaction
unitary regime, where the size of a Cooper pair is comparable to the
separation between Cooper pairs. In GPST the QCF\ are given as the
stochastic average of products of Grassman stochastic position fields, which
are related via Fourier transforms to the stochastic average of products of
Grassman stochastic momentum fields. Using the no-correlation theorem, GPST
shows that the stochastic average of the products of Grassman stochastic
momentum fields at a later time (or lower temperature) is related linearly
to the stochastic average of the products of Grassman stochastic momentum
fields at an earlier time (or higher temperature), and furthermore that the
matrix elements involved in the linear relations are all c-numbers.
Expressions for these matrix elements corresponding to a small time
increment (or a small temperature change) have been obtained analytically,
providing the formulae needed for numerical studies of the evolution that
are planned for a future publication. The initial conditions envisaged
include those for a non-interacting fermionic gas at zero temperature to
study the time evolution of the creation of a Cooper pair when the
interaction is switched on via Feshbach resonance methods. Other initial
conditions described include a high temperature gas (where the effect of the
interactions can be ignored in the initial state), and where the evolution
will be studied as the temperature is lowered until either a BEC state with
small Cooper pairs or a BCS state with large Cooper pairs forms, depending
on the fermion-fermion coupling constant. The behaviour for the case where
the coupling constant is very large (the unitary regime) would be of
particular interest. Full derivations of the expressions have been presented
in the Appendices. \bigskip

\section{Acknowledgements}

This research was supported by the Centre for Quantum Technology Theory,
School of Science, Computing and Engineering Technology, Swinburne
University, Melbourne, Victoria 3122, Australia.\pagebreak

\section{Appendix A - Details for Grassmann Field Theory}

\label{Appendix A - Details GFT}

In this Appendix some details for the Grassmann field theory Section are
provided.

\subsection{Gaussian-Markoff Noise Terms}

The basic stochastic average properties of the Gaussian-Markoff random noise
terms are

\begin{eqnarray}
\overline{\Gamma _{{\small a}}(t_{1})}\, &=&0  \nonumber \\
\overline{\Gamma _{{\small a}}(t_{1})\Gamma _{b}(t_{2})}\, &=&\delta _{%
{\small ab}}\delta (t_{1}-t_{2})  \nonumber \\
\overline{\Gamma _{a}(t_{1})\Gamma _{b}(t_{2})\Gamma _{c}(t_{3})} &=&0 
\nonumber \\
\overline{\Gamma _{a}(t_{1})\Gamma _{b}(t_{2})\Gamma _{c}(t_{3})\Gamma
_{d}(t_{4})} &=&\overline{\Gamma _{a}(t_{1})\Gamma _{b}(t_{2}})\overline{%
\Gamma _{c}(t_{3})\Gamma _{d}(t_{4}})+\overline{\Gamma _{a}(t_{1})\Gamma
_{c}(t_{3}})\overline{\Gamma _{b}(t_{2})\Gamma _{d}(t_{4}})  \nonumber \\
&&+\overline{\Gamma _{a}(t_{1})\Gamma _{d}(t_{4}})\overline{\Gamma
_{b}(t_{2})\Gamma _{c}(t_{3}})  \nonumber \\
&&  \label{Eq.GaussianMarkoffProps}
\end{eqnarray}%
showing that the stochastic averages of a single $\Gamma $ is zero and the
stochastic average of the product of two $\Gamma $'s is zero if they are
different and delta function correlated in the time difference if they are
the same. In addition, the stochastic averages of products of odd numbers of 
$\Gamma $\ are zero and stochastic averages of products of even numbers of $%
\Gamma $\ are the sums of products of stochastic averages of pairs of $%
\Gamma $. \smallskip

\subsection{Determining the B Matrix - Time Evolution Case}

\label{SubSection - B Matrix Time Evn}

Using the notation in Section \ref{SubSect - Ito Stochastic Position Field
Eqns} and after substituting for $\delta (r-s)$, the non-zero elements of
the diffusion matrix in Eq. (\ref{Eq.FFPEInteractFermiFldModel}) are given by%
\begin{eqnarray}
D_{u1,d1}(\psi (s),\psi (r)) &=&-\,\frac{ig}{\hbar V}\sum\limits_{q}\theta
_{u1}^{q}(s)\;\theta _{d1}^{q}(r)\qquad  \nonumber \\
D_{d1,u1}(\psi (s),\psi (r)) &=&-\,\frac{ig}{\hbar V}\sum\limits_{q}\theta
_{d1}^{q}(s)\;\theta _{u1}^{q}(r)  \nonumber \\
D_{u2,d2}(\psi (s),\psi (r)) &=&+\,\frac{ig}{\hbar V}\sum\limits_{q}\theta
_{u2}^{q}(s)\;\theta _{d2}^{q}(r)\qquad  \nonumber \\
D_{d2,u2}(\psi (s),\psi (r)) &=&+\,\frac{ig}{\hbar V}\sum\limits_{q}\theta
_{d2}^{q}(s)\;\theta _{u2}^{q}(r)  \nonumber \\
&&  \label{Eq.DiffMatrixTimeEvn}
\end{eqnarray}%
where 
\begin{eqnarray}
\theta _{d1}^{q}(s) &=&\psi _{d1}(s)\,\exp (+iqs)\qquad \theta
_{u1}^{q}(s)=\psi _{u1}(s)\,\exp (-iqs)  \nonumber \\
\theta _{d2}^{q}(s) &=&\psi _{d2}(s)\,\exp (-iqs)\qquad \theta
_{u2}^{q}(s)=\psi _{u2}(s)\,\exp (+iqs)  \label{Eq.ThetaDiffnTimeEvn}
\end{eqnarray}

We write the diffusion matrix in the form 
\begin{equation}
D_{\alpha A\,\beta B}[\psi (s,t),s;\psi
(r,t),r]=\dsum\limits_{q}\dsum\limits_{\gamma C,\delta D}M_{\alpha A,\gamma
C;\beta B,\delta D}\;\theta _{\gamma C}^{q}(s)\,\theta _{\delta D}^{q}(r)
\label{Eq. DiffMatrixTimeMForm}
\end{equation}%
where the symmetric matrix $M$ is 
\begin{equation}
\lbrack M]=-\,\frac{ig}{\hbar V}\,\left[ 
\begin{tabular}{lllll}
$(\alpha A,\gamma C)\downarrow \backslash \;(\beta B,\delta D)\rightarrow $
& $u1,d1$ & $d1,u1$ & $u2,d2$ & $d2,u2$ \\ 
$u1,d1$ & 0 & 1 & 0 & 0 \\ 
$d1,u1$ & 1 & 0 & 0 & 0 \\ 
$u2,d2$ & 0 & 0 & 0 & -1 \\ 
$d2,u2$ & 0 & 0 & -1 & 0%
\end{tabular}%
\right]  \label{Eq.MMatrix}
\end{equation}

Using Takagi \cite{Takagi25} factorisation $M=KK^{T}$ the matrix $M$ is
given by%
\begin{equation}
M_{\alpha A,\gamma C;\beta B,\delta D}=\dsum\limits_{\xi }K_{\alpha A,\gamma
C;\xi }\,K_{\beta B,\delta D;\xi }  \label{Eq.TakagiM}
\end{equation}%
where 
\begin{equation}
\lbrack K]=\sqrt{\frac{-\,ig}{\hbar V}}\,\left[ 
\begin{tabular}{lllll}
$(\alpha A,\gamma C)\downarrow \backslash \;(\xi )\rightarrow $ & $u1,d1$ & $%
d1,u1$ & $u2,d2$ & $d2,u2$ \\ 
$u1,d1$ & +$\frac{\text{1}}{\sqrt{\text{2}}}$ & +$\frac{\text{i}}{\sqrt{%
\text{2}}}$ & 0 & 0 \\ 
$d1,u1$ & +$\frac{\text{1}}{\sqrt{\text{2}}}$ & -$\,\frac{\text{i}}{\sqrt{%
\text{2}}}$ & 0 & 0 \\ 
$u2,d2$ & 0 & 0 & +$\frac{\text{1}}{\sqrt{\text{2}}}$ & +$\frac{\text{i}}{%
\sqrt{\text{2}}}$ \\ 
$d2,u2$ & 0 & 0 & -$\frac{\text{1}}{\sqrt{\text{2}}}$ & +$\frac{\text{i}}{%
\sqrt{\text{2}}}$%
\end{tabular}%
\right]  \label{Eq.KMatrixTimeEvn}
\end{equation}

Thus we have 
\begin{eqnarray}
&&D_{\alpha A\,\beta B}[\psi (s,t),s;\psi (r,t),r]  \nonumber \\
&=&\dsum\limits_{q}\dsum\limits_{\xi }\,\left( \dsum\limits_{\gamma
C}K_{\alpha A,\gamma C;\xi }\,\theta _{\gamma C}^{q}(s)\right) \,\left(
\dsum\limits_{\delta D}K_{\beta B,\delta D;\xi }\theta _{\delta
D}^{q}(r)\right)  \nonumber \\
&=&\dsum\limits_{q,\xi }B_{q,\xi }^{\alpha A}[\psi (s,t),s]\,B_{q,\xi
}^{\beta B}[\psi (r,t),r]  \label{Eq.DiffFactnTimeEvn}
\end{eqnarray}%
where 
\begin{equation}
B_{q,\xi }^{\alpha A}[\psi (s,t),s]\,=\dsum\limits_{\gamma C}K_{\alpha
A,\gamma C;\xi }\,\theta _{\gamma C}^{q}(s)  \label{Eq.BMatrixTimeEvn}
\end{equation}%
Hence we have written the diffusion matrix in the form $D=BB^{T}$ as
required to determine the Ito SFE.

On substituting for $K_{\alpha A,\gamma C;\xi }$ and $\theta _{\gamma
C}^{q}(s)$ from Eqs. (\ref{Eq.KMatrixTimeEvn}) and (\ref%
{Eq.ThetaDiffnTimeEvn}) we obtain the Ito SFE in the form given in Eq. (\ref%
{Eq.ThetaMatrixTempEvn}).\medskip\ 

\subsection{Determining the B Matrix - Temperature Evolution Case}

\label{SubSection - B Matrix Temp Evn}

The temperature evolution case is treated similarly to the time evolution
case in the previous Section.

However in this case the diffusion matrix is 
\begin{eqnarray}
D_{u1,d1}(\psi (s),\psi (r)) &=&-\frac{g}{2V}\sum\limits_{q}\theta
_{u1}^{q}(s)\;\theta _{d1}^{q}(r)\qquad  \nonumber \\
D_{d1,u1}(\psi (s),\psi (r)) &=&-\frac{g}{2V}\sum\limits_{q}\theta
_{d1}^{q}(s)\;\theta _{u1}^{q}(r)  \nonumber \\
D_{u2,d2}(\psi (s),\psi (r)) &=&-\,\frac{g}{2V}\sum\limits_{q}\theta
_{u2}^{q}(s)\;\theta _{d2}^{q}(r)\qquad  \nonumber \\
D_{d2,u2}(\psi (s),\psi (r)) &=&-\,\frac{g}{2V}\sum\limits_{q}\theta
_{d2}^{q}(s)\;\theta _{u2}^{q}(r)  \nonumber \\
&&  \label{Eq.DiffMatrixTempEvn}
\end{eqnarray}%
and the $M$ matrix is 
\begin{equation}
\lbrack M]=-\,\frac{g}{2V}\,\left[ 
\begin{tabular}{lllll}
$(\alpha A,\gamma C)\downarrow \backslash \;(\beta B,\delta D)\rightarrow $
& $u1,d1$ & $d1,u1$ & $u2,d2$ & $d2,u2$ \\ 
$u1,d1$ & 0 & 1 & 0 & 0 \\ 
$d1,u1$ & 1 & 0 & 0 & 0 \\ 
$u2,d2$ & 0 & 0 & 0 & 1 \\ 
$d2,u2$ & 0 & 0 & 1 & 0%
\end{tabular}%
\right]  \label{Eq.MMatrixTemp}
\end{equation}%
which results in the $K$ matrix 
\begin{equation}
\lbrack K]=\sqrt{\frac{-\,g}{2V}}\,\left[ 
\begin{tabular}{lllll}
$(\alpha A,\gamma C)\downarrow \backslash \;(\xi )\rightarrow $ & $u1,d1$ & $%
d1,u1$ & $u2,d2$ & $d2,u2$ \\ 
$u1,d1$ & +$\frac{\text{1}}{\sqrt{\text{2}}}$ & +$\frac{\text{i}}{\sqrt{%
\text{2}}}$ & 0 & 0 \\ 
$d1,u1$ & +$\frac{\text{1}}{\sqrt{\text{2}}}$ & -$\,\frac{\text{i}}{\sqrt{%
\text{2}}}$ & 0 & 0 \\ 
$u2,d2$ & 0 & 0 & +$\frac{\text{1}}{\sqrt{\text{2}}}$ & +$\frac{\text{i}}{%
\sqrt{\text{2}}}$ \\ 
$d2,u2$ & 0 & 0 & +$\frac{\text{1}}{\sqrt{\text{2}}}$ & -$\frac{\text{i}}{%
\sqrt{\text{2}}}$%
\end{tabular}%
\right]  \label{Eq.KMatrixTempEvn}
\end{equation}

On substituting for $K_{\alpha A,\gamma C;\xi }$ and $\theta _{\gamma
C}^{q}(s)$ from Eqs. (\ref{Eq.KMatrixTempEvn}) and (\ref%
{Eq.ThetaDiffnTimeEvn}) we obtain the Ito SFE in the form given in Eq. (\ref%
{Eq.ItoSFETimeEvoln}).\medskip

\subsection{Stochastic Momentum Fields}

The inverse transformation between the stochastic position and momentum
fields is%
\begin{eqnarray}
\tilde{\phi}_{\alpha }(k) &=&\frac{1}{\sqrt{V}}\int ds\;\exp (-ik\cdot s)\;%
\tilde{\psi}_{\alpha }(s)  \nonumber \\
\tilde{\phi}_{\alpha }^{+}(k) &=&\frac{1}{\sqrt{V}}\int ds\;\exp (+ik\cdot
s)\;\tilde{\psi}_{\alpha }^{+}(s)  \label{Eq.StochMtmFldsInverse}
\end{eqnarray}%
where the $t$ or $\beta $ dependence is implicit.\pagebreak

\section{Appendix B - Stochastic and QCF Equations}

\label{Appendix B - Srochastic Averaging and QCF Eqns}

In this Appendix the details are set out of applying stochastic averaging to
give first order changes in the QCF describing the size of one Cooper pair
and the position correlations for two Cooper pairs.

\subsection{One Cooper Pair - Time Evolution}

For the case of time evolution the QCF at time $t+\delta t$ is related to
that at time $t$ via 
\begin{eqnarray}
&&\overline{\widetilde{\phi }_{d}(k_{1},t+\delta t)\,\widetilde{\phi }%
_{u}(k_{2},t+\delta t)\;\widetilde{\phi }_{u}^{+}(k_{3},t+\delta t)\,%
\widetilde{\phi }_{d}^{+}(k_{4},t+\delta t)}  \nonumber \\
&=&\left[ 
\begin{array}{c}
\dsum\limits_{l_{1},q_{1}}(F_{d,d}(k_{1},l_{1};q_{1},\delta t)\,\widetilde{%
\phi }_{d}(l_{1},t)+F_{d,u}(k_{1},l_{1};q_{1},\delta t)\,\widetilde{\phi }%
_{u}(l_{1},t)) \\ 
\times \dsum\limits_{l_{2},q_{2}}(F_{u,d}(k_{2},l_{2};q_{2},\delta t)\,%
\widetilde{\phi }_{d}(l_{2},t)+F_{u,u}(k_{2},l_{2};q_{2},\delta t)\,%
\widetilde{\phi }_{u}(l_{2},t)) \\ 
\times \dsum\limits_{l_{3},q_{3}}(F_{u,d}^{+}(k_{3},l_{3};q_{3},\delta t)\,%
\widetilde{\phi }_{d}^{+}(l_{3},t)+F_{u,u}^{+}(k_{3},l_{3};q_{3},\delta t)\,%
\widetilde{\phi }_{u}^{+}(l_{3},t)) \\ 
\times \dsum\limits_{l_{4},q_{4}}(F_{d,d}^{+}(k_{4},l_{4};q_{4},\delta t)\,%
\widetilde{\phi }_{d}^{+}(l_{4},t)+F_{d,u}^{+}(k_{4},l_{4};q_{4},\delta t)\,%
\widetilde{\phi }_{u}^{+}(l_{4},t))%
\end{array}%
\right] _{StochAver}  \nonumber \\
&&  \label{Eq.QCFTimeEvnStep1}
\end{eqnarray}%
after substituting from Eq.(\ref{Eq.ItoSMtmTimeEvln}). Apart from the sums
over $l_{1,}q_{1},...,l_{4},q_{4}$ there are $16$ terms for which the
stochastic averages have to be evaluated. If the two terms in each of the
first $2$ \ factors are listed as $1,2$ and the two terms in each of the
second $2$ factors are listed as $a,b$ then the $16$ terms can be listed as $%
11aa,12aa,21aa,22aa,11ab,12ab,21ab,22ab,....,22bb$.

In Ito stochastic equations we then use the factorisation result that
depends on the $F$, $F^{+}$ involving times later than $t$ (see Eq. (9.43)
in Ref. \cite{Dalton15a} 
\begin{eqnarray}
&&\left[ 
\begin{array}{c}
F_{\alpha ,\beta }(k_{a},l_{a};q_{a},\delta t)\,\widetilde{\phi }_{\beta
}(l_{a},t)\,F_{\gamma ,\delta }(k_{b},l_{b};q_{b},\delta t)\,\widetilde{\phi 
}_{\delta }(l_{b},t)\, \\ 
\times \,F_{\epsilon ,\eta }^{+}(k_{c},l_{c};q_{c},\delta t)\,\widetilde{%
\phi }_{\eta }^{+}(l_{c},t)\,F_{\theta ,\kappa
}^{+}(k_{d},l_{d};q_{d},\delta t)\,\widetilde{\phi }_{\kappa }^{+}(l_{d},t)\,%
\end{array}%
\right] _{Stoch\,Aver}  \nonumber \\
&=&\overline{F_{\alpha ,\beta }(k_{a},l_{a};q_{a},\delta t)\,F_{\gamma
,\delta }(k_{b},l_{b};q_{b},\delta t)\,F_{\epsilon ,\eta
}^{+}(k_{c},l_{c};q_{c},\delta t)\,F_{\theta ,\kappa
}^{+}(k_{d},l_{d};q_{d},\delta t)}  \nonumber \\
&&\times \overline{\,\widetilde{\phi }_{\beta }(l_{a},t)\,\,\widetilde{\phi }%
_{\delta }(l_{b},t)\,\widetilde{\phi }_{\eta }^{+}(l_{c},t)\,\widetilde{\phi 
}_{\kappa }^{+}(l_{d},t)}  \label{Eq.StochAverFactorisn}
\end{eqnarray}

We wish to evaluate the various

$\overline{F_{\alpha ,\beta }(k_{a},l_{a};q_{a},\delta t)\,F_{\gamma ,\delta
}(k_{b},l_{b};q_{b},\delta t)\,F_{\epsilon ,\eta
}^{+}(k_{c},l_{c};q_{c},\delta t)\,F_{\theta ,\kappa
}^{+}(k_{d},l_{d};q_{d},\delta t)}$ correct to order $\delta t$. We note
that the $F_{d,d},F_{u,u},F_{d,d}^{+},F_{u,u}^{+}$ factors are
non-stochastic and linear in $\delta t$. Also the $%
F_{d,u},F_{u,d},F_{d,u}^{+},F_{u,d}^{+}$ factors are stochastic and linear
in the stochastic Wiener increments $\delta \widetilde{\omega }%
_{u,d}^{q},\delta \widetilde{\omega }_{d,u}^{q},\delta \widetilde{\omega }%
_{u+,d+}^{q},\delta \widetilde{\omega }_{d+,u+}^{q}$. From Eq.(\ref%
{Eq.PropsWienerIncrem}) the stochastic average of a factor involving an odd
number $1,3$ of the $\delta \widetilde{\omega }_{u,d}^{q},\delta \widetilde{%
\omega }_{d,u}^{q},\delta \widetilde{\omega }_{u+,d+}^{q},\delta \widetilde{%
\omega }_{d+,u+}^{q}$ with any non-stochastic factor is zero, if there are $%
2 $ of these increments the stochastic average is proportional to $\delta t$
and if there are $4$ of these increments the stochastic average is
proportional to $\delta t^{2}$(apart from cases where there is zero Kronecka
delta $\delta _{{\small ab}}$ involved). For the case of two stochastic
Wiener increments we note from Eq. (\ref{Eq.PropsWienerIncrem}) that the
non-zero stochastic averages of the products are 
\begin{eqnarray}
\overline{\delta \widetilde{\omega }_{u,d}^{q}\,\delta \widetilde{\omega }%
_{u,d}^{q}} &=&\overline{\delta \widetilde{\omega }_{d,u}^{q}\,\delta 
\widetilde{\omega }_{d,u}^{q}}=\delta t  \nonumber \\
\overline{\delta \widetilde{\omega }_{u+,d+}^{q}\,\delta \widetilde{\omega }%
_{u+,d+}^{q}} &=&\overline{\delta \widetilde{\omega }_{d+,u+}^{q}\,\delta 
\widetilde{\omega }_{d+,u+}^{q}}=\delta t  \label{Eq.NonZeroStochIncr}
\end{eqnarray}

With these considerations in mind, the non-zero cases are as follows:%
\begin{eqnarray}
12ba &:&\overline{F_{d,d}(k_{1},l_{1};q_{1},\delta
t)F_{u,u}(k_{2},l_{2};q_{2},\delta t)F_{u,u}^{+}(k_{3},l_{3};q_{3},\delta
t)\,F_{d,d}^{+}(k_{4},l_{4};q_{4},\delta t)\,}  \nonumber \\
&=&\delta _{q_{1},0}\;\delta _{k_{1},l_{1}}\left( 1+\frac{i}{\hbar }\left\{ 
\frac{\hbar ^{2}k_{1}^{2}}{2m}\right\} \delta t\right) \delta
_{q_{2},0}\;\delta _{k_{2},l_{2}}\left( 1+\frac{i}{\hbar }\left\{ \frac{%
\hbar ^{2}k_{2}^{2}}{2m}\right\} \delta t\right)  \nonumber \\
&&\times \delta _{q_{3},0}\;\delta _{k_{3},l_{3}}\left( 1-\frac{i}{\hbar }%
\left\{ \frac{\hbar ^{2}k_{3}^{2}}{2m}\right\} \delta t\right) \delta
_{q_{4},0}\;\delta _{k_{4},l_{4}}\left( 1-\frac{i}{\hbar }\left\{ \frac{%
\hbar ^{2}k_{4}^{2}}{2m}\right\} \delta t\right)  \nonumber \\
&\approx &\delta _{q_{1},0}\;\delta _{k_{1},l_{1}}\,\delta
_{q_{2},0}\;\delta _{k_{2},l_{2}}\,\delta _{q_{3},0}\;\delta
_{k_{3},l_{3}}\,\delta _{q_{4},0}\;\delta _{k_{4},l_{4}}  \nonumber \\
&&\times \left( 1+\frac{i}{\hbar }\left\{ \frac{\hbar ^{2}k_{1}^{2}}{2m}+%
\frac{\hbar ^{2}k_{2}^{2}}{2m}-\frac{\hbar ^{2}k_{3}^{2}}{2m}-\frac{\hbar
^{2}k_{4}^{2}}{2m}\right\} \delta t\right)  \label{Eq.StochAver1221}
\end{eqnarray}%
where only terms linear in $\delta t$ have been retained.%
\begin{eqnarray}
12ab &:&\overline{F_{d,d}(k_{1},l_{1};q_{1},\delta
t)F_{u,u}(k_{2},l_{2};q_{2},\delta t)F_{u,d}^{+}(k_{3},l_{3};q_{3},\delta
t)\,F_{d,u}^{+}(k_{4},l_{4};q_{4},\delta t)\,}  \nonumber \\
&=&\delta _{q_{1},0}\;\delta _{k_{1},l_{1}}\left( 1+\frac{i}{\hbar }\left\{ 
\frac{\hbar ^{2}k_{1}^{2}}{2m}\right\} \delta t\right) \delta
_{q_{2},0}\;\delta _{k_{2},l_{2}}\left( 1+\frac{i}{\hbar }\left\{ \frac{%
\hbar ^{2}k_{2}^{2}}{2m}\right\} \delta t\right)  \nonumber \\
&&\times \overline{\lambda \;\delta _{(k_{3}+q_{3}),l_{3}}\;\left\{ \delta 
\widetilde{\omega }_{u+,d+}^{q_{3}}+i\delta \widetilde{\omega }%
_{d+,u+}^{q_{3}}\right\} \,\lambda \;\delta _{(k_{4}-q_{4}),l_{4}}\;\left\{
-\delta \widetilde{\omega }_{u+,d+}^{q_{4}}+i\delta \widetilde{\omega }%
_{d+,u+}^{q_{4}}\right\} }  \nonumber \\
&=&\delta _{q_{1},0}\;\delta _{k_{1},l_{1}}\,\delta _{q_{2},0}\;\delta
_{k_{2},l_{2}}\;\lambda \;\delta _{(k_{3}+q_{3}),l_{3}}\;\lambda \;\delta
_{(k_{4}-q_{4}),l_{4}}  \nonumber \\
&&\times \left( 1+\frac{i}{\hbar }\left\{ \frac{\hbar ^{2}k_{1}^{2}}{2m}%
\right\} \delta t\right) \left( 1+\frac{i}{\hbar }\left\{ \frac{\hbar
^{2}k_{2}^{2}}{2m}\right\} \delta t\right)  \nonumber \\
&&\times \left\{ -\delta _{q_{3},q_{4}}\delta t-\delta _{q_{3},q_{4}}\delta
t\right\}  \nonumber \\
&\approx &-2\lambda ^{2}\,\{\delta _{q_{1},0}\;\delta _{k_{1},l_{1}}\,\delta
_{q_{2},0}\;\delta _{k_{2},l_{2}}\;\delta _{(k_{3}+q_{3}),l_{3}}\;\delta
_{(k_{4}-q_{4}),l_{4}}\,\delta _{q_{3},q_{4}}\}\,\delta t
\label{Eq.StochAver1212}
\end{eqnarray}%
where the Wiener increment properties in Eq.(\ref{Eq.PropsWienerIncrem}) has
been used and only terms linear in $\delta t$ have been retained.%
\begin{eqnarray}
21ba &:&\overline{F_{d,u}(k_{1},l_{1};q_{1},\delta
t)F_{u,d}(k_{2},l_{2};q_{2},\delta t)F_{u,u}^{+}(k_{3},l_{3};q_{3},\delta
t)\,F_{d,d}^{+}(k_{4},l_{4};q_{4},\delta t)\,}  \nonumber \\
&=&\overline{(+\lambda \;\delta _{(k_{1}+q_{1}),l_{1}}\;\left\{ \delta 
\widetilde{\omega }_{u,d}^{q_{1}}-i\delta \widetilde{\omega }%
_{d,u}^{q_{1}}\right\} )\;(+\lambda \;\delta _{(k_{2}-q_{2}),l_{2}}\;\left\{
\delta \widetilde{\omega }_{u,d}^{q_{2}}+i\delta \widetilde{\omega }%
_{d,u}^{q_{2}}\right\} }  \nonumber \\
&&\times \delta _{q_{3},0}\;\delta _{k_{3},l_{3}}\left( 1-\frac{i}{\hbar }%
\left\{ \frac{\hbar ^{2}k_{3}^{2}}{2m}\right\} \delta t\right) \,\delta
_{q_{4},0}\;\delta _{k_{4},l_{4}}\left( 1-\frac{i}{\hbar }\left\{ \frac{%
\hbar ^{2}k_{4}^{2}}{2m}\right\} \delta t\right)  \nonumber \\
&=&\lambda \;\delta _{(k_{1}+q_{1}),l_{1}}\,\lambda \;\delta
_{(k_{2}-q_{2}),l_{2}}\;\delta _{q_{3},0}\;\delta _{k_{3},l_{3}}\,\delta
_{q_{4},0}\;\delta _{k_{4},l_{4}}  \nonumber \\
&&\times \left( 1-\frac{i}{\hbar }\left\{ \frac{\hbar ^{2}k_{3}^{2}}{2m}%
\right\} \delta t\right) \left( 1-\frac{i}{\hbar }\left\{ \frac{\hbar
^{2}k_{4}^{2}}{2m}\right\} \delta t\right)  \nonumber \\
&&\times \{\delta _{q_{1},q_{2}}\delta t+\delta _{q_{1},q_{2}}\delta t\} 
\nonumber \\
&\approx &2\lambda ^{2}\,\{\delta _{(k_{1}+q_{1}),l_{1}}\,\delta
_{(k_{2}-q_{2}),l_{2}}\;\delta _{q_{3},0}\;\delta _{k_{3},l_{3}}\,\delta
_{q_{4},0}\;\delta _{k_{4},l_{4}}\,\delta _{q_{1},q_{2}}\}\delta t
\label{Eq.StochAver2121}
\end{eqnarray}%
where only terms linear in $\delta t$ have been retained.

Substituting the results in Eqs.(\ref{Eq.StochAver1221}), (\ref%
{Eq.StochAver1212}) and (\ref{Eq.StochAver2121}) into Eq. (\ref%
{Eq.QCFTimeEvnStep1}) and performing the sums over $q_{1},q_{2},q_{3},q_{4}$
gives 
\begin{eqnarray}
&&\overline{\widetilde{\phi }_{d}(k_{1},t+\delta t)\,\widetilde{\phi }%
_{u}(k_{2},t+\delta t)\;\widetilde{\phi }_{u}^{+}(k_{3},t+\delta t)\,%
\widetilde{\phi }_{d}^{+}(k_{4},t+\delta t)}  \nonumber \\
&=&\dsum\limits_{l_{1}l_{2}l_{3}l_{4}}\delta _{k_{1},l_{1}}\,\delta
_{k_{2},l_{2}}\,\delta _{k_{3},l_{3}}\,\delta _{k_{4},l_{4}}\left( 1+\frac{i%
}{\hbar }\left\{ \frac{\hbar ^{2}k_{1}^{2}}{2m}+\frac{\hbar ^{2}k_{2}^{2}}{2m%
}-\frac{\hbar ^{2}k_{3}^{2}}{2m}-\frac{\hbar ^{2}k_{4}^{2}}{2m}\right\}
\delta t\right)  \nonumber \\
&&\times \overline{\widetilde{\phi }_{d}(l_{1},t)\,\widetilde{\phi }%
_{u}(l_{2},t)\;\widetilde{\phi }_{u}^{+}(l_{3},t)\,\widetilde{\phi }%
_{d}^{+}(l_{4},t)}  \nonumber \\
&&-2\lambda ^{2}\,\dsum\limits_{l_{1}l_{2}l_{3}l_{4}}\{\delta
_{k_{1},l_{1}}\,\delta _{k_{2},l_{2}}\,\delta
_{(k_{3}+k_{4}),(l_{3}+l_{4})}\}\,\delta t\times \overline{\widetilde{\phi }%
_{d}(l_{1},t)\,\widetilde{\phi }_{u}(l_{2},t)\;\widetilde{\phi }%
_{d}^{+}(l_{3},t)\,\widetilde{\phi }_{u}^{+}(l_{4},t)}  \nonumber \\
&&+2\lambda ^{2}\,\dsum\limits_{l_{1}l_{2}l_{3}l_{4}}\{\delta
_{(k_{1}+k_{2}),(l_{1}+l_{2})}\,\delta _{k_{3},l_{3}}\,\delta
_{k_{4},l_{4}}\}\,\delta t\times \overline{\widetilde{\phi }_{u}(l_{1},t)\,%
\widetilde{\phi }_{d}(l_{2},t)\;\widetilde{\phi }_{u}^{+}(l_{3},t)\,%
\widetilde{\phi }_{d}^{+}(l_{4},t)}  \nonumber \\
&=&\dsum\limits_{l_{1}l_{2}l_{3}l_{4}}\delta _{k_{1},l_{1}}\,\delta
_{k_{2},l_{2}}\,\delta _{k_{3},l_{3}}\,\delta _{k_{4},l_{4}}\left( 1+\frac{i%
}{\hbar }\left\{ \frac{\hbar ^{2}k_{1}^{2}}{2m}+\frac{\hbar ^{2}k_{2}^{2}}{2m%
}-\frac{\hbar ^{2}k_{3}^{2}}{2m}-\frac{\hbar ^{2}k_{4}^{2}}{2m}\right\}
\delta t\right)  \nonumber \\
&&\times \overline{\widetilde{\phi }_{d}(l_{1},t)\,\widetilde{\phi }%
_{u}(l_{2},t)\;\widetilde{\phi }_{u}^{+}(l_{3},t)\,\widetilde{\phi }%
_{d}^{+}(l_{4},t)}  \nonumber \\
&&+2\lambda ^{2}\,\dsum\limits_{l_{1}l_{2}l_{3}l_{4}}\{\delta
_{k_{1},l_{1}}\,\delta _{k_{2},l_{2}}\,\delta
_{(k_{3}+k_{4}),(l_{3}+l_{4})}\}\,\delta t\times \overline{\widetilde{\phi }%
_{d}(l_{1},t)\,\widetilde{\phi }_{u}(l_{2},t)\;\widetilde{\phi }%
_{u}^{+}(l_{3},t)\,\widetilde{\phi }_{d}^{+}(l_{4},t)}  \nonumber \\
&&-2\lambda ^{2}\,\dsum\limits_{l_{1}l_{2}l_{3}l_{4}}\{\delta
_{(k_{1}+k_{2}),(l_{1}+l_{2})}\,\delta _{k_{3},l_{3}}\,\delta
_{k_{4},l_{4}}\}\,\delta t\times \overline{\widetilde{\phi }_{d}(l_{1},t)\,%
\widetilde{\phi }_{u}(l_{2},t)\;\widetilde{\phi }_{u}^{+}(l_{3},t)\,%
\widetilde{\phi }_{d}^{+}(l_{4},t)}  \nonumber \\
&&  \label{Eq.QCFSinglePairTimeEvnChange}
\end{eqnarray}%
where we have used $\dsum\limits_{q_{3}q_{4}}\delta
_{(k_{3}+q_{3}),l_{3}}\;\delta _{(k_{4}-q_{4}),l_{4}}\,\delta
_{q_{3},q_{4}}=\delta _{(k_{3}+k_{4}),(l_{3}+l_{4})}$ and $%
\dsum\limits_{q_{1}}\delta _{q_{1},0}=1$. Also, in the second last line we
have interchanged $l_{3}$ and $l_{4}$ and used the anti-commuting feature of
the Grassmann stochastic momentum fields $\widetilde{\phi }_{d}^{+}(l_{3},t)$
and $\,\widetilde{\phi }_{u}^{+}(l_{4},t)$ to place these in the opposite
order in the stochastic average $\overline{\widetilde{\phi }_{d}(l_{1},t)\,%
\widetilde{\phi }_{u}(l_{2},t)\;\widetilde{\phi }_{d}^{+}(l_{3},t)\,%
\widetilde{\phi }_{u}^{+}(l_{4},t)}$, with similar steps for the last line.

\subsection{One Cooper Pair - Temperature Evolution}

For the case of temperature evolution the QCF at time $\beta +\delta \beta $
is related to that at time $t$ via 
\begin{eqnarray}
&&\overline{\widetilde{\phi }_{d}(k_{1},\beta +\delta \beta )\,\widetilde{%
\phi }_{u}(k_{2},\beta +\delta \beta )\;\widetilde{\phi }_{u}^{+}(k_{3},%
\beta +\delta \beta )\,\widetilde{\phi }_{d}^{+}(k_{4},\beta +\delta \beta )}
\nonumber \\
&=&\left[ 
\begin{array}{c}
\dsum\limits_{l_{1},q_{1}}(F_{d,d}(k_{1},l_{1};q_{1},\delta \beta )\,%
\widetilde{\phi }_{d}(l_{1},\beta )+F_{d,u}(k_{1},l_{1};q_{1},\delta \beta
)\,\widetilde{\phi }_{u}(l_{1},\beta )) \\ 
\times \dsum\limits_{l_{2},q_{2}}(F_{u,d}(k_{2},l_{2};q_{2},\delta \beta )\,%
\widetilde{\phi }_{d}(l_{2},\beta )+F_{u,u}(k_{2},l_{2};q_{2},\delta \beta
)\,\widetilde{\phi }_{u}(l_{2},\beta )) \\ 
\times \dsum\limits_{l_{3},q_{3}}(F_{u,d}^{+}(k_{3},l_{3};q_{3},\delta \beta
)\,\widetilde{\phi }_{d}^{+}(l_{3},\beta
)+F_{u,u}^{+}(k_{3},l_{3};q_{3},\delta \beta )\,\widetilde{\phi }%
_{u}^{+}(l_{3},\beta )) \\ 
\times \dsum\limits_{l_{4},q_{4}}(F_{d,d}^{+}(k_{4},l_{4};q_{4},\delta \beta
)\,\widetilde{\phi }_{d}^{+}(l_{4},\beta
)+F_{d,u}^{+}(k_{4},l_{4};q_{4},\delta \beta )\,\widetilde{\phi }%
_{u}^{+}(l_{4},\beta ))%
\end{array}%
\right] _{StochAver}  \nonumber \\
&&  \label{Eq.QCFTempEvnStep1}
\end{eqnarray}%
after substituting from Eq.(\ref{Eq.ItoSMtmFTemper}). Apart from the sums
over $l_{1,}q_{1},...,l_{4},q_{4}$ there are $16$ terms for which the
stochastic averages have to be evaluated. .If the two terms in each of the
first $2$ \ factors are listed as $1,2$ and the two terms in each of the
second $2$ factors are listed as $a,b$ then the $16$ terms can be listed as $%
11aa,12aa,21aa,22aa,11ab,12ab,21ab,22ab,....,22bb$.

In Ito stochastic equations we then use the factorisation result that
depends on the $F$, $F^{+}$ involving times later than $t$ (see Eq. (9.43)
in Ref. \cite{Dalton15a} 
\begin{eqnarray}
&&\left[ 
\begin{array}{c}
F_{\alpha ,\beta }(k_{a},l_{a};q_{a},\delta \beta )\,\widetilde{\phi }%
_{\beta }(l_{1},\beta )\,F_{\gamma ,\delta }(k_{b},l_{b};q_{b},\delta \beta
)\,\widetilde{\phi }_{\delta }(l_{b},\beta )\, \\ 
\times \,F_{\epsilon ,\eta }^{+}(k_{c},l_{c};q_{c},\delta \beta )\,%
\widetilde{\phi }_{\eta }^{+}(l_{c},\beta )\,F_{\theta ,\kappa
}^{+}(k_{d},l_{d};q_{d},\delta \beta )\,\widetilde{\phi }_{\kappa
}^{+}(l_{d},\beta )\,%
\end{array}%
\right] _{Stoch\,Aver}  \nonumber \\
&=&\overline{F_{\alpha ,\beta }(k_{a},l_{a};q_{a},\delta \beta )\,F_{\gamma
,\delta }(k_{b},l_{b};q_{b},\delta \beta )\,F_{\epsilon ,\eta
}^{+}(k_{c},l_{c};q_{c},\delta \beta )\,F_{\theta ,\kappa
}^{+}(k_{d},l_{d};q_{d},\delta \beta )}  \nonumber \\
&&\times \overline{\,\widetilde{\phi }_{\beta }(l_{a},\beta )\,\,\widetilde{%
\phi }_{\delta }(l_{b},\beta )\,\widetilde{\phi }_{\eta }^{+}(l_{c},\beta )\,%
\widetilde{\phi }_{\kappa }^{+}(l_{d},\beta )}
\end{eqnarray}

We wish to evaluate the various

$\overline{F_{\alpha ,\beta }(k_{a},l_{a};q_{a},\delta \beta )\,F_{\gamma
,\delta }(k_{b},l_{b};q_{b},\delta \beta )\,F_{\epsilon ,\eta
}^{+}(k_{c},l_{c};q_{c},\delta \beta )\,F_{\theta ,\kappa
}^{+}(k_{d},l_{d};q_{d},\delta \beta )}$ correct to order $\delta \beta $.
We note that the $F_{d,d},F_{u,u},F_{d,d}^{+},F_{u,u}^{+}$ factors are
non-stochastic and linear in $\delta \beta $. Also the $%
F_{d,u},F_{u,d},F_{d,u}^{+},F_{u,d}^{+}$ factors are stochastic and linear
in the stochastic Wiener increments $\delta \widetilde{\omega }%
_{u,d}^{q},\delta \widetilde{\omega }_{d,u}^{q},\delta \widetilde{\omega }%
_{u+,d+}^{q},\delta \widetilde{\omega }_{d+,u+}^{q}$. From Eq.(\ref%
{Eq.PropsWienerIncrem}) the stochastic average of a factor involving an odd
number $1,3$ of the $\delta \widetilde{\omega }_{u,d}^{q},\delta \widetilde{%
\omega }_{d,u}^{q},\delta \widetilde{\omega }_{u+,d+}^{q},\delta \widetilde{%
\omega }_{d+,u+}^{q}$ with any non-stochastic factor is zero, if there are $%
2 $ of these increments the stochastic average is proportional to $\delta
\beta $ and if there are $4$ of these increments the stochastic average is
proportional to $\delta \beta ^{2}$(apart from cases where there is zero
Kronecka delta $\delta _{{\small ab}}$ involved). For the case of two
stochastic Wiener increments we note from Eq. (\ref{Eq.PropsWienerIncrem})
that the non-zero stochastic averages of the products are 
\begin{eqnarray}
\overline{\delta \widetilde{\omega }_{u,d}^{q}\,\delta \widetilde{\omega }%
_{u,d}^{q}} &=&\overline{\delta \widetilde{\omega }_{d,u}^{q}\,\delta 
\widetilde{\omega }_{d,u}^{q}}=\delta \beta  \nonumber \\
\overline{\delta \widetilde{\omega }_{u+,d+}^{q}\,\delta \widetilde{\omega }%
_{u+,d+}^{q}} &=&\overline{\delta \widetilde{\omega }_{d+,u+}^{q}\,\delta 
\widetilde{\omega }_{d+,u+}^{q}}=\delta \beta
\end{eqnarray}

With these considerations in mind, the non-zero cases are as follows:%
\begin{eqnarray}
12ba &:&\overline{F_{d,d}(k_{1},l_{1};q_{1},\delta \beta
)F_{u,u}(k_{2},l_{2};q_{2},\delta \beta
)F_{u,u}^{+}(k_{3},l_{3};q_{3},\delta \beta
)\,F_{d,d}^{+}(k_{4},l_{4};q_{4},\delta \beta )\,}  \nonumber \\
&=&\delta _{q_{1},0}\;\delta _{k_{1},l_{1}}\left( 1+\frac{1}{2}\left\{ \frac{%
\hbar ^{2}k_{1}^{2}}{2m}\right\} \delta \beta \right) \delta
_{q_{2},0}\;\delta _{k_{2},l_{2}}\left( 1+\frac{1}{2}\left\{ \frac{\hbar
^{2}k_{2}^{2}}{2m}\right\} \delta \beta \right)  \nonumber \\
&&\times \delta _{q_{3},0}\;\delta _{k_{3},l_{3}}\left( 1+\frac{1}{2}\left\{ 
\frac{\hbar ^{2}k_{3}^{2}}{2m}\right\} \delta \beta \right) \delta
_{q_{4},0}\;\delta _{k_{4},l_{4}}\left( 1+\frac{1}{2}\left\{ \frac{\hbar
^{2}k_{4}^{2}}{2m}\right\} \delta \beta \right)  \nonumber \\
&\approx &\delta _{q_{1},0}\;\delta _{k_{1},l_{1}}\,\delta
_{q_{2},0}\;\delta _{k_{2},l_{2}}\,\delta _{q_{3},0}\;\delta
_{k_{3},l_{3}}\,\delta _{q_{4},0}\;\delta _{k_{4},l_{4}}  \nonumber \\
&&\times \left( 1+\frac{1}{2}\left\{ \frac{\hbar ^{2}k_{1}^{2}}{2m}+\frac{%
\hbar ^{2}k_{2}^{2}}{2m}+\frac{\hbar ^{2}k_{3}^{2}}{2m}+\frac{\hbar
^{2}k_{4}^{2}}{2m}\right\} \delta \beta \right)  \label{Eq.StochAverTemp1221}
\end{eqnarray}%
where only terms linear in $\delta t$ have been retained.%
\begin{eqnarray}
12ab &:&\overline{F_{d,d}(k_{1},l_{1};q_{1},\delta \beta
)F_{u,u}(k_{2},l_{2};q_{2},\delta \beta
)F_{u,d}^{+}(k_{3},l_{3};q_{3},\delta \beta
)\,F_{d,u}^{+}(k_{4},l_{4};q_{4},\delta \beta )\,}  \nonumber \\
&=&\delta _{q_{1},0}\;\delta _{k_{1},l_{1}}\left( 1+\frac{1}{2}\left\{ \frac{%
\hbar ^{2}k_{1}^{2}}{2m}\right\} \delta \beta \right) \delta
_{q_{2},0}\;\delta _{k_{2},l_{2}}\left( 1+\frac{1}{2}\left\{ \frac{\hbar
^{2}k_{2}^{2}}{2m}\right\} \delta \beta \right)  \nonumber \\
&&\times \overline{\eta \;\delta _{(k_{3}+q_{3}),l_{3}}\;\left\{ \delta 
\widetilde{\omega }_{u+,d+}^{q_{3}}+i\delta \widetilde{\omega }%
_{d+,u+}^{q_{3}}\right\} \,\eta \;\delta _{(k_{4}-q_{4}),l_{4}}\;\left\{
\delta \widetilde{\omega }_{u+,d+}^{q_{4}}-i\delta \widetilde{\omega }%
_{d+,u+}^{q_{4}}\right\} }  \nonumber \\
&=&\delta _{q_{1},0}\;\delta _{k_{1},l_{1}}\,\delta _{q_{2},0}\;\delta
_{k_{2},l_{2}}\;\eta \;\delta _{(k_{3}+q_{3}),l_{3}}\;\eta \;\delta
_{(k_{4}-q_{4}),l_{4}}  \nonumber \\
&&\times \left( 1+\frac{1}{2}\left\{ \frac{\hbar ^{2}k_{1}^{2}}{2m}\right\}
\delta \beta \right) \left( 1+\frac{1}{2}\left\{ \frac{\hbar ^{2}k_{2}^{2}}{%
2m}\right\} \delta \beta \right)  \nonumber \\
&&\times \left\{ +\delta _{q_{3},q_{4}}\delta \beta +\delta
_{q_{3},q_{4}}\delta \beta \right\}  \nonumber \\
&\approx &+2\eta ^{2}\,\{\delta _{q_{1},0}\;\delta _{k_{1},l_{1}}\,\delta
_{q_{2},0}\;\delta _{k_{2},l_{2}}\;\delta _{(k_{3}+q_{3}),l_{3}}\;\delta
_{(k_{4}-q_{4}),l_{4}}\,\delta _{q_{3},q_{4}}\}\,\delta \beta
\label{Eq.StochAverTemp1212}
\end{eqnarray}%
where the Wiener increment properties in Eq.(\ref{Eq.PropsWienerIncrem}) has
been used and only terms linear in $\delta t$ have been retained.%
\begin{eqnarray}
21ba &:&\overline{F_{d,u}(k_{1},l_{1};q_{1},\delta \beta
)F_{u,d}(k_{2},l_{2};q_{2},\delta \beta
)F_{u,u}^{+}(k_{3},l_{3};q_{3},\delta \beta
)\,F_{d,d}^{+}(k_{4},l_{4};q_{4},\delta \beta )\,}  \nonumber \\
&=&\overline{(+\eta \;\delta _{(k_{1}+q_{1}),l_{1}}\;\left\{ \delta 
\widetilde{\omega }_{u,d}^{q_{1}}-i\delta \widetilde{\omega }%
_{d,u}^{q_{1}}\right\} )\;(+\eta \;\delta _{(k_{2}-q_{2}),l_{2}}\;\left\{
\delta \widetilde{\omega }_{u,d}^{q_{2}}+i\delta \widetilde{\omega }%
_{d,u}^{q_{2}}\right\} }  \nonumber \\
&&\times \delta _{q_{3},0}\;\delta _{k_{3},l_{3}}\left( 1+\frac{1}{2}\left\{ 
\frac{\hbar ^{2}k_{3}^{2}}{2m}\right\} \delta \beta \right) \,\delta
_{q_{4},0}\;\delta _{k_{4},l_{4}}\left( 1+\frac{1}{2}\left\{ \frac{\hbar
^{2}k_{4}^{2}}{2m}\right\} \delta \beta \right)  \nonumber \\
&=&\eta \;\delta _{(k_{1}+q_{1}),l_{1}}\,\eta \;\delta
_{(k_{2}-q_{2}),l_{2}}\;\delta _{q_{3},0}\;\delta _{k_{3},l_{3}}\,\delta
_{q_{4},0}\;\delta _{k_{4},l_{4}}  \nonumber \\
&&\times \left( 1+\frac{1}{2}\left\{ \frac{\hbar ^{2}k_{3}^{2}}{2m}\right\}
\delta \beta \right) \left( 1+\frac{1}{2}\left\{ \frac{\hbar ^{2}k_{4}^{2}}{%
2m}\right\} \delta \beta \right)  \nonumber \\
&&\times \{\delta _{q_{1},q_{2}}\delta \beta +\delta _{q_{1},q_{2}}\delta
\beta \}  \nonumber \\
&\approx &2\eta ^{2}\,\{\delta _{(k_{1}+q_{1}),l_{1}}\,\delta
_{(k_{2}-q_{2}),l_{2}}\;\delta _{q_{3},0}\;\delta _{k_{3},l_{3}}\,\delta
_{q_{4},0}\;\delta _{k_{4},l_{4}}\,\delta _{q_{1},q_{2}}\}\delta \beta
\label{Eq.StochAverTemp2121}
\end{eqnarray}%
where only terms linear in $\delta \beta $ have been retained.

Substituting the results in Eqs.(\ref{Eq.StochAverTemp1221}), (\ref%
{Eq.StochAverTemp1212}) and (\ref{Eq.StochAverTemp2121}) into Eq. (\ref%
{Eq.QCFTempEvnStep1}) and performing the sums over $q_{1},q_{2},q_{3},q_{4}$
gives 
\begin{eqnarray}
&&\overline{\widetilde{\phi }_{d}(k_{1},\beta +\delta \beta )\,\widetilde{%
\phi }_{u}(k_{2},\beta +\delta \beta )\;\widetilde{\phi }_{u}^{+}(k_{3},%
\beta +\delta \beta )\,\widetilde{\phi }_{d}^{+}(k_{4},\beta +\delta \beta )}
\nonumber \\
&=&\dsum\limits_{l_{1}l_{2}l_{3}l_{4}}\delta _{k_{1},l_{1}}\,\delta
_{k_{2},l_{2}}\,\delta _{k_{3},l_{3}}\,\delta _{k_{4},l_{4}}\left( 1+\frac{1%
}{2}\left\{ \frac{\hbar ^{2}k_{1}^{2}}{2m}+\frac{\hbar ^{2}k_{2}^{2}}{2m}+%
\frac{\hbar ^{2}k_{3}^{2}}{2m}+\frac{\hbar ^{2}k_{4}^{2}}{2m}\right\} \delta
\beta \right)  \nonumber \\
&&\times \overline{\widetilde{\phi }_{d}(l_{1},\beta )\,\widetilde{\phi }%
_{u}(l_{2},\beta )\;\widetilde{\phi }_{u}^{+}(l_{3},\beta )\,\widetilde{\phi 
}_{d}^{+}(l_{4},\beta )}  \nonumber \\
&&+2\eta ^{2}\,\dsum\limits_{l_{1}l_{2}l_{3}l_{4}}\{\delta
_{k_{1},l_{1}}\,\delta _{k_{2},l_{2}}\,\delta
_{(k_{3}+k_{4}),(l_{3}+l_{4})}\}\,\delta \beta \times \overline{\widetilde{%
\phi }_{d}(l_{1},\beta )\,\widetilde{\phi }_{u}(l_{2},\beta )\;\widetilde{%
\phi }_{d}^{+}(l_{3},\beta )\,\widetilde{\phi }_{u}^{+}(l_{4},\beta )} 
\nonumber \\
&&+2\eta ^{2}\,\dsum\limits_{l_{1}l_{2}l_{3}l_{4}}\{\delta
_{(k_{1}+k_{2}),(l_{1}+l_{2})}\,\delta _{k_{3},l_{3}}\,\delta
_{k_{4},l_{4}}\}\,\delta \beta \times \overline{\widetilde{\phi }%
_{u}(l_{1},\beta )\,\widetilde{\phi }_{d}(l_{2},\beta )\;\widetilde{\phi }%
_{u}^{+}(l_{3},\beta )\,\widetilde{\phi }_{d}^{+}(l_{4},\beta )}  \nonumber
\\
&=&\dsum\limits_{l_{1}l_{2}l_{3}l_{4}}\delta _{k_{1},l_{1}}\,\delta
_{k_{2},l_{2}}\,\delta _{k_{3},l_{3}}\,\delta _{k_{4},l_{4}}\left( 1+\frac{1%
}{2}\left\{ \frac{\hbar ^{2}k_{1}^{2}}{2m}+\frac{\hbar ^{2}k_{2}^{2}}{2m}+%
\frac{\hbar ^{2}k_{3}^{2}}{2m}+\frac{\hbar ^{2}k_{4}^{2}}{2m}\right\} \delta
\beta \right)  \nonumber \\
&&\times \overline{\widetilde{\phi }_{d}(l_{1},\beta )\,\widetilde{\phi }%
_{u}(l_{2},\beta )\;\widetilde{\phi }_{u}^{+}(l_{3},\beta )\,\widetilde{\phi 
}_{d}^{+}(l_{4},\beta )}  \nonumber \\
&&-2\eta ^{2}\,\dsum\limits_{l_{1}l_{2}l_{3}l_{4}}\{\delta
_{k_{1},l_{1}}\,\delta _{k_{2},l_{2}}\,\delta
_{(k_{3}+k_{4}),(l_{3}+l_{4})}\}\,\delta \beta \times \overline{\widetilde{%
\phi }_{d}(l_{1},\beta )\,\widetilde{\phi }_{u}(l_{2},\beta )\;\widetilde{%
\phi }_{u}^{+}(l_{3},\beta )\,\widetilde{\phi }_{d}^{+}(l_{4},\beta )} 
\nonumber \\
&&-2\eta ^{2}\,\dsum\limits_{l_{1}l_{2}l_{3}l_{4}}\{\delta
_{(k_{1}+k_{2}),(l_{1}+l_{2})}\,\delta _{k_{3},l_{3}}\,\delta
_{k_{4},l_{4}}\}\,\delta \beta \times \overline{\widetilde{\phi }%
_{d}(l_{1},\beta )\,\widetilde{\phi }_{u}(l_{2},\beta )\;\widetilde{\phi }%
_{u}^{+}(l_{3},\beta )\,\widetilde{\phi }_{d}^{+}(l_{4},\beta )}  \nonumber
\\
&&  \label{Eq.QCFOnePairTempEvnFirstOrderChange}
\end{eqnarray}%
where we have used $\dsum\limits_{q_{3}q_{4}}\delta
_{(k_{3}+q_{3}),l_{3}}\;\delta _{(k_{4}-q_{4}),l_{4}}\,\delta
_{q_{3},q_{4}}=\delta _{(k_{3}+k_{4}),(l_{3}+l_{4})}$ and $%
\dsum\limits_{q_{1}}\delta _{q_{1},0}=1$. Also, in the second last line we
have interchanged $l_{3}$ and $l_{4}$ and used the anti-commuting feature of
the Grassmann stochastic momentum fields $\widetilde{\phi }%
_{d}^{+}(l_{3},\beta )$ and $\,\widetilde{\phi }_{u}^{+}(l_{4},\beta )$ to
place these in the opposite order in the stochastic average $\overline{%
\widetilde{\phi }_{d}(l_{1},\beta )\,\widetilde{\phi }_{u}(l_{2},\beta )\;%
\widetilde{\phi }_{d}^{+}(l_{3},\beta )\,\widetilde{\phi }%
_{u}^{+}(l_{4},\beta )}$, with similar steps for the last line. \medskip

\subsection{Two Cooper Pairs - Time Evolution}

\label{SubSection - Two Cooper Pairs Time Evoln}

For the case of time evolution the QCF for two Cooper pairs at time $%
t+\delta t$ is related to that at time $t$ via 
\begin{eqnarray}
&&\left[ 
\begin{array}{c}
\widetilde{\phi }_{d}(k_{1},t+\delta t)\,\widetilde{\phi }%
_{d}(k_{2},t+\delta t)\;\widetilde{\phi }_{u}(k_{3},t+\delta t)\,\widetilde{%
\phi }_{u}(k_{4},t+\delta t) \\ 
\times \;\widetilde{\phi }_{u}^{+}(k_{5},t+\delta t)\,\widetilde{\phi }%
_{u}^{+}(k_{6},t+\delta t)\widetilde{\phi }_{d}^{+}(k_{7},t+\delta t)\,%
\widetilde{\phi }_{d}^{+}(k_{8},t+\delta t)%
\end{array}%
\right] _{StochAver}  \nonumber \\
&=&\left[ 
\begin{array}{c}
\dsum\limits_{l_{1},q_{1}}(F_{d,d}(k_{1},l_{1};q_{1},\delta t)\,\widetilde{%
\phi }_{d}(l_{1},t)+F_{d,u}(k_{1},l_{1};q_{1},\delta t)\,\widetilde{\phi }%
_{u}(l_{1},t)) \\ 
\times \dsum\limits_{l_{2},q_{2}}(F_{d,d}(k_{2},l_{2};q_{2},\delta t)\,%
\widetilde{\phi }_{d}(l_{2},t)+F_{d,u}(k_{2},l_{2};q_{2},\delta t)\,%
\widetilde{\phi }_{u}(l_{2},t)) \\ 
\times \dsum\limits_{l_{3},q_{3}}(F_{u,d}(k_{3},l_{3};q_{3},\delta t)\,%
\widetilde{\phi }_{d}(l_{3},t)+F_{u,u}(k_{3},l_{3};q_{3},\delta t)\,%
\widetilde{\phi }_{u}(l_{3},t)) \\ 
\times \dsum\limits_{l_{4},q_{4}}(F_{u,d}(k_{4},l_{4};q_{4},\delta t)\,%
\widetilde{\phi }_{d}(l_{4},t)+F_{u,u}(k_{4},l_{4};q_{4},\delta t)\,%
\widetilde{\phi }_{u}(l_{4},t)) \\ 
\times \dsum\limits_{l_{5},q_{5}}(F_{u,d}^{+}(k_{5},l_{5};q_{5},\delta t)\,%
\widetilde{\phi }_{d}^{+}(l_{5},t)+F_{u,u}^{+}(k_{5},l_{5};q_{5},\delta t)\,%
\widetilde{\phi }_{u}^{+}(l_{5},t)) \\ 
\times \dsum\limits_{l_{6},q_{6}}(F_{u,d}^{+}(k_{6},l_{6};q_{6},\delta t)\,%
\widetilde{\phi }_{d}^{+}(l_{6},t)+F_{u,u}^{+}(k_{6},l_{6};q_{6},\delta t)\,%
\widetilde{\phi }_{u}^{+}(l_{6},t)) \\ 
\times \dsum\limits_{l_{7},q_{7}}(F_{d,d}^{+}(k_{7},l_{7};q_{7},\delta t)\,%
\widetilde{\phi }_{d}^{+}(l_{7},t)+F_{d,u}^{+}(k_{7},l_{7};q_{7},\delta t)\,%
\widetilde{\phi }_{u}^{+}(l_{7},t)) \\ 
\times \dsum\limits_{l_{8},q_{8}}(F_{d,d}^{+}(k_{8},l_{8};q_{8},\delta t)\,%
\widetilde{\phi }_{d}^{+}(l_{8},t)+F_{d,u}^{+}(k_{8},l_{8};q_{8},\delta t)\,%
\widetilde{\phi }_{u}^{+}(l_{8},t))%
\end{array}%
\right] _{StochAver}  \nonumber \\
&&  \label{Eq.QCFTwoPairsTimeEvnStep1}
\end{eqnarray}%
after substituting from Eq.(\ref{Eq.ItoSMtmTimeEvln}). Apart from the sums
over $l_{1,}q_{1},...,l_{8},q_{8}$ there are $256$ terms for which the
stochastic averages have to be evaluated. If the two terms in each of the
first $4$ \ factors are listed as $1,2$ and the two terms in each of the
second $4$ factors are listed as $a,b$ then the $256$ terms can be listed as

$1111aaaa,1211aaaa,2111aaaa,2211aaaa,....,2222aaaa,$

$%
1111abaa,1211abaa,2111abaa,2211abaa,....,2222abaa,....,1111bbbb,....2222bbbb. 
$.

In Ito stochastic equations we then use the factorisation result that
depends on the $F$, $F^{+}$ involving times later than $t$ (see Eq. (9.43)
in Ref. \cite{Dalton15a} 
\begin{eqnarray}
&&\left[ 
\begin{array}{c}
F_{\alpha ,\beta }(k_{a},l_{a};q_{a},\delta t)\,\widetilde{\phi }_{\beta
}(l_{a},t)\,F_{\gamma ,\delta }(k_{b},l_{b};q_{b},\delta t)\,\widetilde{\phi 
}_{\delta }(l_{b},t)\, \\ 
\times F_{\epsilon ,\eta }(k_{c},l_{c};q_{c},\delta t)\,\widetilde{\phi }%
_{\eta }(l_{c},t)\,F_{\theta ,\kappa }(k_{d},l_{d};q_{d},\delta t)\,%
\widetilde{\phi }_{\kappa }(l_{d},t) \\ 
\times F_{\pi ,\xi }^{+}(k_{e},l_{e};q_{e},\delta t)\,\widetilde{\phi }_{\xi
}^{+}(l_{e},t)\,F_{\zeta ,\tau }^{+}(k_{f},l_{f};q_{f},\delta t)\,\widetilde{%
\phi }_{\tau }^{+}(l_{f},t) \\ 
\times \,F_{\nu ,\mu }^{+}(k_{g},l_{g};q_{g},\delta t)\,\widetilde{\phi }%
_{\mu }^{+}(l_{g},t)\,F_{\omega ,\chi }^{+}(k_{h},l_{h};q_{h},\delta t)\,%
\widetilde{\phi }_{\chi }^{+}(l_{h},t)%
\end{array}%
\right] _{StochAver}\,  \nonumber \\
&=&\left[ 
\begin{array}{c}
F_{\alpha ,\beta }(k_{a},l_{a};q_{a},\delta t)\,F_{\gamma ,\delta
}(k_{b},l_{b};q_{b},\delta t) \\ 
\times \,F_{\epsilon ,\eta }(k_{c},l_{c};q_{c},\delta t)\,F_{\theta ,\kappa
}(k_{d},l_{d};q_{d},\delta t) \\ 
\times F_{\pi ,\xi }^{+}(k_{e},l_{e};q_{e},\delta t)\,F_{\zeta ,\tau
}^{+}(k_{f},l_{f};q_{f},\delta t) \\ 
\times F_{\nu ,\mu }^{+}(k_{g},l_{g};q_{g},\delta t)\,F_{\omega ,\chi
}^{+}(k_{h},l_{h};q_{h},\delta t)%
\end{array}%
\right] _{StochAver}  \nonumber \\
&&\times \overline{\,\widetilde{\phi }_{\beta }(l_{a},t)\,\,\widetilde{\phi }%
_{\delta }(l_{b},t)\,\widetilde{\phi }_{\eta }(l_{c},t)\,\widetilde{\phi }%
_{\kappa }(l_{d},t)\,\widetilde{\phi }_{\xi }^{+}(l_{e},t)\,\widetilde{\phi }%
_{\tau }^{+}(l_{f},t)\,\,\widetilde{\phi }_{\mu }^{+}(l_{g},t)\,\widetilde{%
\phi }_{\chi }^{+}(l_{h},t)}  \nonumber \\
&&  \label{Eq.QuantumCorrelnFnGeneralB}
\end{eqnarray}

We wish to evaluate the various

$\overline{F_{\alpha ,\beta }(k_{a},l_{a};q_{a},\delta t)\,F_{\gamma ,\delta
}(k_{b},l_{b};q_{b},\delta t)\,...F_{\nu ,\mu }^{+}(k_{g},l_{g};q_{g},\delta
t)\,F_{\omega ,\chi }^{+}(k_{h},l_{h};q_{h},\delta t)}$ correct to order $%
\delta t$. We note that the $F_{d,d},F_{u,u},F_{d,d}^{+},F_{u,u}^{+}$
factors are non-stochastic and linear in $\delta t$. Also the $%
F_{d,u},F_{u,d},F_{d,u}^{+},F_{u,d}^{+}$ factors are stochastic and linear
in the stochastic Wiener increments $\delta \widetilde{\omega }%
_{u,d}^{q},\delta \widetilde{\omega }_{d,u}^{q},\delta \widetilde{\omega }%
_{u+,d+}^{q},\delta \widetilde{\omega }_{d+,u+}^{q}$. From Eq.(\ref%
{Eq.PropsWienerIncrem}) the stochastic average of a factor involving an odd
number $1,3,5,7$ of the $\delta \widetilde{\omega }_{u,d}^{q},\delta 
\widetilde{\omega }_{d,u}^{q},\delta \widetilde{\omega }_{u+,d+}^{q},\delta 
\widetilde{\omega }_{d+,u+}^{q}$ with any non-stochastic factor is zero, if
there are $2$ of these increments the stochastic average is proportional to $%
\delta t$, if there are $4$ of these increments the stochastic average is
proportional to $\delta t^{2}$ and if there are $6$ the stochastic average
is proportional to $\delta t^{3}$ (apart from cases where there is zero
Kronecka delta $\delta _{{\small ab}}$ involved). For the case of two
stochastic Wiener increments we note from Eq. (\ref{Eq.PropsWienerIncrem})
that the non-zero stochastic averages of the products of $F^{\prime }s$ \
and $F^{+\prime }s$ are 
\begin{eqnarray}
\overline{F_{d,u}(k_{a},l_{a};q_{a},\delta
t)F_{u,d}(k_{b},l_{b};q_{b},\delta t)} &=&+2\lambda ^{2}\,\delta
_{q_{a},q_{b}}\,\delta _{(k_{a}+q_{a}),l_{a}}\,\delta
_{(k_{b}-q_{b}),l_{b}}\;\delta t  \nonumber \\
\overline{F_{u,d}^{+}(k_{a},l_{a};q_{a},\delta
t)F_{d,u}^{+}(k_{b},l_{b};q_{b},\delta t)} &=&-2\lambda ^{2}\,\delta
_{q_{a},q_{b}}\,\delta _{(k_{a}+q_{a}),l_{a}}\,\delta
_{(k_{b}-q_{b}),l_{b}}\;\delta t  \nonumber \\
&&  \label{Eq.StochAverProdFs}
\end{eqnarray}

As we are looking for contributions that are only linear in $\delta t$, the
only terms that will result in this have:

(1) Only non-stochastic $F^{\prime }s$ from the first $4$ factors and only
non-stochastic $F^{+\prime }s$ from the second $4$ factors. There is only
one term - $1122bbaa.$

(2) Only non-stochastic $F^{\prime }s$ from the first $4$ factors and only
one stochastic $F_{u,d}^{+}$ and one stochastic $F_{d,u}^{+}$ from the
second $4$ factors.There are four terms - $1122abba,%
\,1122abab,1122baba,1122baab.$

(3) Only one stochastic $F_{u,d}$ and one stochastic $F_{d,u}$ from the
first $4$ factors and only non-stochastic $F^{+\prime }s$ from the second $4$
factors.There are four terms - $1221bbaa,1212bbaa,2121bbaa,2112bbaa.$

With these considerations in mind, the non-zero cases are as follows:

For the terms of type (1):%
\begin{eqnarray}
1122bbaa &:&\left[ 
\begin{array}{c}
F_{d,d}(k_{1},l_{1};q_{1},\delta t)F_{d,d}(k_{2},l_{2};q_{2},\delta
t)F_{u,u}(k_{3},l_{3};q_{3},\delta t)F_{u,u}(k_{4},l_{4};q_{4},\delta t) \\ 
\times \;F_{u,u}^{+}(k_{5},l_{5};q_{5},\delta
t)\,F_{u,u}^{+}(k_{6},l_{6};q_{6},\delta
t)\,F_{d,d}^{+}(k_{7},l_{7};q_{7},\delta
t)\,F_{d,d}^{+}(k_{8},l_{8};q_{8},\delta t)%
\end{array}%
\right] _{StochAver}  \nonumber \\
&=&\delta _{q_{1},0}\;\delta _{k_{1},l_{1}}\left( 1+\frac{i}{\hbar }\left\{ 
\frac{\hbar ^{2}k_{1}^{2}}{2m}\right\} \delta t\right) \delta
_{q_{2},0}\;\delta _{k_{2},l_{2}}\left( 1+\frac{i}{\hbar }\left\{ \frac{%
\hbar ^{2}k_{2}^{2}}{2m}\right\} \delta t\right)  \nonumber \\
&&\times \delta _{q_{3},0}\;\delta _{k_{3},l_{3}}\left( 1+\frac{i}{\hbar }%
\left\{ \frac{\hbar ^{2}k_{3}^{2}}{2m}\right\} \delta t\right) \delta
_{q_{4},0}\;\delta _{k_{4},l_{4}}\left( 1+\frac{i}{\hbar }\left\{ \frac{%
\hbar ^{2}k_{4}^{2}}{2m}\right\} \delta t\right)  \nonumber \\
&&\times \delta _{q_{5},0}\;\delta _{k_{5},l_{5}}\left( 1-\frac{i}{\hbar }%
\left\{ \frac{\hbar ^{2}k_{5}^{2}}{2m}\right\} \delta t\right) \delta
_{q_{6},0}\;\delta _{k_{6},l_{6}}\left( 1-\frac{i}{\hbar }\left\{ \frac{%
\hbar ^{2}k_{6}^{2}}{2m}\right\} \delta t\right)  \nonumber \\
&&\times \delta _{q_{7},0}\;\delta _{k_{7},l_{7}}\left( 1-\frac{i}{\hbar }%
\left\{ \frac{\hbar ^{2}k_{7}^{2}}{2m}\right\} \delta t\right) \delta
_{q_{8},0}\;\delta _{k_{8},l_{8}}\left( 1-\frac{i}{\hbar }\left\{ \frac{%
\hbar ^{2}k_{8}^{2}}{2m}\right\} \delta t\right)  \nonumber \\
&\approx &\delta _{q_{1},0}\;\delta _{k_{1},l_{1}}\,\delta
_{q_{2},0}\;\delta _{k_{2},l_{2}}\,\delta _{q_{3},0}\;\delta
_{k_{3},l_{3}}\,\delta _{q_{4},0}\;\delta _{k_{4},l_{4}}  \nonumber \\
&&\times \,\delta _{q_{5},0}\;\delta _{k_{5},l_{5}}\,\delta
_{q_{6},0}\;\delta _{k_{6},l_{6}}\,\delta _{q_{7},0}\;\delta
_{k_{7},l_{7}}\,\delta _{q_{8},0}\;\delta _{k_{8},l_{8}}  \nonumber \\
&&\times \left( 1+\frac{i}{\hbar }\left\{ \frac{\hbar ^{2}k_{1}^{2}}{2m}+%
\frac{\hbar ^{2}k_{2}^{2}}{2m}+\frac{\hbar ^{2}k_{3}^{2}}{2m}+\frac{\hbar
^{2}k_{4}^{2}}{2m}-\frac{\hbar ^{2}k_{5}^{2}}{2m}-\frac{\hbar ^{2}k_{6}^{2}}{%
2m}-\frac{\hbar ^{2}k_{7}^{2}}{2m}-\frac{\hbar ^{2}k_{8}^{2}}{2m}\right\}
\delta t\right)  \nonumber \\
&&  \label{Eq.StochAver1122bbaa}
\end{eqnarray}%
where only terms linear in $\delta t$ have been retained.

For the terms of type (2):%
\begin{eqnarray}
1122abba &:&\left[ 
\begin{array}{c}
F_{d,d}(k_{1},l_{1};q_{1},\delta t)F_{d,d}(k_{2},l_{2};q_{2},\delta
t)F_{u,u}(k_{3},l_{3};q_{3},\delta t)F_{u,u}(k_{4},l_{4};q_{4},\delta t) \\ 
\times \;F_{u,d}^{+}(k_{5},l_{5};q_{5},\delta
t)\,F_{u,u}^{+}(k_{6},l_{6};q_{6},\delta
t)\,F_{d,u}^{+}(k_{7},l_{7};q_{7},\delta
t)\,F_{d,d}^{+}(k_{8},l_{8};q_{8},\delta t)%
\end{array}%
\right] _{StochAver}  \nonumber \\
&=&\delta _{q_{1},0}\;\delta _{k_{1},l_{1}}\left( 1+\frac{i}{\hbar }\left\{ 
\frac{\hbar ^{2}k_{1}^{2}}{2m}\right\} \delta t\right) \delta
_{q_{2},0}\;\delta _{k_{2},l_{2}}\left( 1+\frac{i}{\hbar }\left\{ \frac{%
\hbar ^{2}k_{2}^{2}}{2m}\right\} \delta t\right)  \nonumber \\
&&\times \delta _{q_{3},0}\;\delta _{k_{3},l_{3}}\left( 1+\frac{i}{\hbar }%
\left\{ \frac{\hbar ^{2}k_{3}^{2}}{2m}\right\} \delta t\right) \delta
_{q_{4},0}\;\delta _{k_{4},l_{4}}\left( 1+\frac{i}{\hbar }\left\{ \frac{%
\hbar ^{2}k_{4}^{2}}{2m}\right\} \delta t\right)  \nonumber \\
&&\times \delta _{q_{6},0}\;\delta _{k_{6},l_{6}}\left( 1-\frac{i}{\hbar }%
\left\{ \frac{\hbar ^{2}k_{6}^{2}}{2m}\right\} \delta t\right) \delta
_{q_{8},0}\;\delta _{k_{8},l_{8}}\left( 1-\frac{i}{\hbar }\left\{ \frac{%
\hbar ^{2}k_{8}^{2}}{2m}\right\} \delta t\right)  \nonumber \\
&&\times (-2\lambda ^{2})\,\delta _{q_{5},q_{7}}\,\delta
_{(k_{5}+q_{5}),l_{5}}\,\delta _{(k_{7}-q_{7}),l_{7}}\;\delta t  \nonumber \\
&\approx &(-2\lambda ^{2})\,  \nonumber \\
&&\times \delta _{q_{1},0}\;\delta _{k_{1},l_{1}}\delta _{q_{2},0}\;\delta
_{k_{2},l_{2}}\delta _{q_{3},0}\;\delta _{k_{3},l_{3}}\delta
_{q_{4},0}\;\delta _{k_{4},l_{4}}  \nonumber \\
&&\times \delta _{q_{6},0}\;\delta _{k_{6},l_{6}}\delta _{q_{8},0}\;\delta
_{k_{8},l_{8}}\;\delta _{q_{5},q_{7}}\,\delta _{(k_{5}+q_{5}),l_{5}}\,\delta
_{(k_{7}-q_{7}),l_{7}}\;\delta t  \nonumber \\
&&  \label{Eq.StochAver1122abba}
\end{eqnarray}%
\begin{eqnarray}
1122abab &:&\left[ 
\begin{array}{c}
F_{d,d}(k_{1},l_{1};q_{1},\delta t)F_{d,d}(k_{2},l_{2};q_{2},\delta
t)F_{u,u}(k_{3},l_{3};q_{3},\delta t)F_{u,u}(k_{4},l_{4};q_{4},\delta t) \\ 
\times \;F_{u,d}^{+}(k_{5},l_{5};q_{5},\delta
t)\,F_{u,u}^{+}(k_{6},l_{6};q_{6},\delta
t)\,F_{d,d}^{+}(k_{7},l_{7};q_{7},\delta
t)\,F_{d,u}^{+}(k_{8},l_{8};q_{8},\delta t)%
\end{array}%
\right] _{StochAver}  \nonumber \\
&=&\delta _{q_{1},0}\;\delta _{k_{1},l_{1}}\left( 1+\frac{i}{\hbar }\left\{ 
\frac{\hbar ^{2}k_{1}^{2}}{2m}\right\} \delta t\right) \delta
_{q_{2},0}\;\delta _{k_{2},l_{2}}\left( 1+\frac{i}{\hbar }\left\{ \frac{%
\hbar ^{2}k_{2}^{2}}{2m}\right\} \delta t\right)  \nonumber \\
&&\times \delta _{q_{3},0}\;\delta _{k_{3},l_{3}}\left( 1+\frac{i}{\hbar }%
\left\{ \frac{\hbar ^{2}k_{3}^{2}}{2m}\right\} \delta t\right) \delta
_{q_{4},0}\;\delta _{k_{4},l_{4}}\left( 1+\frac{i}{\hbar }\left\{ \frac{%
\hbar ^{2}k_{4}^{2}}{2m}\right\} \delta t\right)  \nonumber \\
&&\times \delta _{q_{6},0}\;\delta _{k_{6},l_{6}}\left( 1-\frac{i}{\hbar }%
\left\{ \frac{\hbar ^{2}k_{6}^{2}}{2m}\right\} \delta t\right) \delta
_{q_{7},0}\;\delta _{k_{7},l_{7}}\left( 1-\frac{i}{\hbar }\left\{ \frac{%
\hbar ^{2}k_{7}^{2}}{2m}\right\} \delta t\right)  \nonumber \\
&&\times (-2\lambda ^{2})\,\delta _{q_{5},q_{8}}\,\delta
_{(k_{5}-q_{5}),l_{5}}\,\delta _{(k_{8}+q_{8}),l_{8}}\;\delta t  \nonumber \\
&\approx &(-2\lambda ^{2})\,  \nonumber \\
&&\times \delta _{q_{1},0}\;\delta _{k_{1},l_{1}}\delta _{q_{2},0}\;\delta
_{k_{2},l_{2}}\delta _{q_{3},0}\;\delta _{k_{3},l_{3}}\delta
_{q_{4},0}\;\delta _{k_{4},l_{4}}  \nonumber \\
&&\times \delta _{q_{6},0}\;\delta _{k_{6},l_{6}}\delta _{q_{7},0}\;\delta
_{k_{7},l_{7}}\;\delta _{q_{5},q_{8}}\,\delta _{(k_{5}+q_{5}),l_{5}}\,\delta
_{(k_{8}-q_{8}),l_{8}}\;\delta t \\
&&  \label{Eq.StochAver1122abab}
\end{eqnarray}%
\begin{eqnarray}
1122baba &:&\left[ 
\begin{array}{c}
F_{d,d}(k_{1},l_{1};q_{1},\delta t)F_{d,d}(k_{2},l_{2};q_{2},\delta
t)F_{u,u}(k_{3},l_{3};q_{3},\delta t)F_{u,u}(k_{4},l_{4};q_{4},\delta t) \\ 
\times \;F_{u,u}^{+}(k_{5},l_{5};q_{5},\delta
t)\,F_{u,d}^{+}(k_{6},l_{6};q_{6},\delta
t)\,F_{d,u}^{+}(k_{7},l_{7};q_{7},\delta
t)\,F_{d,d}^{+}(k_{8},l_{8};q_{8},\delta t)%
\end{array}%
\right] _{StochAver}  \nonumber \\
&=&\delta _{q_{1},0}\;\delta _{k_{1},l_{1}}\left( 1+\frac{i}{\hbar }\left\{ 
\frac{\hbar ^{2}k_{1}^{2}}{2m}\right\} \delta t\right) \delta
_{q_{2},0}\;\delta _{k_{2},l_{2}}\left( 1+\frac{i}{\hbar }\left\{ \frac{%
\hbar ^{2}k_{2}^{2}}{2m}\right\} \delta t\right)  \nonumber \\
&&\times \delta _{q_{3},0}\;\delta _{k_{3},l_{3}}\left( 1+\frac{i}{\hbar }%
\left\{ \frac{\hbar ^{2}k_{3}^{2}}{2m}\right\} \delta t\right) \delta
_{q_{4},0}\;\delta _{k_{4},l_{4}}\left( 1+\frac{i}{\hbar }\left\{ \frac{%
\hbar ^{2}k_{4}^{2}}{2m}\right\} \delta t\right)  \nonumber \\
&&\times \delta _{q_{5},0}\;\delta _{k_{5},l_{5}}\left( 1-\frac{i}{\hbar }%
\left\{ \frac{\hbar ^{2}k_{5}^{2}}{2m}\right\} \delta t\right) \delta
_{q_{8},0}\;\delta _{k_{8},l_{8}}\left( 1-\frac{i}{\hbar }\left\{ \frac{%
\hbar ^{2}k_{8}^{2}}{2m}\right\} \delta t\right)  \nonumber \\
&&\times (-2\lambda ^{2})\,\delta _{q_{6},q_{7}}\,\delta
_{(k_{6}+q_{6}),l_{6}}\,\delta _{(k_{7}-q_{7}),l_{7}}\;\delta t  \nonumber \\
&\approx &(-2\lambda ^{2})\,  \nonumber \\
&&\times \delta _{q_{1},0}\;\delta _{k_{1},l_{1}}\delta _{q_{2},0}\;\delta
_{k_{2},l_{2}}\delta _{q_{3},0}\;\delta _{k_{3},l_{3}}\delta
_{q_{4},0}\;\delta _{k_{4},l_{4}}  \nonumber \\
&&\delta _{q_{5},0}\;\delta _{k_{5},l_{5}}\delta _{q_{8},0}\;\delta
_{k_{8},l_{8}}\;\delta _{q_{6},q_{7}}\,\delta _{(k_{6}+q_{6}),l_{6}}\,\delta
_{(k_{7}-q_{7}),l_{7}}\;\delta t  \nonumber \\
&&  \label{Eq.StochAver1122baba}
\end{eqnarray}%
\begin{eqnarray}
1122baab &:&\left[ 
\begin{array}{c}
F_{d,d}(k_{1},l_{1};q_{1},\delta t)F_{d,d}(k_{2},l_{2};q_{2},\delta
t)F_{u,u}(k_{3},l_{3};q_{3},\delta t)F_{u,u}(k_{4},l_{4};q_{4},\delta t) \\ 
\times \;F_{u,u}^{+}(k_{5},l_{5};q_{5},\delta
t)\,F_{u,d}^{+}(k_{6},l_{6};q_{6},\delta
t)\,F_{d,d}^{+}(k_{7},l_{7};q_{7},\delta
t)\,F_{d,u}^{+}(k_{8},l_{8};q_{8},\delta t)%
\end{array}%
\right] _{StochAver}  \nonumber \\
&=&\delta _{q_{1},0}\;\delta _{k_{1},l_{1}}\left( 1+\frac{i}{\hbar }\left\{ 
\frac{\hbar ^{2}k_{1}^{2}}{2m}\right\} \delta t\right) \delta
_{q_{2},0}\;\delta _{k_{2},l_{2}}\left( 1+\frac{i}{\hbar }\left\{ \frac{%
\hbar ^{2}k_{2}^{2}}{2m}\right\} \delta t\right)  \nonumber \\
&&\times \delta _{q_{3},0}\;\delta _{k_{3},l_{3}}\left( 1+\frac{i}{\hbar }%
\left\{ \frac{\hbar ^{2}k_{3}^{2}}{2m}\right\} \delta t\right) \delta
_{q_{4},0}\;\delta _{k_{4},l_{4}}\left( 1+\frac{i}{\hbar }\left\{ \frac{%
\hbar ^{2}k_{4}^{2}}{2m}\right\} \delta t\right)  \nonumber \\
&&\times \delta _{q_{5},0}\;\delta _{k_{5},l_{5}}\left( 1-\frac{i}{\hbar }%
\left\{ \frac{\hbar ^{2}k_{5}^{2}}{2m}\right\} \delta t\right) \delta
_{q_{7},0}\;\delta _{k_{7},l_{7}}\left( 1-\frac{i}{\hbar }\left\{ \frac{%
\hbar ^{2}k_{7}^{2}}{2m}\right\} \delta t\right)  \nonumber \\
&&\times (-2\lambda ^{2})\,\delta _{q_{6},q_{8}}\,\delta
_{(k_{6}+q_{6}),l_{6}}\,\delta _{(k_{8}-q_{8}),l_{8}}\;\delta t  \nonumber \\
&\approx &(-2\lambda ^{2})\,  \nonumber \\
&&\times \delta _{q_{1},0}\;\delta _{k_{1},l_{1}}\delta _{q_{2},0}\;\delta
_{k_{2},l_{2}}\delta _{q_{3},0}\;\delta _{k_{3},l_{3}}\delta
_{q_{4},0}\;\delta _{k_{4},l_{4}}  \nonumber \\
&&\times \delta _{q_{5},0}\;\delta _{k_{5},l_{5}}\delta _{q_{7},0}\;\delta
_{k_{7},l_{7}}\;\delta _{q_{6},q_{8}}\,\delta _{(k_{6}+q_{6}),l_{6}}\,\delta
_{(k_{8}-q_{8}),l_{8}}\;\delta t  \nonumber \\
&&  \label{Eq.StochAver1122baab}
\end{eqnarray}%
where only terms linear in $\delta t$ have been retained.

For terms of type (3): 
\begin{eqnarray}
1221bbaa &:&\left[ 
\begin{array}{c}
F_{d,d}(k_{1},l_{1};q_{1},\delta t)F_{d,u}(k_{2},l_{2};q_{2},\delta
t)F_{u,u}(k_{3},l_{3};q_{3},\delta t)F_{u,d}(k_{4},l_{4};q_{4},\delta t) \\ 
\times \;F_{u,u}^{+}(k_{5},l_{5};q_{5},\delta
t)\,F_{u,u}^{+}(k_{6},l_{6};q_{6},\delta
t)\,F_{d,d}^{+}(k_{7},l_{7};q_{7},\delta
t)\,F_{d,d}^{+}(k_{8},l_{8};q_{8},\delta t)%
\end{array}%
\right] _{StochAver}  \nonumber \\
&=&\delta _{q_{1},0}\;\delta _{k_{1},l_{1}}\left( 1+\frac{i}{\hbar }\left\{ 
\frac{\hbar ^{2}k_{1}^{2}}{2m}\right\} \delta t\right) \delta
_{q_{3},0}\;\delta _{k_{3},l_{3}}\left( 1+\frac{i}{\hbar }\left\{ \frac{%
\hbar ^{2}k_{3}^{2}}{2m}\right\} \delta t\right)  \nonumber \\
&&\times \delta _{q_{5},0}\;\delta _{k_{5},l_{5}}\left( 1-\frac{i}{\hbar }%
\left\{ \frac{\hbar ^{2}k_{5}^{2}}{2m}\right\} \delta t\right) \delta
_{q_{6},0}\;\delta _{k_{6},l_{6}}\left( 1-\frac{i}{\hbar }\left\{ \frac{%
\hbar ^{2}k_{6}^{2}}{2m}\right\} \delta t\right)  \nonumber \\
&&\times \delta _{q_{7},0}\;\delta _{k_{7},l_{7}}\left( 1-\frac{i}{\hbar }%
\left\{ \frac{\hbar ^{2}k_{7}^{2}}{2m}\right\} \delta t\right) \delta
_{q_{8},0}\;\delta _{k_{8},l_{8}}\left( 1-\frac{i}{\hbar }\left\{ \frac{%
\hbar ^{2}k_{8}^{2}}{2m}\right\} \delta t\right)  \nonumber \\
&&\times (+2\lambda ^{2})\,\delta _{q_{2},q_{4}}\,\delta
_{(k_{2}+q_{2}),l_{2}}\,\delta _{(k_{4}-q_{4}),l_{4}}\;\delta t  \nonumber \\
&\approx &(+2\lambda ^{2})\,  \nonumber \\
&&\times \delta _{q_{1},0}\;\delta _{k_{1},l_{1}}\,\delta _{q_{3},0}\;\delta
_{k_{3},l_{3}}\,\delta _{q_{5},0}\;\delta _{k_{5},l_{5}}\,\delta
_{q_{6},0}\;\delta _{k_{6},l_{6}}\,  \nonumber \\
&&\times \delta _{q_{7},0}\;\delta _{k_{7},l_{7}}\,\delta _{q_{8},0}\;\delta
_{k_{8},l_{8}}\;\,\delta _{q_{2},q_{4}}\,\delta
_{(k_{2}+q_{2}),l_{2}}\,\delta _{(k_{4}-q_{4}),l_{4}}\;\delta t  \nonumber \\
&&  \label{Eq.StochAver1221bbaa}
\end{eqnarray}%
\begin{eqnarray}
1212bbaa &:&\left[ 
\begin{array}{c}
F_{d,d}(k_{1},l_{1};q_{1},\delta t)F_{d,u}(k_{2},l_{2};q_{2},\delta
t)F_{u,d}(k_{3},l_{3};q_{3},\delta t)F_{u,u}(k_{4},l_{4};q_{4},\delta t) \\ 
\times \;F_{u,u}^{+}(k_{5},l_{5};q_{5},\delta
t)\,F_{u,u}^{+}(k_{6},l_{6};q_{6},\delta
t)\,F_{d,d}^{+}(k_{7},l_{7};q_{7},\delta
t)\,F_{d,d}^{+}(k_{8},l_{8};q_{8},\delta t)%
\end{array}%
\right] _{StochAver}  \nonumber \\
&=&\delta _{q_{1},0}\;\delta _{k_{1},l_{1}}\left( 1+\frac{i}{\hbar }\left\{ 
\frac{\hbar ^{2}k_{1}^{2}}{2m}\right\} \delta t\right) \delta
_{q_{4},0}\;\delta _{k_{4},l_{4}}\left( 1+\frac{i}{\hbar }\left\{ \frac{%
\hbar ^{2}k_{4}^{2}}{2m}\right\} \delta t\right)  \nonumber \\
&&\times \delta _{q_{5},0}\;\delta _{k_{5},l_{5}}\left( 1-\frac{i}{\hbar }%
\left\{ \frac{\hbar ^{2}k_{5}^{2}}{2m}\right\} \delta t\right) \delta
_{q_{6},0}\;\delta _{k_{6},l_{6}}\left( 1-\frac{i}{\hbar }\left\{ \frac{%
\hbar ^{2}k_{6}^{2}}{2m}\right\} \delta t\right)  \nonumber \\
&&\times \delta _{q_{7},0}\;\delta _{k_{7},l_{7}}\left( 1-\frac{i}{\hbar }%
\left\{ \frac{\hbar ^{2}k_{7}^{2}}{2m}\right\} \delta t\right) \delta
_{q_{8},0}\;\delta _{k_{8},l_{8}}\left( 1-\frac{i}{\hbar }\left\{ \frac{%
\hbar ^{2}k_{8}^{2}}{2m}\right\} \delta t\right)  \nonumber \\
&&\times (+2\lambda ^{2})\,\delta _{q_{2},q_{3}}\,\delta
_{(k_{2}+q_{2}),l_{2}}\,\delta _{(k_{3}-q_{3}),l_{3}}\;\delta t  \nonumber \\
&\approx &(+2\lambda ^{2})\,  \nonumber \\
&&\times \delta _{q_{1},0}\;\delta _{k_{1},l_{1}}\,\delta _{q_{4},0}\;\delta
_{k_{4},l_{4}}\,\delta _{q_{5},0}\;\delta _{k_{5},l_{5}}\,\delta
_{q_{6},0}\;\delta _{k_{6},l_{6}}\,  \nonumber \\
&&\times \delta _{q_{7},0}\;\delta _{k_{7},l_{7}}\,\delta _{q_{8},0}\;\delta
_{k_{8},l_{8}}\;\,\delta _{q_{2},q_{3}}\,\delta
_{(k_{2}+q_{2}),l_{2}}\,\delta _{(k_{3}-q_{3}),l_{3}}\;\delta t  \nonumber \\
&&  \label{Eq.StochAver1212bbaa}
\end{eqnarray}%
\begin{eqnarray}
2121bbaa &:&\left[ 
\begin{array}{c}
F_{d,u}(k_{1},l_{1};q_{1},\delta t)F_{d,d}(k_{2},l_{2};q_{2},\delta
t)F_{u,u}(k_{3},l_{3};q_{3},\delta t)F_{u,d}(k_{4},l_{4};q_{4},\delta t) \\ 
\times \;F_{u,u}^{+}(k_{5},l_{5};q_{5},\delta
t)\,F_{u,u}^{+}(k_{6},l_{6};q_{6},\delta
t)\,F_{d,d}^{+}(k_{7},l_{7};q_{7},\delta
t)\,F_{d,d}^{+}(k_{8},l_{8};q_{8},\delta t)%
\end{array}%
\right] _{StochAver}  \nonumber \\
&=&\delta _{q_{2},0}\;\delta _{k_{2},l_{2}}\left( 1+\frac{i}{\hbar }\left\{ 
\frac{\hbar ^{2}k_{2}^{2}}{2m}\right\} \delta t\right) \delta
_{q_{3},0}\;\delta _{k_{3},l_{3}}\left( 1+\frac{i}{\hbar }\left\{ \frac{%
\hbar ^{2}k_{3}^{2}}{2m}\right\} \delta t\right)  \nonumber \\
&&\times \delta _{q_{5},0}\;\delta _{k_{5},l_{5}}\left( 1-\frac{i}{\hbar }%
\left\{ \frac{\hbar ^{2}k_{5}^{2}}{2m}\right\} \delta t\right) \delta
_{q_{6},0}\;\delta _{k_{6},l_{6}}\left( 1-\frac{i}{\hbar }\left\{ \frac{%
\hbar ^{2}k_{6}^{2}}{2m}\right\} \delta t\right)  \nonumber \\
&&\times \delta _{q_{7},0}\;\delta _{k_{7},l_{7}}\left( 1-\frac{i}{\hbar }%
\left\{ \frac{\hbar ^{2}k_{7}^{2}}{2m}\right\} \delta t\right) \delta
_{q_{8},0}\;\delta _{k_{8},l_{8}}\left( 1-\frac{i}{\hbar }\left\{ \frac{%
\hbar ^{2}k_{8}^{2}}{2m}\right\} \delta t\right)  \nonumber \\
&&\times (+2\lambda ^{2})\,\delta _{q_{1},q_{4}}\,\delta
_{(k_{1}+q_{1}),l_{1}}\,\delta _{(k_{4}-q_{4}),l_{4}}\;\delta t  \nonumber \\
&\approx &(+2\lambda ^{2})\,  \nonumber \\
&&\times \delta _{q_{2},0}\;\delta _{k_{2},l_{2}}\,\delta _{q_{3},0}\;\delta
_{k_{3},l_{3}}\,\delta _{q_{5},0}\;\delta _{k_{5},l_{5}}\,\delta
_{q_{6},0}\;\delta _{k_{6},l_{6}}  \nonumber \\
&&\times \,\delta _{q_{7},0}\;\delta _{k_{7},l_{7}}\,\delta
_{q_{8},0}\;\delta _{k_{8},l_{8}}\;\,\delta _{q_{1},q_{4}}\,\delta
_{(k_{1}+q_{1}),l_{1}}\,\delta _{(k_{4}-q_{4}),l_{4}}\;\delta t  \nonumber \\
&&  \label{Eq.StochAver2121bbaa}
\end{eqnarray}%
\begin{eqnarray}
2112bbaa &:&\left[ 
\begin{array}{c}
F_{d,u}(k_{1},l_{1};q_{1},\delta t)F_{d,d}(k_{2},l_{2};q_{2},\delta
t)F_{u,d}(k_{3},l_{3};q_{3},\delta t)F_{u,u}(k_{4},l_{4};q_{4},\delta t) \\ 
\times \;F_{u,u}^{+}(k_{5},l_{5};q_{5},\delta
t)\,F_{u,u}^{+}(k_{6},l_{6};q_{6},\delta
t)\,F_{d,d}^{+}(k_{7},l_{7};q_{7},\delta
t)\,F_{d,d}^{+}(k_{8},l_{8};q_{8},\delta t)%
\end{array}%
\right] _{StochAver}  \nonumber \\
&=&\delta _{q_{2},0}\;\delta _{k_{2},l_{2}}\left( 1+\frac{i}{\hbar }\left\{ 
\frac{\hbar ^{2}k_{2}^{2}}{2m}\right\} \delta t\right) \delta
_{q_{4},0}\;\delta _{k_{4},l_{4}}\left( 1+\frac{i}{\hbar }\left\{ \frac{%
\hbar ^{2}k_{4}^{2}}{2m}\right\} \delta t\right)  \nonumber \\
&&\times \delta _{q_{5},0}\;\delta _{k_{5},l_{5}}\left( 1-\frac{i}{\hbar }%
\left\{ \frac{\hbar ^{2}k_{5}^{2}}{2m}\right\} \delta t\right) \delta
_{q_{6},0}\;\delta _{k_{6},l_{6}}\left( 1-\frac{i}{\hbar }\left\{ \frac{%
\hbar ^{2}k_{6}^{2}}{2m}\right\} \delta t\right)  \nonumber \\
&&\times \delta _{q_{7},0}\;\delta _{k_{7},l_{7}}\left( 1-\frac{i}{\hbar }%
\left\{ \frac{\hbar ^{2}k_{7}^{2}}{2m}\right\} \delta t\right) \delta
_{q_{8},0}\;\delta _{k_{8},l_{8}}\left( 1-\frac{i}{\hbar }\left\{ \frac{%
\hbar ^{2}k_{8}^{2}}{2m}\right\} \delta t\right)  \nonumber \\
&&\times (+2\lambda ^{2})\,\delta _{q_{1},q_{3}}\,\delta
_{(k_{1}+q_{1}),l_{1}}\,\delta _{(k_{3}-q_{3}),l_{3}}\;\delta t  \nonumber \\
&\approx &(+2\lambda ^{2})\,  \nonumber \\
&&\times \delta _{q_{2},0}\;\delta _{k_{2},l_{2}}\,\delta _{q_{4},0}\;\delta
_{k_{4},l_{4}}\,\delta _{q_{5},0}\;\delta _{k_{5},l_{5}}\,\delta
_{q_{6},0}\;\delta _{k_{6},l_{6}}\,  \nonumber \\
&&\times \delta _{q_{7},0}\;\delta _{k_{7},l_{7}}\,\delta _{q_{8},0}\;\delta
_{k_{8},l_{8}}\;\,\delta _{q_{1},q_{3}}\,\delta
_{(k_{1}+q_{1}),l_{1}}\,\delta _{(k_{3}-q_{3}),l_{3}}\;\delta t  \nonumber \\
&&  \label{Eq.StochAver2112bbaa}
\end{eqnarray}%
where the Wiener increment properties in Eq.(\ref{Eq.PropsWienerIncrem}) has
been used and only terms linear in $\delta t$ have been retained.

Substituting the $9$ results in Eqs.(\ref{Eq.StochAver1122bbaa}), (\ref%
{Eq.StochAver1122abba}), .... ,(\ref{Eq.StochAver2112bbaa}) into Eq. (\ref%
{Eq.QuantumCorrelnFnGeneralB}) gives 
\begin{eqnarray}
&&\left[ 
\begin{array}{c}
\widetilde{\phi }_{d}(k_{1},t+\delta t)\,\widetilde{\phi }%
_{d}(k_{2},t+\delta t)\;\widetilde{\phi }_{u}(k_{3},t+\delta t)\,\widetilde{%
\phi }_{u}(k_{4},t+\delta t) \\ 
\times \;\widetilde{\phi }_{u}^{+}(k_{5},t+\delta t)\,\widetilde{\phi }%
_{u}^{+}(k_{6},t+\delta t)\widetilde{\phi }_{d}^{+}(k_{7},t+\delta t)\,%
\widetilde{\phi }_{d}^{+}(k_{8},t+\delta t)%
\end{array}%
\right] _{StochAver}  \nonumber \\
&=&\dsum\limits_{l_{1},q_{1}}\dsum\limits_{l_{2},q_{2}}\dsum%
\limits_{l_{3},q_{3}}\dsum\limits_{l_{4},q_{4}}\dsum\limits_{l_{5},q_{5}}%
\dsum\limits_{l_{6},q_{6}}\dsum\limits_{l_{7},q_{7}}\dsum%
\limits_{l_{8},q_{8}}  \nonumber \\
&&\times \left[ 
\begin{array}{c}
(\delta _{q_{1},0}\;\delta _{k_{1},l_{1}}\,\delta _{q_{2},0}\;\delta
_{k_{2},l_{2}}\,\delta _{q_{3},0}\;\delta _{k_{3},l_{3}}\,\delta
_{q_{4},0}\;\delta _{k_{4},l_{4}}\, \\ 
\times \delta _{q_{5},0}\;\delta _{k_{5},l_{5}}\,\delta _{q_{6},0}\;\delta
_{k_{6},l_{6}}\,\delta _{q_{7},0}\;\delta _{k_{7},l_{7}}\,\delta
_{q_{8},0}\;\delta _{k_{8},l_{8}}) \\ 
\times \left( 1+\frac{i}{\hbar }\left\{ \frac{\hbar ^{2}k_{1}^{2}}{2m}+\frac{%
\hbar ^{2}k_{2}^{2}}{2m}+\frac{\hbar ^{2}k_{3}^{2}}{2m}+\frac{\hbar
^{2}k_{4}^{2}}{2m}-\frac{\hbar ^{2}k_{5}^{2}}{2m}-\frac{\hbar ^{2}k_{6}^{2}}{%
2m}-\frac{\hbar ^{2}k_{7}^{2}}{2m}-\frac{\hbar ^{2}k_{8}^{2}}{2m}\right\}
\delta t\right) \\ 
\times \overline{\,\widetilde{\phi }_{d}(l_{1},t)\,\,\widetilde{\phi }%
_{d}(l_{2},t)\,\widetilde{\phi }_{u}(l_{3},t)\,\widetilde{\phi }%
_{u}(l_{4},t)\,\widetilde{\phi }_{d}^{+}(l_{5},t)\,\widetilde{\phi }%
_{u}^{+}(l_{6},t)\,\,\widetilde{\phi }_{u}^{+}(l_{7},t)\,\widetilde{\phi }%
_{d}^{+}(l_{8},t)} \\ 
-((2\lambda ^{2})\,\delta _{q_{1},0}\;\delta _{k_{1},l_{1}}\delta
_{q_{2},0}\;\delta _{k_{2},l_{2}}\delta _{q_{3},0}\;\delta
_{k_{3},l_{3}}\delta _{q_{4},0}\;\delta _{k_{4},l_{4}} \\ 
\times \delta _{q_{6},0}\;\delta _{k_{6},l_{6}}\delta _{q_{8},0}\;\delta
_{k_{8},l_{8}}\;\delta _{q_{5},q_{7}}\,\delta _{(k_{5}+q_{5}),l_{5}}\,\delta
_{(k_{7}-q_{7}),l_{7}}\;\delta t) \\ 
\times \overline{\,\widetilde{\phi }_{d}(l_{1},t)\,\,\widetilde{\phi }%
_{d}(l_{2},t)\,\widetilde{\phi }_{u}(l_{3},t)\,\widetilde{\phi }%
_{u}(l_{4},t)\,\widetilde{\phi }_{d}^{+}(l_{5},t)\,\widetilde{\phi }%
_{u}^{+}(l_{6},t)\,\,\widetilde{\phi }_{u}^{+}(l_{7},t)\,\widetilde{\phi }%
_{d}^{+}(l_{8},t)} \\ 
-((2\lambda ^{2})\,\delta _{q_{1},0}\;\delta _{k_{1},l_{1}}\delta
_{q_{2},0}\;\delta _{k_{2},l_{2}}\delta _{q_{3},0}\;\delta
_{k_{3},l_{3}}\delta _{q_{4},0}\;\delta _{k_{4},l_{4}} \\ 
\times \delta _{q_{6},0}\;\delta _{k_{6},l_{6}}\delta _{q_{7},0}\;\delta
_{k_{7},l_{7}}\;\delta _{q_{5},q_{8}}\,\delta _{(k_{5}+q_{5}),l_{5}}\,\delta
_{(k_{8}-q_{8}),l_{8}}\;\delta t) \\ 
\times \overline{\,\widetilde{\phi }_{d}(l_{1},t)\,\,\widetilde{\phi }%
_{d}(l_{2},t)\,\widetilde{\phi }_{u}(l_{3},t)\,\widetilde{\phi }%
_{u}(l_{4},t)\,\widetilde{\phi }_{d}^{+}(l_{5},t)\,\widetilde{\phi }%
_{u}^{+}(l_{6},t)\,\,\widetilde{\phi }_{d}^{+}(l_{7},t)\,\widetilde{\phi }%
_{u}^{+}(l_{8},t)} \\ 
-((2\lambda ^{2})\,\delta _{q_{1},0}\;\delta _{k_{1},l_{1}}\delta
_{q_{2},0}\;\delta _{k_{2},l_{2}}\delta _{q_{3},0}\;\delta
_{k_{3},l_{3}}\delta _{q_{4},0}\;\delta _{k_{4},l_{4}} \\ 
\times \delta _{q_{5},0}\;\delta _{k_{5},l_{5}}\delta _{q_{8},0}\;\delta
_{k_{8},l_{8}}\;\delta _{q_{6},q_{7}}\,\delta _{(k_{6}+q_{6}),l_{6}}\,\delta
_{(k_{7}-q_{7}),l_{7}}\;\delta t) \\ 
\times \overline{\,\widetilde{\phi }_{d}(l_{1},t)\,\,\widetilde{\phi }%
_{d}(l_{2},t)\,\widetilde{\phi }_{u}(l_{3},t)\,\widetilde{\phi }%
_{u}(l_{4},t)\,\widetilde{\phi }_{u}^{+}(l_{5},t)\,\widetilde{\phi }%
_{d}^{+}(l_{6},t)\,\,\widetilde{\phi }_{u}^{+}(l_{7},t)\,\widetilde{\phi }%
_{d}^{+}(l_{8},t)} \\ 
-((2\lambda ^{2})\,\delta _{q_{1},0}\;\delta _{k_{1},l_{1}}\delta
_{q_{2},0}\;\delta _{k_{2},l_{2}}\delta _{q_{3},0}\;\delta
_{k_{3},l_{3}}\delta _{q_{4},0}\;\delta _{k_{4},l_{4}} \\ 
\times \delta _{q_{5},0}\;\delta _{k_{5},l_{5}}\delta _{q_{7},0}\;\delta
_{k_{7},l_{7}}\;\delta _{q_{6},q_{8}}\,\delta _{(k_{6}+q_{6}),l_{6}}\,\delta
_{(k_{8}-q_{8}),l_{8}}\;\delta t) \\ 
\times \overline{\,\widetilde{\phi }_{d}(l_{1},t)\,\,\widetilde{\phi }%
_{d}(l_{2},t)\,\widetilde{\phi }_{u}(l_{3},t)\,\widetilde{\phi }%
_{u}(l_{4},t)\,\widetilde{\phi }_{u}^{+}(l_{5},t)\,\widetilde{\phi }%
_{d}^{+}(l_{6},t)\,\,\widetilde{\phi }_{d}^{+}(l_{7},t)\,\widetilde{\phi }%
_{u}^{+}(l_{8},t)} \\ 
+((2\lambda ^{2})\,\delta _{q_{1},0}\;\delta _{k_{1},l_{1}}\,\delta
_{q_{3},0}\;\delta _{k_{3},l_{3}}\,\delta _{q_{5},0}\;\delta
_{k_{5},l_{5}}\,\delta _{q_{6},0}\;\delta _{k_{6},l_{6}}\, \\ 
\times \delta _{q_{7},0}\;\delta _{k_{7},l_{7}}\,\delta _{q_{8},0}\;\delta
_{k_{8},l_{8}}\;\,\delta _{q_{2},q_{4}}\,\delta
_{(k_{2}+q_{2}),l_{2}}\,\delta _{(k_{4}-q_{4}),l_{4}}\;\delta t) \\ 
\times \overline{\,\widetilde{\phi }_{d}(l_{1},t)\,\,\widetilde{\phi }%
_{u}(l_{2},t)\,\widetilde{\phi }_{u}(l_{3},t)\,\widetilde{\phi }%
_{d}(l_{4},t)\,\widetilde{\phi }_{u}^{+}(l_{5},t)\,\widetilde{\phi }%
_{u}^{+}(l_{6},t)\,\,\widetilde{\phi }_{d}^{+}(l_{7},t)\,\widetilde{\phi }%
_{d}^{+}(l_{8},t)} \\ 
+((2\lambda ^{2})\,\delta _{q_{1},0}\;\delta _{k_{1},l_{1}}\,\delta
_{q_{4},0}\;\delta _{k_{4},l_{4}}\,\delta _{q_{5},0}\;\delta
_{k_{5},l_{5}}\,\delta _{q_{6},0}\;\delta _{k_{6},l_{6}}\, \\ 
\times \delta _{q_{7},0}\;\delta _{k_{7},l_{7}}\,\delta _{q_{8},0}\;\delta
_{k_{8},l_{8}}\;\,\delta _{q_{2},q_{3}}\,\delta
_{(k_{2}+q_{2}),l_{2}}\,\delta _{(k_{3}-q_{3}),l_{3}}\;\delta t) \\ 
\times \overline{\,\widetilde{\phi }_{d}(l_{1},t)\,\,\widetilde{\phi }%
_{u}(l_{2},t)\,\widetilde{\phi }_{d}(l_{3},t)\,\widetilde{\phi }%
_{u}(l_{4},t)\,\widetilde{\phi }_{u}^{+}(l_{5},t)\,\widetilde{\phi }%
_{u}^{+}(l_{6},t)\,\,\widetilde{\phi }_{d}^{+}(l_{7},t)\,\widetilde{\phi }%
_{d}^{+}(l_{8},t)} \\ 
+((2\lambda ^{2})\,\delta _{q_{2},0}\;\delta _{k_{2},l_{2}}\,\delta
_{q_{3},0}\;\delta _{k_{3},l_{3}}\,\delta _{q_{5},0}\;\delta
_{k_{5},l_{5}}\,\delta _{q_{6},0}\;\delta _{k_{6},l_{6}}\, \\ 
\times \delta _{q_{7},0}\;\delta _{k_{7},l_{7}}\,\delta _{q_{8},0}\;\delta
_{k_{8},l_{8}}\;\,\delta _{q_{1},q_{4}}\,\delta
_{(k_{1}+q_{1}),l_{1}}\,\delta _{(k_{4}-q_{4}),l_{4}}\;\delta t) \\ 
\times \overline{\,\widetilde{\phi }_{u}(l_{1},t)\,\,\widetilde{\phi }%
_{d}(l_{2},t)\,\widetilde{\phi }_{u}(l_{3},t)\,\widetilde{\phi }%
_{d}(l_{4},t)\,\widetilde{\phi }_{u}^{+}(l_{5},t)\,\widetilde{\phi }%
_{u}^{+}(l_{6},t)\,\,\widetilde{\phi }_{d}^{+}(l_{7},t)\,\widetilde{\phi }%
_{d}^{+}(l_{8},t)} \\ 
+((2\lambda ^{2})\,\delta _{q_{2},0}\;\delta _{k_{2},l_{2}}\,\delta
_{q_{4},0}\;\delta _{k_{4},l_{4}}\,\delta _{q_{5},0}\;\delta
_{k_{5},l_{5}}\,\delta _{q_{6},0}\;\delta _{k_{6},l_{6}}\, \\ 
\times \delta _{q_{7},0}\;\delta _{k_{7},l_{7}}\,\delta _{q_{8},0}\;\delta
_{k_{8},l_{8}}\;\,\delta _{q_{1},q_{3}}\,\delta
_{(k_{1}+q_{1}),l_{1}}\,\delta _{(k_{3}-q_{3}),l_{3}}\;\delta t) \\ 
\times \overline{\,\widetilde{\phi }_{u}(l_{1},t)\,\,\widetilde{\phi }%
_{d}(l_{2},t)\,\widetilde{\phi }_{d}(l_{3},t)\,\widetilde{\phi }%
_{u}(l_{4},t)\,\widetilde{\phi }_{u}^{+}(l_{5},t)\,\widetilde{\phi }%
_{u}^{+}(l_{6},t)\,\,\widetilde{\phi }_{d}^{+}(l_{7},t)\,\widetilde{\phi }%
_{d}^{+}(l_{8},t)}%
\end{array}%
\right]  \nonumber \\
&&  \label{Eq.QCFTwoCooperMtmTimeEvnResultA}
\end{eqnarray}%
and performing the sums over $%
q_{1},q_{2},q_{3},q_{4},q_{5},q_{6},q_{7},q_{8} $ gives 
\begin{eqnarray}
&&\left[ 
\begin{array}{c}
\widetilde{\phi }_{d}(k_{1},t+\delta t)\,\widetilde{\phi }%
_{d}(k_{2},t+\delta t)\;\widetilde{\phi }_{u}(k_{3},t+\delta t)\,\widetilde{%
\phi }_{u}(k_{4},t+\delta t) \\ 
\times \;\widetilde{\phi }_{u}^{+}(k_{5},t+\delta t)\,\widetilde{\phi }%
_{u}^{+}(k_{6},t+\delta t)\widetilde{\phi }_{d}^{+}(k_{7},t+\delta t)\,%
\widetilde{\phi }_{d}^{+}(k_{8},t+\delta t)%
\end{array}%
\right] _{StochAver}  \nonumber \\
&=&\dsum\limits_{l_{1}}\dsum\limits_{l_{2}}\dsum\limits_{l_{3}}\dsum%
\limits_{l_{4}}\dsum\limits_{l_{5}}\dsum\limits_{l_{6}}\dsum\limits_{l_{7}}%
\dsum\limits_{l_{8}}  \nonumber \\
&&\times \left[ 
\begin{array}{c}
(\delta _{k_{1},l_{1}}\,\delta _{k_{2},l_{2}}\,\delta _{k_{3},l_{3}}\,\delta
_{k_{4},l_{4}}\,\delta _{k_{5},l_{5}}\,\delta _{k_{6},l_{6}}\,\delta
_{k_{7},l_{7}}\,\delta _{k_{8},l_{8}}) \\ 
\times \left( 1+\frac{i}{\hbar }\left\{ \frac{\hbar ^{2}k_{1}^{2}}{2m}+\frac{%
\hbar ^{2}k_{2}^{2}}{2m}+\frac{\hbar ^{2}k_{3}^{2}}{2m}+\frac{\hbar
^{2}k_{4}^{2}}{2m}-\frac{\hbar ^{2}k_{5}^{2}}{2m}-\frac{\hbar ^{2}k_{6}^{2}}{%
2m}-\frac{\hbar ^{2}k_{7}^{2}}{2m}-\frac{\hbar ^{2}k_{8}^{2}}{2m}\right\}
\delta t\right) \\ 
\times \overline{\,\widetilde{\phi }_{d}(l_{1},t)\,\,\widetilde{\phi }%
_{d}(l_{2},t)\,\widetilde{\phi }_{u}(l_{3},t)\,\widetilde{\phi }%
_{u}(l_{4},t)\,\widetilde{\phi }_{u}^{+}(l_{5},t)\,\widetilde{\phi }%
_{u}^{+}(l_{6},t)\,\,\widetilde{\phi }_{d}^{+}(l_{7},t)\,\widetilde{\phi }%
_{d}^{+}(l_{8},t)} \\ 
+((-2\lambda ^{2})\,\delta _{k_{1},l_{1}}\delta _{k_{2},l_{2}}\delta
_{k_{3},l_{3}}\delta _{k_{4},l_{4}}\delta _{k_{6},l_{6}}\delta
_{k_{8},l_{8}}\;\delta _{(k_{5}+k_{7}),(l_{5}+l_{7})}\;\delta t) \\ 
\times \overline{\,\widetilde{\phi }_{d}(l_{1},t)\,\,\widetilde{\phi }%
_{d}(l_{2},t)\,\widetilde{\phi }_{u}(l_{3},t)\,\widetilde{\phi }%
_{u}(l_{4},t)\,\widetilde{\phi }_{d}^{+}(l_{5},t)\,\widetilde{\phi }%
_{u}^{+}(l_{6},t)\,\,\widetilde{\phi }_{u}^{+}(l_{7},t)\,\widetilde{\phi }%
_{d}^{+}(l_{8},t)} \\ 
+((-2\lambda ^{2})\,\delta _{k_{1},l_{1}}\delta _{k_{2},l_{2}}\delta
_{k_{3},l_{3}}\delta _{k_{4},l_{4}}\delta _{k_{6},l_{6}}\delta
_{k_{7},l_{7}}\;\delta _{(k_{5}+k_{8}),(l_{5}+l_{8})}\;\delta t) \\ 
\times \overline{\,\widetilde{\phi }_{d}(l_{1},t)\,\,\widetilde{\phi }%
_{d}(l_{2},t)\,\widetilde{\phi }_{u}(l_{3},t)\,\widetilde{\phi }%
_{u}(l_{4},t)\,\widetilde{\phi }_{d}^{+}(l_{5},t)\,\widetilde{\phi }%
_{u}^{+}(l_{6},t)\,\,\widetilde{\phi }_{d}^{+}(l_{7},t)\,\widetilde{\phi }%
_{u}^{+}(l_{8},t)} \\ 
+((-2\lambda ^{2})\,\delta _{k_{1},l_{1}}\delta _{k_{2},l_{2}}\delta
_{k_{3},l_{3}}\delta _{k_{4},l_{4}}\delta _{k_{5},l_{5}}\delta
_{k_{8},l_{8}}\;\delta _{(k_{6}+k_{7}),(l_{6}+l_{7})}\;\delta t) \\ 
\times \overline{\,\widetilde{\phi }_{d}(l_{1},t)\,\,\widetilde{\phi }%
_{d}(l_{2},t)\,\widetilde{\phi }_{u}(l_{3},t)\,\widetilde{\phi }%
_{u}(l_{4},t)\,\widetilde{\phi }_{u}^{+}(l_{5},t)\,\widetilde{\phi }%
_{d}^{+}(l_{6},t)\,\,\widetilde{\phi }_{u}^{+}(l_{7},t)\,\widetilde{\phi }%
_{d}^{+}(l_{8},t)} \\ 
+((-2\lambda ^{2})\,\delta _{k_{1},l_{1}}\delta _{k_{2},l_{2}}\delta
_{k_{3},l_{3}}\delta _{k_{4},l_{4}}\delta _{k_{5},l_{5}}\delta
_{k_{7},l_{7}}\;\delta _{(k_{6}+k_{8}),(l_{6}+l_{8})}\;\delta t) \\ 
\times \overline{\,\widetilde{\phi }_{d}(l_{1},t)\,\,\widetilde{\phi }%
_{d}(l_{2},t)\,\widetilde{\phi }_{u}(l_{3},t)\,\widetilde{\phi }%
_{u}(l_{4},t)\,\widetilde{\phi }_{u}^{+}(l_{5},t)\,\widetilde{\phi }%
_{d}^{+}(l_{6},t)\,\,\widetilde{\phi }_{d}^{+}(l_{7},t)\,\widetilde{\phi }%
_{u}^{+}(l_{8},t)} \\ 
+((+2\lambda ^{2})\,\delta _{k_{1},l_{1}}\,\delta _{k_{3},l_{3}}\,\delta
_{k_{5},l_{5}}\,\delta _{k_{6},l_{6}}\,\delta _{k_{7},l_{7}}\,\delta
_{k_{8},l_{8}}\;\delta _{(k_{2}+k_{4}),(l_{2}+l_{4})}\,\;\delta t) \\ 
\times \overline{\,\widetilde{\phi }_{d}(l_{1},t)\,\,\widetilde{\phi }%
_{u}(l_{2},t)\,\widetilde{\phi }_{u}(l_{3},t)\,\widetilde{\phi }%
_{d}(l_{4},t)\,\widetilde{\phi }_{u}^{+}(l_{5},t)\,\widetilde{\phi }%
_{u}^{+}(l_{6},t)\,\,\widetilde{\phi }_{d}^{+}(l_{7},t)\,\widetilde{\phi }%
_{d}^{+}(l_{8},t)} \\ 
+((+2\lambda ^{2})\,\delta _{k_{1},l_{1}}\,\delta _{k_{4},l_{4}}\,\delta
_{k_{5},l_{5}}\,\delta _{k_{6},l_{6}}\,\delta _{k_{7},l_{7}}\,\delta
_{k_{8},l_{8}}\;\,\delta _{(k_{2}+k_{3}),(l_{2}+l_{3})}\;\delta t) \\ 
\times \overline{\,\widetilde{\phi }_{d}(l_{1},t)\,\,\widetilde{\phi }%
_{u}(l_{2},t)\,\widetilde{\phi }_{d}(l_{3},t)\,\widetilde{\phi }%
_{u}(l_{4},t)\,\widetilde{\phi }_{u}^{+}(l_{5},t)\,\widetilde{\phi }%
_{u}^{+}(l_{6},t)\,\,\widetilde{\phi }_{d}^{+}(l_{7},t)\,\widetilde{\phi }%
_{d}^{+}(l_{8},t)} \\ 
+((+2\lambda ^{2})\,\delta _{k_{2},l_{2}}\,\delta _{k_{3},l_{3}}\,\delta
_{k_{5},l_{5}}\,\delta _{k_{6},l_{6}}\,\delta _{k_{7},l_{7}}\,\delta
_{k_{8},l_{8}}\;\,\delta _{(k_{1}+k_{4}),(l_{1}+l_{4})}\;\delta t) \\ 
\times \overline{\,\widetilde{\phi }_{u}(l_{1},t)\,\,\widetilde{\phi }%
_{d}(l_{2},t)\,\widetilde{\phi }_{u}(l_{3},t)\,\widetilde{\phi }%
_{d}(l_{4},t)\,\widetilde{\phi }_{u}^{+}(l_{5},t)\,\widetilde{\phi }%
_{u}^{+}(l_{6},t)\,\,\widetilde{\phi }_{d}^{+}(l_{7},t)\,\widetilde{\phi }%
_{d}^{+}(l_{8},t)} \\ 
+((+2\lambda ^{2})\,\delta _{k_{2},l_{2}}\,\delta _{k_{4},l_{4}}\,\delta
_{k_{5},l_{5}}\,\delta _{k_{6},l_{6}}\,\delta _{k_{7},l_{7}}\,\delta
_{k_{8},l_{8}}\;\,\delta _{(k_{1}+k_{3}),(l_{1}+l_{3})}\;\delta t) \\ 
\times \overline{\,\widetilde{\phi }_{u}(l_{1},t)\,\,\widetilde{\phi }%
_{d}(l_{2},t)\,\widetilde{\phi }_{d}(l_{3},t)\,\widetilde{\phi }%
_{u}(l_{4},t)\,\widetilde{\phi }_{u}^{+}(l_{5},t)\,\widetilde{\phi }%
_{u}^{+}(l_{6},t)\,\,\widetilde{\phi }_{d}^{+}(l_{7},t)\,\widetilde{\phi }%
_{d}^{+}(l_{8},t)}%
\end{array}%
\right]  \nonumber \\
&&  \label{Eq.QCFTwoCooperMtmTimeEvnResultB}
\end{eqnarray}%
where we have used $\dsum\limits_{q_{3}q_{4}}\delta
_{(k_{3}+q_{3}),l_{3}}\;\delta _{(k_{4}-q_{4}),l_{4}}\,\delta
_{q_{3},q_{4}}=\delta _{(k_{3}+k_{4}),(l_{3}+l_{4})}$ and $%
\dsum\limits_{q_{1}}\delta _{q_{1},0}=1$.

As can be seen, all the QCF involved contain two stochastic momentem fields
for each of $\widetilde{\phi }_{d},\widetilde{\phi }_{u},\widetilde{\phi }%
_{d}^{+},\widetilde{\phi }_{u}^{+}$.. so by re-arranging these Grassmann
fields using their anti-commutation properties and then re-ordering the
indices $l_{1},l_{2},...,l_{8}$ and introducing the notation $%
X(d\,k_{1},d\,k_{2},u\,k_{3},u%
\,k_{4},u^{+}k_{5},u^{+}k_{6},d^{+}k_{7},d^{+}k_{8})$ from Eq.(\ref%
{Eq.MtmStochCFTwoCooper}) we have 
\begin{eqnarray}
&&X(d\,k_{1},d\,k_{2},u\,k_{3},u%
\,k_{4},u^{+}k_{5},u^{+}k_{6},d^{+}k_{7},d^{+}k_{8})_{t+\delta t}  \nonumber
\\
&=&\dsum\limits_{l_{1}}\dsum\limits_{l_{2}}\dsum\limits_{l_{3}}\dsum%
\limits_{l_{4}}\dsum\limits_{l_{5}}\dsum\limits_{l_{6}}\dsum\limits_{l_{7}}%
\dsum\limits_{l_{8}}  \nonumber \\
&&\left[ 
\begin{array}{c}
(\delta _{k_{1},l_{1}}\,\delta _{k_{2},l_{2}}\,\delta _{k_{3},l_{3}}\,\delta
_{k_{4},l_{4}}\,\delta _{k_{5},l_{5}}\,\delta _{k_{6},l_{6}}\,\delta
_{k_{7},l_{7}}\,\delta _{k_{8},l_{8}}) \\ 
\times \left( 1+\frac{i}{\hbar }\left\{ \frac{\hbar ^{2}k_{1}^{2}}{2m}+\frac{%
\hbar ^{2}k_{2}^{2}}{2m}+\frac{\hbar ^{2}k_{3}^{2}}{2m}+\frac{\hbar
^{2}k_{4}^{2}}{2m}-\frac{\hbar ^{2}k_{5}^{2}}{2m}-\frac{\hbar ^{2}k_{6}^{2}}{%
2m}-\frac{\hbar ^{2}k_{7}^{2}}{2m}-\frac{\hbar ^{2}k_{8}^{2}}{2m}\right\}
\delta t\right) \\ 
\\ 
+(-2\lambda ^{2})\,\delta _{k_{1},l_{1}}\delta _{k_{2},l_{2}}\delta
_{k_{3},l_{3}}\delta _{k_{4},l_{4}}\,\delta t \\ 
\times \left( 
\begin{array}{c}
\delta _{k_{6},l_{5}}\delta _{k_{8},l_{8}}\;\delta
_{(k_{5}+k_{7}),(l_{6}+l_{7})}+\delta _{k_{6},l_{5}}\delta
_{k_{7},l_{7}}\;\delta _{(k_{5}+k_{8}),(l_{6}+l_{8})} \\ 
+\delta _{k_{5},l_{6}}\delta _{k_{8},l_{8}}\;\delta
_{(k_{6}+k_{7}),(l_{5}+l_{7})}+\delta _{k_{5},l_{6}}\delta
_{k_{7},l_{7}}\;\delta _{(k_{6}+k_{8}),(l_{5}+l_{8})}%
\end{array}%
\right) \\ 
\\ 
+(2\lambda ^{2})\,\delta _{k_{5},l_{5}}\,\delta _{k_{6},l_{6}}\,\delta
_{k_{7},l_{7}}\,\delta _{k_{8},l_{8}}\,\delta t \\ 
\times \left( 
\begin{array}{c}
\delta _{k_{1},l_{1}}\delta _{k_{3},l_{4}}\;\delta
_{(k_{2}+k_{4}),(l_{2}+l_{3})}+\delta _{k_{1},l_{1}}\delta
_{k_{4},l_{3}}\;\delta _{(k_{2}+k_{3}),(l_{2}+l_{4})} \\ 
+\delta _{k_{2},l_{2}}\delta _{k_{3},l_{4}}\;\delta
_{(k_{1}+k_{4}),(l_{1}+l_{3})}+\delta _{k_{2},l_{2}}\delta
_{k_{4},l_{3}}\;\delta _{(k_{1}+k_{3}),(l_{1}+l_{4})}%
\end{array}%
\right)%
\end{array}%
\right]  \nonumber \\
&&\times
X(d\,l_{1},d\,l_{2},u\,l_{3},u%
\,l_{4},u^{+}l_{5},u^{+}l_{6},d^{+}l_{7},d^{+}l_{8})_{t}  \nonumber \\
&&  \label{Eq.QCFOTwoPairsTempEvnFirstOrderChange}
\end{eqnarray}%
Thus the first order change in the QCF for two Cooper pairs depends linearly
on $\delta t$. \pagebreak


\begin{thebibliography}{99}
\bibitem{Chin16a} C.Chin, \textit{Nat. Sci. Rev. }\textbf{3}, 168 (2016).

\bibitem{Chin10a} C. Chin, R. Grimm, P. Julienne and E. Tiesinga\textit{,
Rev. Mod. Phys. }\textbf{82}, 1225 (2010).

\bibitem{Ketterle08a} W. Ketterle and M. W. Zwierlein, in \textit{Ultracold
Fermi Gases, Proc. International School of Physics, Varenna 2006, }(eds) M.
Inguscio, W. Ketterle and C. Salomon, (IOS Press, Amsterdam 2008).

\bibitem{Zwerger12a} W. Zwerger (ed), \textit{The BEC-BCS Crossover and the
Unitary Fermi Gas }(Springer Lecture Notes in Physics \textbf{836 }, 2012).

\bibitem{Randeria14a} M. Randeria and E. Taylor, \textit{Ann. Rev.
Condens.Matter Phys.} 5, 209 (2014).

\bibitem{Strinati18a} C. Strinati, P. Pieri, G. Ropke, PP. Schuck and M.
Urban, \textit{Phys. Rep. }\textbf{738}, 1 (2018).

\bibitem{Pitaevski03a} L. Pitaevski and S. Stringari, \textit{Bose-Einstein
Condensation }(Clarendon Press, Oxford, 2003).

\bibitem{Leggett80a} A. J. Leggett, In \textit{Modern Trends in the Theory
of Condensed Matter, }(eds) A.Pekalski \textit{et. al.,}Springer-Verlag,
Berlin 1980) p14.

\bibitem{Torma15a} P. T\"{o}rm\"{a}, in \textit{Quantum Gas Experiments
Exploring Many-Body States, Cold Atoms}, (eds) P. T\"{o}rm\"{a}, K.
Sengstock, (Imperial

College Press, London, 2015).

\bibitem{Strinati12a} C. Strinati, In W. Zwerger (ed), \textit{The BEC-BCS
Crossover and the Unitary Fermi Gas }(Springer Lecture Notes in Physics 
\textbf{836 }, 2012) p99..

\bibitem{Nozieres85a} P. Nozieres and S. Schmitt-Rink\textit{\ , J. Low
Temp. Phys. }\textbf{59}, 195 (1985).

\bibitem{Tajima17a} H. Tajima, P. van Wyk, R. Hanai, D. Kagamihara, D.
Inotani, M. Horikoshi and Y. Ohashi, \textit{Phys. Rev. A }\textbf{95},
043625 (2017).

\bibitem{Nishida07a} Y. Nishida and D. T. Son\textit{, Phys. Rev. A }\textbf{%
75}, 063617 (2007).

\bibitem{Mulkerin22a} B. C. Mulkerin, X.--C. Yao, Y. Ohashi, X.-J. Liu and
H. Hui, \textit{ArXiv 2201.04798 [cond-mat.quant-gas] }(2022).

\bibitem{Bulgac12a} A. Bulgac, M. Forbes\textit{\ }and P. Magierski,\textit{%
\ }In W. Zwerger (ed), \textit{The BEC-BCS Crossover and the Unitary Fermi
Gas }(Springer Lecture Notes in Physics \textbf{836 }, 2012) p305.

\bibitem{Pang16a} T. Pang, \textit{An Introduction to Monte-Carlo Methods }%
(IOP Concise Physics, IOP Publishers, Bristol,2016).

\bibitem{Richie20a} A. Richie-Halford, J. E. Drut and A. Bulgac, \textit{%
Phys. Rev. Letts }\textbf{125}, 060403 (2020).

\bibitem{Jensen20a} S. Jensen, C.N. Gilbreth and Y. Alhassid, \textit{Phys.
Rev. Letts }\textbf{124}, 090604 (2020).

\bibitem{Kohn99a} W.Kohn, \textit{Rev. Mod. Phys. }\textbf{71}, 1253 (1999).

\bibitem{Balian81a} R. Balian and M. Vereroni, \textit{Phys. Rev. Letts }%
\textbf{47}, 1353 (1981).

\bibitem{Hoinka17a} S. Hoinka, P. Dyke, M. G. Lingham, G. M. Braun and C. J.
Vale, \textit{Nat. Phys. }\textbf{13}, 943 (2017).

\bibitem{Horikoshi17a} M. Horikoshi, M. Koashi, H. Tajima, Y. Ohashi and M.
Kuwata-Gonokami, \textit{Phys. Rev. X }\textbf{7}, 041004 (2017).

\bibitem{GardinerZoller} C. W. Gardiner and P. Zoller, \textit{Quantum
Noise: A Handbook of Markovian and Non-Markovian Quantum Stochastic Methods
with Applications to Quantum Optics}, (Springer Series in Synergetics,
Springer, Berlin, 2004)

\bibitem{WallsMilburn} D. F. Walls and G. J. Milburn, \textit{Quantum
Optics, }(Springer, Berlin, 2008).

\bibitem{BarnettRadmore} S. M. Barnett and P. Radmore, \textit{Methods in
Theoretical Quantum Optics, }(Oxford University Press, Oxford, UK, 1997).

\bibitem{Dalton15a} B. J. Dalton, J. Jeffers and S. M. Barnett,.\textit{%
Phase Space Methods for Degenerate Quantum Gases }(Oxford University Press,
Oxford, UK, 2015).

\bibitem{DrummondGardiner} P. D. Drummond and C. W. Gardiner, \textit{J.
Phys. A: Math. Gen. }\textbf{13}, 2353 (1980).

\bibitem{CahillGlauber} K. E. Cahill and R. J. Glauber, \textit{Phys. Rev. A}
\textbf{59}, 1538 (1999).

\bibitem{Corney06a} J. F. Corney and P. D. Drummond, \textit{J. Phys. A:
Math. Gen. }\textbf{39}, 269 (2006).

\bibitem{Corney04a} J. F. Corney and P. D. Drummond, \textit{Phys. Rev.
Letts }\textbf{93}, 260401 (2004).

\bibitem{Rosales15a} L.E. C. Rosales-Zarate and P. D. Drummond, \textit{New
J. Phys. }\textbf{17}\textit{, }032002 (2015)\textit{. }

\bibitem{Joseph18a} R. R. Joseph, L. E. C. Rosales-Zarate and P. D.
Drummond, \textit{J. Phys. A: Math. Gen. }\textbf{51}, 245302 (2018).

\bibitem{Joseph21a} R. R. Joseph, P. D. Drummond and L. E. C.
Rosales-Zarate, \textit{Phys. Rev. A} \textbf{104}, 062208 (2021).

\bibitem{Plimak01a} L. Plimak, M. J. Collett and M. K. Olsen, \textit{Phys.
Rev. A} \textbf{64}, 063409 (2001).

\bibitem{Dalton16a} B. J. Dalton, J. Jeffers and S. M. Barnett,.\textit{Ann.
Phys.}, \textbf{370}, 12 (2016).

\bibitem{Dalton17a} B. J. Dalton, J. Jeffers and S. M. Barnett, \textit{Ann.
Phys. }\textbf{377}, 268 (2017).

\bibitem{Dalton20a} B. J. Dalton, J. Jeffers and S. M. Barnett, \textit{Ann.
Phys. }\textbf{422}, 168309 (2020).

\bibitem{Polyakov16a} E. A. Polyakov, \textit{Phys. Rev. A} \textbf{94},
062104 (2016).

\bibitem{Kidwani20a} N. M. Kidwani and B. J. Dalton, \textit{J. Phys. Comm. }%
\textbf{4}, 015015 (2020).

\bibitem{Takagi25} T. Takagi, \textit{Japan J. Math. }\textbf{1}, 83 (1925).
\end{thebibliography}
\end{document}